 \documentclass[review,preprint,3p,times]{elsarticle}

\usepackage{amssymb}
\usepackage{amsmath}
\usepackage{multirow}
\usepackage{makecell}
\usepackage{booktabs}

\usepackage{subfigure}
\usepackage{caption}
\DeclareCaptionLabelFormat{cont}{#1~#2\alph{ContinuedFloat}}
\usepackage{nomencl}
\makenomenclature
\usepackage{etoolbox}
\renewcommand\nomgroup[1]{%
  \item[\bfseries
  \ifstrequal{#1}{C}{\textit{For compressible fluid dynamics}}{%
  \ifstrequal{#1}{L}{\textit{For local neural operator}}{%
  \ifstrequal{#1}{W}{\textit{For weight initialization}}{}}}%
]}

\journal{arXiv}

\begin{document}

\begin{frontmatter}



\title{On the locality of local neural operator in learning fluid dynamics}


\author[label1]{Ximeng Ye\corref{equal}}
\ead{yeximeng@stu.xjtu.edu.cn}

\author[label2]{Hongyu Li$^{*,}$\corref{equal}}
\ead{lihongyu@stu.xjtu.edu.cn}
\author[label2]{Jingjie Huang}
\ead{hjj0415@stu.xjtu.edu.cn}

\author[label1]{Guoliang Qin$^{*,}$}
\ead{glqin@xjtu.edu.cn}

\cortext[mycorrespondingauthor]{Corresponding author}
\cortext[equal]{These authors contributed equally to this work.}

\affiliation[label1]{organization={School of Energy and Power Engineering, Xi’an Jiaotong University},
            city={Xi'an},
            postcode={710049},
            state={Shaanxi},
            country={China}}

\affiliation[label2]{organization={State Key Lab for Strength and Vibration of Mechanical Structures, Department of Engineering Mechanics, Xi’an Jiaotong University},
            city={Xi'an},
            postcode={710049},
            state={Shaanxi},
            country={China}}

\begin{abstract}
This paper launches a thorough discussion on the locality of local neural operator (LNO), which is the core that enables LNO great flexibility on varied computational domains in solving transient partial differential equations (PDEs).
We investigate the locality of LNO by looking into its receptive field and receptive range, carrying a main concern about how the locality acts in LNO training and applications.
In a large group of LNO training experiments for learning fluid dynamics, it is found that an initial receptive range compatible with the learning task is crucial for LNO to perform well.
On the one hand, an over-small receptive range is fatal and usually leads LNO to numerical oscillation; 
on the other hand, an over-large receptive range hinders LNO from achieving the best accuracy.
We deem rules found in this paper general when applying LNO to learn and solve transient PDEs in diverse fields. 
Practical examples of applying the pre-trained LNOs in flow prediction are presented to confirm the findings further. 
Overall, with the architecture properly designed with a compatible receptive range, the pre-trained LNO shows commendable accuracy and efficiency in solving practical cases. 
\end{abstract}




\end{frontmatter}


\section{Introduction\label{sec1}}
These years, emerging deep learning methods provide new options for numerical analysis of physics, such as computational fluid dynamics.
People concerned about this topic are now forging towards a common objective of achieving the possible revolution on fast and reliable numerical estimations for solutions of partial differential equations (PDEs) \cite{Higgins2021}.

Researchers are trying different routes to approach the destination. 
One kind of method is developed on conventional numerical solving frameworks. Neural networks provide flexible discretization or interpolation and thus accelerate the PDE-solving process \cite{Dmitrii2021,Bar-Sinai2019}. 
The other group of approaches attempts to fully substitute conventional solvers with neural networks, which could be bolder and more promising. 
One distinctive way is to use neural networks to approximate the solution function of PDEs directly. 
Specifically, the neural networks input the position and output the field values such as velocities, temperature, etc. 
The most representative method of this category is physics-informed neural network (PINN) \cite{Raissi2020,Raissi2019}.
They introduce prior knowledge of physics into neural networks and regard the training process as the PDE-solving process, showing an astonishing ability to overcome the curse of dimensionality \cite{Han2018}. 
There are quite a lot of studies continuously improving PINNs on data processing \cite{ChenZhao2021,Dabrowski2022}, differentiation for gradients \cite{Chiu2022,Navaneeth2022}, domain decomposition \cite{Jagtap2020,Kharazmi2021}, etc. 
Another approach is neural operator that learns the solution operator of PDEs, such as deep operator network (DeepONet) \cite{LuLu2021,Goswami2022a,He2023,Wang2022a} and Fourier neural operator (FNO) \cite{LiZongyi2020,LiZongyi2021}. 
Neural operators input known functions (initial physical fields, the distribution of physical parameters, etc.) and output desired physical fields, naturally making part of the conditions changeable (e.g., the boundary value or the initial condition in transient problems) and extending the reusable range of the pre-trained neural networks. 
It is reported that the pre-trained neural operator solves Navier-Stokes (N-S) equations more than hundreds of times faster than the conventional numerical methods \cite{LiZongyi2020}.
However, the problems to be solved are assumed in a certain computational domain same as the training samples. 
A newly proposed framework named local neural operator (LNO) \cite{LiHongyu2022} further lifts the limitation. 
The pre-trained LNOs are reusable in variable computational domains when approximating the time-marching operators of transient PDEs. 
Specifically, LNO is designed as a local and shift-invariant mapping unit. It handles various computational domains with different boundary conditions (BCs) by shifting and cooperating with specific boundary treatments. 
In this way, the reusable scope of pre-trained neural networks is extended to a practical level as conventional numerical schemes. 
The success of LNO relies highly on an essential characteristic, the ‘locality’, which is designed to imitate the local-related nature of physics. 
In this work, we take a closer look at the locality to monitor, understand, and explain how it acts in LNO learning and affects the performance of LNO.

Many transient partial differential equations, as the mathematical description for the physical laws of the real world, possess a local-related feature. 
For example, hyperbolic equations such as the wave equation and the convection equation are undoubtedly local, as the initial disturbance propagates at a finite speed \cite{Evans2010} that bounds finite domains of dependence and influence for a certain point within a finite time interval.
The other type of transient PDEs, the parabolic equations represented by the diffusion equation, have an infinite domain of dependence mathematically, yet its solution under a point source (also known as the heat kernel) is rapidly decreasing \cite{Stein2003}, which means the interaction between two far-apart points is weak and the influence mainly concentrates in the local area. 
In general, most transient PDEs describing the dynamics of physics are, to some extent, local, reminding us that this feature could and should be considered when designing a neural network for solving these PDEs.

There are many neural networks designed with the locality inborn. 
Originally, the primary idea of convolutional neural network (CNN) by Y. Lecun et al. \cite{Lecun1998} is the local receptive fields, i.e., the input and output values of one convolutional layer are only related locally.
As CNNs spread widely to a number of areas, there are lots of neural networks mainly comprised of convolutions with the locality. 
In the fields of numerical analysis of physics, for example, the CNN-based methods are developed in solving basic PDEs \cite{Winovich2019,Ren2022}, equations describing fluid dynamics \cite{DeAvilaBelbute-Peres2020,Guo2016,Lee2019}, equations for multi-physics problems \cite{Kim2020};
the CNN-based methods successfully predicted the fields of strain and stress for the deformation of solids, which also could be regarded as PDE solving \cite{He2023,YangZ2021,YangZ2021a}. 
However, the primary idea of CNN, the locality, is not distinctively raised and analyzed in these works. 
The recently appeared conception of LNO \cite{LiHongyu2022} proposed that the locality of neural networks is essential for approximating reusable time-marching operators.
However, they just accomplished experimental tests of the locally designed neural networks. 
Hence, a thorough analysis of the locality of neural networks in solving PDEs is still an open and valuable topic.

Intuitively, there should be a correspondence between the locality of neural operator and the locality of PDEs to properly use the priori information of physics in the learning process. 
This viewpoint reminds us that the architecture of LNO should be designed according to the locality of the task to be learned. This paper investigates the locality of LNO in an effort to answer the question:
\emph{what is the principle of designing LNOs for specific transient PDEs regarding their locality? }
To do this, we must first measure the locality with proper indicators. Then, we conduct plenty of experiments to train and validate LNOs with different architectures on exampled learning tasks of fluid dynamics to investigate how the locality acts in LNO learning and how it affects the feasibility and accuracy of LNOs approximating transient PDEs. 
In this process, the discoveries could be summarized as guidance for architecting LNOs, for not only learning fluid dynamics but also for learning local operators of any other transient system governed by PDEs.

This paper is organized as follows. 
In Section \ref{sec2}, the governing equations of fluid dynamics to be learned and the basic methodologies of LNO are introduced. 
Section \ref{sec3} starts from the concept of receptive field to define two measurements for the locality of LNO, i.e., the maximum receptive range and the effective receptive range. 
With the two measurements equipped, \ref{sec4} investigates how the locality changes during training and how it impacts the performance of LNO in learning fluid dynamics. 
The pre-trained LNOs are further applied to two numerical examples in Section \ref{sec5}.
Finally, the conclusions are drawn in Section \ref{sec6}.
A nomenclature of symbols used throughout this paper is provided before appendices for convenient reference.

\section{Preliminaries\label{sec2}}
\subsection{Equations for compressible fluid dynamics\label{sec2:1}}

This work considers a 2D compressible fluid flow, which is described by the continuity equation, the N-S equation, and the energy equation as
\begin{equation}
\left\{\begin{matrix}
 \frac{\partial \rho}{\partial t}+\nabla \cdot(\rho \boldsymbol{v})=0\\
 \frac{\partial \rho \boldsymbol{v}}{\partial t}+\nabla \cdot(\rho \boldsymbol{v} \boldsymbol{v})=-\nabla p+\nabla \cdot \boldsymbol{\tau}\\
\frac{\partial \rho E}{\partial t}+\nabla \cdot[(\rho E+p) \boldsymbol{v}]=\nabla \cdot(\boldsymbol{\tau} \cdot \boldsymbol{v})+\nabla \cdot \kappa \nabla T
\end{matrix}\right.
,\quad \textup{in}~\Omega.
\label{eq:NSequation}
\end{equation}
Here $\rho(\boldsymbol{x},t)\in \mathbb{R}^+$, 
$T(\boldsymbol{x},t)\in \mathbb{R}^+$, 
$\boldsymbol{v}(\boldsymbol{x},t)=\{v_x(\boldsymbol{x},t),v_y(\boldsymbol{x},t)\}\in\mathbb{R}^2$, 
$\boldsymbol{x}\in\Omega\subset\mathbb{R}^2$,
$t\in\mathbb{R}^+$, are respectively density, temperature, velocities, which are independent fields to be solved defined on the computational domain $\Omega$. 
$p(\boldsymbol{x},t)\in\mathbb{R}$ is the pressure, and $\boldsymbol{\tau}(\boldsymbol{x},t)\in\mathbb{R}^{2\times2}$ is the tensor of viscous stress as
\begin{equation}
p=\rho RT,\quad \textup{in}~\Omega,
\label{eq:equation_pressure}
\end{equation}
\begin{equation}
\boldsymbol{\tau}=-\frac{2}{3}\mu(\nabla\cdot\boldsymbol{v})\boldsymbol{I}+2\mu(\nabla\boldsymbol{v}+\nabla\boldsymbol{v}^T),\quad \textup{in}~\Omega.
\label{eq:equation_stress}
\end{equation}
$E(\boldsymbol{x},t)\in\mathbb{R}$ is the total energy:
\begin{equation}
E=\frac{|\boldsymbol{v}|^2}{2}+C_\mathrm{v} T,\quad \textup{in}~\Omega.
\label{eq:equation_energy}
\end{equation}
With $p,\boldsymbol{\tau},E$ in Eq. (\ref{eq:NSequation}) substituted by Eqs. (\ref{eq:equation_pressure}-\ref{eq:equation_energy}), there come out the governing equations describing the motion of compressible fluids with $\rho,T,\boldsymbol{v}$ as variables to be solved. 
The equations are parameterized by the viscosity $\mu$, the gas constant $R$, the heat capacity $C_\mathrm{v}$, and the thermal conductivity $\kappa$.

Usually, the equations of compressible fluid dynamics are presented in a dimensionless form with $\rho,T,\boldsymbol{v}$ substituted by $\rho/\rho^{\prime}$, $T/T^{\prime}$, $\boldsymbol{v}/v^{\prime}=\{v_x/v^{\prime},v_y/v^{\prime}\}$, respectively.
Here $\rho^{\prime},T^{\prime},v^{\prime}$ are the characteristic values of density, temperature, and velocity. 
The parameters are also transformed to the dimensionless form as the Reynolds number $Re$, the Mach number $Ma$, the Prandtl number $Pr$, and the specific heat ratio $\gamma$ by
\begin{equation}
\mu=\frac{\rho^{\prime}v^{\prime}d^{\prime}}{Re},\quad R=\frac{v^{\prime2}}{\gamma Ma^2T^{\prime}},\quad C_\mathrm{v}=\frac{R}{\gamma-1},\quad \kappa=\frac{\gamma\mu C_\mathrm{v}}{Pr},
\end{equation}
where $d^{\prime}$ is the characteristic length. Then, the equations for fluid dynamics herein are overall parameterized with $Re$ and $Ma$ while $Pr=0.72$ and $\gamma=1.4$ are fixed in this work.
To simplify the formulations, the denominators of the dimensionless variables are omitted, i.e., we denote the dimensionless density, temperature, velocity as $\rho,T,\boldsymbol{v}$ in the rest of this paper.
And for convenience, these variables are combined into the vector $\boldsymbol{u}=\{\rho,T,v_x,v_y \}$ and $\boldsymbol{u}(\boldsymbol{x},t)$ is abbreviated as $\boldsymbol{u}_t (\boldsymbol{x})$. 

Transient PDEs are usually solved in time series, that is, with a given initial condition $\boldsymbol{u}_0$ and a determined time interval $\Delta t$, calculate (whether explicitly or implicitly) to obtain $\boldsymbol{u}_{\Delta t}$.
This operation is repeated until the physical field $\boldsymbol{u}_t$ of any desired $t$ is obtained.
This solving process could be regarded as a time-marching operator with input function $\boldsymbol{u}_t$ and output function $\boldsymbol{u}_{t+\Delta t}$ which is determined by the equation (parameterized by $Re$ and $Ma$) and interval $\Delta t$. 
For the sake of convenience, in the rest of the paper, whenever a learning task or time-marching operator is mentioned, it corresponds to the equations parameterized by ($Re,Ma,\Delta t$).

\subsection{Range of dependence for fluid dynamics\label{sec2:2}}
Then, we evaluate the locality of the equations for describing fluid dynamics with the domain of dependence. 
We measure the size of the domain of dependence with the maximum distance between the related points. 
We name the distance as ‘range of dependence’. 
Within a small time interval $\Delta t$, we assume that the change in the speed of information propagation can be ignored. 
Then, the domain of dependence and the domain of influence for a fixed point are mirror images symmetric about the point.

The governing equations of fluid dynamics are split into inviscid and viscous steps as a simplification for the present analysis to separately measure the contribution of different terms, which is inspired by the conception of the projection method (or time-splitting method) \cite{Guermond2006,Shen2011}.
Removing the viscous terms ($\nabla\cdot\boldsymbol{\tau}$, $\nabla\cdot\left(\boldsymbol{\tau}\cdot\boldsymbol{v}\right)$) and heat conduction terms ($\nabla\cdot\kappa\nabla T$) from Eq. (\ref{eq:NSequation}) brings the equation for the inviscid step.
Then, removing the convection terms ($\nabla\cdot\left(\rho\boldsymbol{v}\right)$, $\nabla\cdot\left(\rho\boldsymbol{vv}\right)$, $\nabla\cdot\left[\left(\rho E+p\right)\boldsymbol{v}\right]$) and the pressure gradient ($\nabla p$) from Eq. (\ref{eq:NSequation}) carries the equation for the viscous step.

The equation of the inviscid step equals the governing equation for inviscid compressible flows. 
According to \cite{Kovasznay1953,Thompson1897,Kim2004}, there are three independent modes within the motion of the inviscid flow,
including the vorticity and entropy mode which propagate at the convective velocity $v$, and the sound wave mode that propagates at speed $v+c$ and $v-c$. Here $c$ is the speed of sound. 
It suggests that, in the inviscid step the information propagates at a speed no more than $|v|+|c|$, and within a finite time step $\Delta t$ the propagation distance is no more than $(|v|+|c|)\Delta t$. In other words, the range of dependence for the inviscid step is 
\begin{equation}
r_{\textup{inv}}(\Delta t)=(|v|+|c|)\Delta t.
\end{equation}
Next, we evaluate the range of dependence for the viscous step.
Set a disturbance of velocity $v_x=\delta (x)$ at the initial moment, then the solution of this problem reflects how strongly each position is influenced by the initial disturbance.
The problem is solved in \ref{secA2}. The solution is in an integral form and we evaluate concrete values numerically as shown in Figure \ref{fig:viscousdist}. 
It is seen that the solution decreases rapidly as the distance to $x=0$ increases, thus it allows us to define the range of dependence in a truncated way. 
Concretely, $r_{\textup{visc}}$ is defined as the distance where the magnitude falls to 1\% of the peak value:
\begin{equation}
v_x(r_{\textup{visc}}(\Delta t),\Delta t)=0.01v_x(0,\Delta t).
\end{equation}
With $r_{\textup{visc}}$ and $r_{\textup{inv}}$ known, the range of dependence for the viscous compressible flow can be estimated by the combination of $r_{\textup{visc}}$ and $r_{\textup{inv}}$ according to the error bound of projection method \cite{Guermond2006} that:
for any $\varepsilon>0$, there exists $\Delta t>0$ that
\begin{equation}
|r_{\textup{dep}}(\tau)-(r_{\textup{visc}}(\tau)+r_{\textup{inv}}(\tau))|<\varepsilon,\quad \forall\tau\leq\Delta t.
\label{eq:DEPrange}
\end{equation}
The proof is in \ref{secA3}.
Thus, we use $r_{\textup{visc}}+r_{\textup{inv}}$ as the estimation for $r_{\textup{dep}}$. 
Table \ref{tab:DEPrange} lists the value of $r_{\textup{visc}}$, $r_{\textup{inv}}$, and $r_{\textup{dep}}$ for different learning tasks ($Ma,Re,\Delta t$).
Here the convection speed $v$ is estimated as 1, and speed of sound is $c=v⁄Ma\approx 1⁄Ma$. 
It is found that $r_{\textup{dep}}$ is larger for smaller $Ma$ (larger speed of sound), smaller $Re$ (larger viscosity), and larger $\Delta t$ (larger time interval).

\begin{table}[!ht]
\footnotesize
\renewcommand{\arraystretch}{0.8}
    \centering
    \caption{$r_{\textup{visc}}$, $r_{\textup{inv}}$, and the range of dependence $r_{\textup{dep}}$ for compressible N-S equations with different $Ma,Re,\Delta t$.
    $\Delta x=1⁄64$ is the grid size of spatial discretization used in this work.\label{tab:DEPrange}}
    \begin{tabular*}{\textwidth}{@{\extracolsep{\fill}}ccccccc}
    \toprule
        $Ma$ & $Re$ & $\Delta t$ & $r_{\textup{inv}}$ & $r_{\textup{visc}}$ & $r_{\textup{dep}}$ & $r_{\textup{dep}}/\Delta x$ \\
        \midrule
        0.2 & 100  & 0.05 & 0.300 & 0.110 & 0.410 & 26.24  \\ 
        0.2 & 100  & 0.03 & 0.180 & 0.086 & 0.266 & 16.99  \\ 
        0.2 & 100  & 0.07 & 0.420 & 0.131 & 0.551 & 35.26  \\ 
        0.2 & 20  & 0.05 & 0.300 & 0.246 & 0.546 & 34.94  \\ 
        0.2 & 500  & 0.05 & 0.300 & 0.050 & 0.350 & 22.37  \\ 
        0.1 & 100  & 0.05 & 0.550 & 0.110 & 0.660 & 42.24  \\ 
        0.4 & 100  & 0.05 & 0.175 & 0.110 & 0.285 & 18.24  \\ 
        \bottomrule
    \end{tabular*}
\end{table}

\subsection{Learn time-marching operator with LNO\label{sec2:3}}
Here we introduce our deep learning methodology to learn the time-marching operator of transient PDEs with LNO, including the concept and definition of LNO, the architecture, the benchmark dataset, and the training and validation process.

\begin{figure}[htbp]%
\centering
\includegraphics[width=0.9\textwidth]{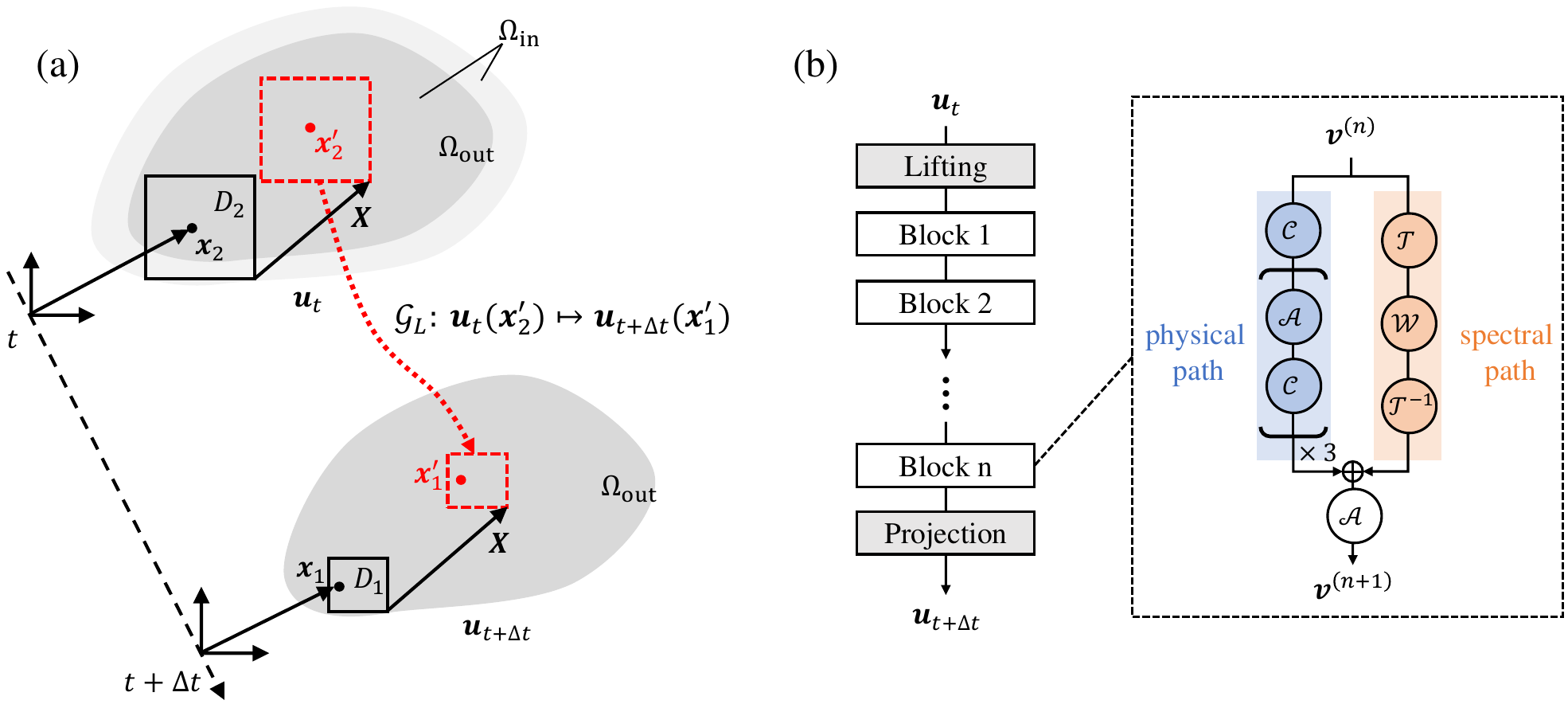}
\caption{LNO for learning time-marching operator. 
(a) The target time-marching operator $\mathcal{G}_\mathrm{L}$ maps function $\boldsymbol{u}_t$ on domain $\Omega_\textup{in}$ to $u_{t+\Delta t}$ on $\Omega_\textup{out}$.
$\Omega_\textup{out}$ and $\Omega_\textup{in}$ are variable.
(b) The architecture of LNO $\mathcal{G}_{\theta}$ for approximating $\mathcal{G}_\mathrm{L}$.}\label{fig:LNOconcept}
\end{figure}

We start with a brief review of the LNO conception. 
According to \cite{LiHongyu2022} as presented in Figure \ref{fig:LNOconcept}(a), LNO is defined as an approximation of the time-marching operator of transient PDEs as 
\begin{gather}
    \mathcal{G}_\mathrm{L}:\boldsymbol{u}_t(\boldsymbol{x}_2^{\prime})\mapsto \boldsymbol{u}_{t+\Delta t}(\boldsymbol{x}_1^{\prime}),\quad t\geq 0,\boldsymbol{x}_1^{\prime}\in\Omega_\textup{out},\boldsymbol{x}_\textup{in}^{\prime}\in\Omega_2, \\
\Omega_\textup{out}=\{\boldsymbol{x}_1+\boldsymbol{X}|\boldsymbol{x}_1\in D_1,\boldsymbol{X}\in\mathcal{X}\},\nonumber \\
\Omega_\textup{in}=\{\boldsymbol{x}_2+\boldsymbol{X}|\boldsymbol{x}_2\in D_2,\boldsymbol{X}\in\mathcal{X}\}.\nonumber
\end{gather}
$D_1$ is the unit output domain, $D_2$ is the input domain derived from $D_1$ with the local-related condition as 
\begin{equation}
\frac{\partial \boldsymbol{u}_{t+\Delta t}(\boldsymbol{x}_1)}{\partial \boldsymbol{u}_{t}(\boldsymbol{x}_2)}=0,\quad \forall \left \| \boldsymbol{x}_1-\boldsymbol{x}_2 \right \|>r. 
\label{eq:localrelate}
\end{equation}
$r$ is a finite positive number, and its minimum $r_{\textup{min}}$ indicates the maximal related-range of LNO.
$\mathcal{G}_\mathrm{L}$ is the target time-marching operator, which is approximated by LNO $\mathcal{G}_{\theta}$ with trainable weights $\theta$.
$\Omega_\textup{out}$, $\Omega_\textup{in}$ are respectively the output and input computational domain. $\boldsymbol{X}\in \mathcal{X}$ is the shifting vector. Note that $\Omega_\textup{out}$ and $\Omega_\textup{in}$ are variable as the set for shifting vectors $\mathcal{X} \subset \mathbb{R}^2$ is variable determined by the specific problem case to be solved. 

The present architecture of LNO is inherited from \cite{LiHongyu2022} with minor modifications, as shown in Figure \ref{fig:LNOconcept}(b). 
 It is comprised of three parts, the lifting layer as the head, the projection layers as the tail, and $n$ blocks as the main body. 
 The head and tail layers are comprised of point-wise operations $\mathcal{P}$ and activations $\mathcal{A}$.
Each block in the main body has two paths of operations.
 One is the physical path. 
 Four convolutional operations $\mathcal{C}$ sandwiched with activations $\mathcal{A}$ are designed to transform the interior tensors in the original physical space. 
 The other is the spectral path. The interior tensors on local subdomains are transformed in the spectral space with $(N-1)^{\textup{th}}$-order Legendre basis.
 Concretely, the interior tensors go through the Legendre transform $\mathcal{T}$, the linear layer $\mathcal{W}$, the inverse Legendre transform $\mathcal{T}^{-1}$. 
 Outputs from the physical and spectral paths are added together and activated by $\mathcal{A}$ as the output of one block.
 We adopt GELU activations \cite{Hendrycks2016} throughout the architecture.
 Please refer to \ref{secA4} for detailed formulas of $\mathcal{A},\mathcal{P},\mathcal{C},\mathcal{T},\mathcal{W},\mathcal{T}^{-1}$and how they compose LNO $\mathcal{G}_{\theta}$.
 Operations $\mathcal{P}$, $\mathcal{C}$, and $\mathcal{W}$ in the architecture are with specific and independent learnable weights.
 Overall, one LNO instance includes two parts: the architecture (determined by hyper-parameters $n,N,K,M$ respectively denoting the number of blocks, the width of local spectral transform, the number of repetitions for geometry decomposition, the number of modes adopted in spectral space) and a group of specific weights (whether they are randomly initialized or well-optimized).

The investigations in this work all lay on a uniform LNO training and validation process with uniform datasets as follows.
The 2-D case with Cartesian discretization of equidistant grid size $\Delta x=1/64$ is considered here.
For an LNO parameterized by ($n,N,K,M$), according to the operations defined in \ref{secA4}, its minimum unit maps tensors of size $\left [\frac{n(K-1)N}{K}+\frac{N}{K}+\frac{n(K-1)N}{K}\right ]\times\left [\frac{n(K-1)N}{K}+\frac{N}{K}+\frac{n(K-1)N}{K}\right ]$ to tensors of $\frac{N}{K}\times \frac{N}{K}$. 
By translations, LNO can map tensors of $\left [\frac{n(K-1)N}{K}+A\frac{N}{K}+\frac{n(K-1)N}{K}\right ]\times\left [\frac{n(K-1)N}{K}+B\frac{N}{K}+\frac{n(K-1)N}{K}\right ]$ to tensors of $\left (A \frac{N}{K}\right )\times\left (B \frac{N}{K}\right )$, where $A,B\in \mathbb{N}^+$.
In this paper, the samples for training and validation are in a square domain expressed as $128\times 128$ matrices. 
With these settings, the specific steps of LNO prediction for training and validation are as follows. 
Firstly, the fields of size $128^2$ are extended to $\left (128+\frac{2n(K-1)}{K} N+C\right )^2$ as the input for LNO inference according to the boundary condition (the present samples are with periodic boundaries to simplify the boundary treatment as much as possible), where $C$ is the extra extending length to make sure the extended size is divisible by $\frac{N}{K}$.
Then, LNO infers and outputs fields of $(128+C)^2$. The output of fields of $128^2$ is obtained by removing the extra extended size $C$.
 In this way, LNO can smoothly do inference in training or validation with the size of domains holding invariant.

The following introduces the training and validation process of LNO, starting from the generation of data samples.
Learning tasks designed here are identified by the parameters ($Re,Ma,\Delta t$) of the time-marching operator of fluid dynamics as mentioned in Section \ref{sec2:1}.
Studies in this paper involve seven learning tasks. With ($Re,Ma,\Delta t$)=(100,0.2,0.05) as the baseline, learning tasks include variant $Re$ as \{20,100,500\}, variant $Ma$ as \{0.1,0.2,0.4\}, and variant $\Delta t$ as \{0.03,0.05,0.07\}. 
Data samples for all the tasks are free flow in a square domain with the size of $2=128\Delta x$ with periodic boundary condition around and are generated by random excitation of velocity:
\begin{equation}
\boldsymbol{u}_0=\left \{ 
\begin{matrix}
    1\\
    1\\
    0.6[\sin\pi x,\sin 2\pi x,\cos \pi x,\cos 2\pi x]\mathbf{\Lambda}_1[\sin \pi y,\sin 2\pi y,\cos \pi y,\cos 2\pi y]^T\\
    0.6[\sin\pi x,\sin 2\pi x,\cos \pi x,\cos 2\pi x]\mathbf{\Lambda}_2[\sin \pi y,\sin 2\pi y,\cos \pi y,\cos 2\pi y]^T
\end{matrix}
 \right \}^T,
 \label{eq:randomIC}
\end{equation}
where $\mathbf{\Lambda}_k=\{\lambda_{k,ij}\}$  ($i,j=1\sim 4,k=1\sim 2$),$\lambda_{k,ij}\sim N(0,1)$;
the factor 0.6 is to limit the magnitude of velocities.
The samples are calculated numerically by discontinuous Galerkin FEM with Runge-Kutta scheme.
Starting from Eq. (\ref{eq:randomIC}), beforehand calculate for a duration of 0.1, the physical fields at this moment are used as the initial condition of data samples. 
Then the fields are calculated until $t=5$ and recorded with time interval $\Delta \tau=0.01$ as one piece of data sample $\{\boldsymbol{u}_{i\Delta \tau}|i\in\mathbb{N},i\leq\frac{5}{\Delta\tau}\}$. 
For each learning task, there is a total of 225 pieces of samples, where 200 of them are for training and the other 25 are for validation. 
The task of ($Re,Ma,\Delta t$)=(20,0.2,0.05) is special since its great viscosity dissipates the disturbance much faster, thus each piece of sample ends at $t=2.5$ and the number of sample pieces is doubled. 

LNOs are trained under the supervision of 200 pieces of samples.
In each training iteration with one series of sampled data $\{\boldsymbol{u}_{t+i\Delta t}|i=0,1,…,rounds\}$ extracted according to a random $t$, LNO $\mathcal{G}_{\theta}$ is trained to minimize the loss function $\mathcal{L}$ defined as
\begin{equation}
\mathcal{L}=\frac{1}{rounds}\sum_{i=1}^{rounds}\left \| \boldsymbol{u}_{t+i\Delta t}-\tilde{\boldsymbol{u}}_{t+i\Delta t} \right \|_2, 
\label{eq:loss}
\end{equation}
where $\boldsymbol{u}_{t+i\Delta t}$ and $\tilde{\boldsymbol{u}}_{t+i\Delta t}$ are both in a discretized form of being in $\mathbb{R}^{4\times 128^2}$, and
\begin{equation}
\tilde{\boldsymbol{u}}_{t+i\Delta t}=\left \{
\begin{matrix}
    \mathcal{G}_{\theta}(\boldsymbol{u}_t),\quad i=1\\
    \mathcal{G}_{\theta}\left (\tilde{\boldsymbol{u}}_{t+(i-1)\Delta t}\right ),\quad i>1
\end{matrix}
\right . .
\end{equation}
This loss function makes the present training scheme similar to that in training recurrent neural networks (RNNs), which benefits the stability of LNO when using the pre-trained models to do long-term recurrent time marching.
$rounds\geq 1$ is the number of recurrent rounds of LNO inferences. 
It is set as 10 in the present study.
In each round, the input should be extended before LNO inference to maintain the size of the output. 
As the samples are flows in a periodic domain, it provides great convenience to extend as wished.

The accuracy of trained LNOs is evaluated by the mean $L_2$ error in predicting solutions of 25 validation samples as 
\begin{equation}
e_t^{\rho}=\frac{1}{25\times 128^2}\sum_{i=1}^{25}\sum_{a=1}^{128^2}|\rho^{(i)}-\tilde{\rho}^{(i)}|_{t,\boldsymbol{x}_a},
\label{eq:error_rho}
\end{equation}
\begin{equation}
e_t^{T}=\frac{1}{25\times 128^2}\sum_{i=1}^{25}\sum_{a=1}^{128^2}|T^{(i)}-\tilde{T}^{(i)}|_{t,\boldsymbol{x}_a},
\label{eq:error_T}
\end{equation}
\begin{equation}
e_t^{v}=\frac{1}{25\times 128^2}\sqrt{(v_x^{(i)}-\tilde{v}_x^{(i)})^2_{t,\boldsymbol{x}_a}+(v_y^{(i)}-\tilde{v}_y^{(i)})^2_{t,\boldsymbol{x}_a}},
\label{eq:error_v}
\end{equation}
where $\rho,T,v_x,v_y$ are the density, temperature, and velocities from the samples;
and the tilde denotes the predicted value by LNO; 
the subscript `$t,x_a$' denote the velocity at time $t$ and position $\boldsymbol{x}_a$; 
the superscript `$(i)$' denotes the $i^\textup{th}$ sample for validation.
Here the errors of density, temperature, and velocity are calculated separately as they have different physical meanings.

To elevate the training performance of LNO, there are two worth-mentioning techniques: variable normalization and weight initialization. 
All the experiments of LNO training in this paper are based on the two techniques. 
Readers refer to \ref{secA5} for more details. 

There is additional information about LNO training to ensure the reproducibility of the present study. 
We use Adam \cite{Kingma2015} as the optimizer with an initial learning rate of 0.001.
The learning rate is manually multiplied by 0.7 every 20 epochs.
Each training course costs 200 epochs with 500 iterations in each epoch, and data samples are organized by bootstrap that, the field series are extracted from sample pieces according to a random $t$.
All the training, validation, and applications in this paper are built upon PyTorch and implemented on one NVIDIA GeForce RTX 2080ti GPU.
The code accompanying this paper is available on GitHub at https://github.com/PPhub-hy/torch-lno-compressible-fluid-dynamics.

\section{Measurement for the locality of LNO\label{sec3}}
How to measure the locality of LNO is the first thing to determine.
This section starts from a basic related concept of the \emph{receptive field} and then raises two measurements, including the \emph{maximum receptive range} and the \emph{effective receptive range}.
The analysis here only depends on LNO itself, thus the measurements are universal for learning not only fluid dynamics but also any other transient PDEs.

\subsection{Receptive field\label{sec3:1}}
For deep CNNs, Ref. \cite{Luo2016} uses receptive fields defined as $\frac{\partial z^l_{0,0}}{\partial z^0_{i,j}}$ to measure how much the input $z_{i,j}^0$ (pixel value in layer 0 at $(i,j)$) contributes to the output $z_{0,0}^l$ (pixel value in layer $l$ at $(0,0)$). 
We extend this expression into a continuous conception for LNO.
The following definition considers $u_t\in\mathbb{R}$ in 1-D space for simplification, while the following definition is easy to generalize to cases of higher dimensions. 
Then, we define the relevance between the input function $u_t$ at $x_2$ to the output function $\tilde{u}_{t+\Delta t}$ at $x_1$ as
\begin{gather}
\label{eq:Fdef}
    F(x_1,x_2)\overset{\textup{def}}{=}\left |\frac{\partial \tilde{u}_{t+\Delta t}(x_1)}{\partial u_t(x_2)}\right |, \quad x_1\in D_1, x_2\in D_2, \\
\textup{where} \quad \tilde{u}_{t+\Delta t}(x_1)=\mathcal{G}_{\theta}(u_t)(x_1).\nonumber
\end{gather}
$D_1$ and $D_2$ are respectively the output and input unit domain. For non-linear operators, $\frac{\partial \tilde{u}_{t+\Delta t}(x_1)}{\partial u_t(x_2)}$ is related to the variable input function $u_t$ which should be regarded as random variable. 
Besides, the initialized weights for LNOs before training are random variables.
Thus, with standard assumptions of these random variables (zero mean and symmetric distribution around 0), we transform $F(x_1,x_2)$ in Eq. (\ref{eq:Fdef}) to a statistical form as 
\begin{equation}
F(x_1,x_2)=\sqrt{\mathrm{Var}\left (\frac{\partial \tilde{u}_{t+\Delta t}(x_1)}{\partial u_{t}(x_2)}\right )}.
\label{eq:Fstatistic}
\end{equation}
Eq. (\ref{eq:Fstatistic}) represents multiple receptive fields regarding $x_2$ for output points $x_1$ in $D_1$.
To simplify the expression, With $x=x_2-x_1$, $x\in D_x$, we sum up $F(x_1,x_2)$ for $x_1$ in $D_1$ and normalize it in $D_x$ as
\begin{equation}
\bar{F}_0(x)=\frac{\bar{F}(x)}{\int_{D_x}\bar{F}(\xi)\mathrm{d}\xi},\quad \textup{where}\quad \bar{F}(x)=\int_{D_1}F(x_1,x+x_1)\mathrm{d}x_1.
\label{eq:F_bardef}
\end{equation}
In practice, usually the input and output fields have more than one channel, e.g., the task of fluid dynamics requires 4 channels ($\rho,T,v_x,v_y$).
In this case, $\bar{F}(x)$ is a field of vectors with 16 components. 
As what we are concerned about is the amplitude distribution of the relevance, $\bar{F}(x)$ is pointwise averaged to a scalar field. 

Thereby, we obtain the receptive field $\bar{F}(x)$ and its normalization $\bar{F}_0(x)$ for describing how LNO locally links the input and output functions. 
Next, two scalars are raised to describe the size and concentration level of the receptive field of LNO.

\subsection{Maximum receptive range\label{sec3:2}}
One indicator for measuring the range of the locality of LNO is the maximum related distance of the receptive field $\bar{F}(x),x\in D_x$ as \emph{maximum receptive range} (MRR),
\begin{equation}
r_{\textup{LNO}}=\max_{x\in D_x}\left \|x\right \|=\max_{x_1\in D_x,x_2\in D_2}\left \|x_1-x_2\right \|.
\label{eq:MRRdef}
\end{equation}
The infinite norm is used for $\|\cdot\|$ in this paper. 
The present $r_{\textup{LNO}}$ is essential for LNO as it equals to the minimal of $r$ in Eq. (\ref{eq:localrelate}) which is the original local-related condition for LNO definition.
In practice, $r_{\textup{LNO}}$ comes along with the LNO architecture and does not change in the training process.

\subsection{Effective receptive range\label{sec3:3}}
To further measure the concentration level of the local-related intensity of LNO, we define \emph{effective receptive range} (ERR) as the standard second central moment of $\bar{F}_0(x)$ that
\begin{equation}
\varrho(\bar{F}_0)=\sqrt{\int_{D_x}(x-\bar{x})^2\bar{F}_0(x)dx},\quad \textup{where}\quad \bar{x}=\int_{D_x}x\bar{F}_0(x)dx.
\label{eq:ERRdef}
\end{equation}
According to Eq. (\ref{eq:ERRdef}), smaller $\varrho$ means that the related intensity in the receptive field is more centering, indicating a smaller effective local-related range of LNO, and vice versa.
Note that the value of $\bar{F}_0(x)$ is relevant to not only the LNO architecture but also the concrete weights, i.e., $\varrho$ alters during training. 
We separately denote ERR for an LNO with randomly initialized weights as $\varrho_{\textup{init}}$ and that of a trained LNO as $\varrho_{\textup{trained}}$.

\subsection{A simple case of LNO with one block\label{sec3:4}}
As an example, this subsection analysis the receptive field (Eq. (\ref{eq:F_bardef})), MRR (Eq. (\ref{eq:MRRdef})) and ERR (Eq. (\ref{eq:ERRdef})) of the 2-D one-block LNO with randomly initialized weights.
Some of the components in LNO do not affect the receptive field. 
One is the non-linear activation function.
We simplify the GELU activations to ReLU.
With two standard assumptions on the input function (it has zero mean and symmetric distribution around 0),
the ReLU activation function changes the variance of the output by a constant factor $\frac{1}{2}$ \cite{He2015} which does not affect the shape of the receptive field. 
The other is the pointwise operations in the lifting and projection layers. 
These pointwise linear operations do not affect the shape of the receptive field.
Single or multiple channels of operations do not affect the shape of the receptive field as well, thus we focus on operations in the block with a single channel.

MRR of LNO is determined by the spectral path and the physical path together. 
MRR of the physical path $r_\textup{phy}=\frac{k-1}{2}l\Delta x=4\Delta x$, where $k=3$ is the kernel size and $l=4$ is the number of layers. 
For the spectral path, functions in one square subdomain with width $N\Delta x$ are completely related as they are sent together to the spectral transform and the follow-up operations. 
Thus, the maximum receptive range is $r_\textup{spect}=N\Delta x$. 
As the physical path and the spectral path are parallelly connected, the overall MRR is the largest MRR of these paths as it covers that of the other. 
Thus, MRR of one-block LNO is 
\begin{equation}
r_{n=1}=\max(r_{\textup{phy},},r_\textup{spect})=\max(4\Delta x,N\Delta x).
\end{equation}
The specific expression of the receptive field $\bar{F}$ is required to calculate the initial ERR $\varrho_\textup{init}$.
For the present 2-D discretized case, let $v_{ij}^{(1)}$ denote field value at $\left(x_2^{(i)},y_2^{(j)}\right)=(i,j)\Delta x$ in the input domain, where $i,j=1,2,…,\frac{(2K-1)N}{K}$, 
and $v_{ab}^{(2)}$ denote field value at $\left(x_1^{(a)},y_1^{(b)}\right)=\left (\frac{(K-1)N}{K}+a,\frac{(K-1)N}{K}+b\right )\Delta x$ in the output domain, where $a,b=1,2,…,\frac{N}{K}$.
The relative position is $\left(x^{(\alpha)},y^{(\beta)}\right)=\left(x_2^{(i)}-x_1^{(a)},y_2^{(j)}-y_1^{(b)}\right)=\left (i-a-\frac{(K-1)N}{K},\notag \right. \left. j-b-\frac{(K-1)N}{K}\right )\Delta x=(\alpha,\beta)\Delta x$, i.e., $(i,j)=\left(\alpha+a+\frac{(K-1)N}{K},\beta+b+\frac{(K-1)N}{K}\right)$ where $\alpha,\beta=-N+1,…,N-1$. Then, according to Eqs.~(\ref{eq:Fstatistic})~(\ref{eq:F_bardef}) the receptive field is
\begin{equation}
    \bar{F}(\alpha,\beta)=\sum_{a,b}\sqrt{\mathrm{Var}\left(\frac{\partial v_{ab}^{(2)}}{\partial v_{ij}^{(1)}}\right)}
    =\sum_{a,b}\sqrt{\mathrm{Var}\left(\frac{\partial v_{ab}^{(2)}}{\partial v^{(1)}_{\left(\alpha+a+\frac{(K-1)N}{K}\right)\left(\beta+b+\frac{(K-1)N}{K}\right)}}\right)}
    \label{eq:F_bar_oneblock1}
\end{equation}

The output of one block is the sum of the physical path and the spectral path that, $v_{ab}^{(2)}=\left[ v_{ab}^{(2)} \right]_\textup{phy}+\left[v_{ab}^{(2)}\right]_\textup{spect}$. 
As $\frac{\partial\left[v_{ab}^{\left(2\right)}\right]_\textup{phy}}{\partial v_{ij}^{\left(1\right)}}$ and $\frac{\partial\left[v_{ab}^{\left(2\right)}\right]_\textup{spect}}{\partial v_{ij}^{\left(1\right)}}$ are solely related to the weights in their own paths, they are independent of each other, then Eq. (\ref{eq:F_bar_oneblock1}) equals that 
\begin{equation}
\bar{F}\left(\alpha,\beta\right)=
\sum_{a,b}\sqrt{\mathrm{Var}\left(\frac{\partial\left[v_{ab}^{\left(2\right)}\right]_\textup{phy}}{\partial v_{\left(\alpha+a+\frac{\left(K-1\right)N}{K}\right)\left(\beta+b+\frac{\left(K-1\right)N}{K}\right)}^{\left(1\right)}}\right)
+\mathrm{Var}\left(\frac{\partial\left[v_{ab}^{\left(2\right)}\right]_\textup{spect}}{\partial v_{\left(\alpha+a+\frac{\left(K-1\right)N}{K}\right)\left(\beta+b+\frac{\left(K-1\right)N}{K}\right)}^{\left(1\right)}}\right)}.
\label{eq:F_bar_oneblock2}
\end{equation}
Since the physical path is comprised of stacked conventional discretized convolutional layers, according to Ref. \cite{Luo2016}, $\mathrm{Var}\left(\frac{\partial\left[v_{ab}^{\left(2\right)}\right]_\textup{phy}}{\partial v_{\left(\alpha+a+\frac{\left(K-1\right)N}{K}\right)\left(\beta+b+\frac{\left(K-1\right)N}{K}\right)}^{\left(1\right)}}\right)$ equals to $\mathrm{\Psi}_{\alpha\beta}$, 
which is the probability of $(\alpha,\beta)=(\sum_{n=1}^{4}{\alpha}_n,\sum_{n=1}^{4}{\beta}_n)$, where $\left\{\alpha_n\right\}, \left\{\beta_n\right\}$ are independent identically distributed (i.i.d.) random variables following a discrete uniform distribution taking values in $\{-1,0,1\}$. 
For the latter term of the spectral path, with the formula (\ref{eq:C7}) in \ref{secA4}, and the randomly initialized weight ${\bar{W}}_{{mm}^\prime}$ that have $\mathrm{Var}\left({\bar{W}}_{{mm}^\prime}\right)=\frac{\mathrm{\Theta}_{N,K,M}}{3M^2}$ according to \ref{secA5},
\begin{equation}
\begin{aligned}
\mathrm{Var}\left(\frac{\partial\left[v_{ab}^{\left(2\right)}\right]_\textup{spect}}{\partial v_{\left(\alpha+a+\frac{\left(K-1\right)N}{K}\right)\left(\beta+b+\frac{\left(K-1\right)N}{K}\right)}^{\left(1\right)}}\right)
&=\mathrm{Var}\left(\frac{1}{K^2}\sum_{m=1}^{M^2}\sum_{m^\prime=1}^{M^2}\sum_{p=0}^{K-1}\sum_{q=0}^{K-1}{\psi_{m\left(a-\frac{p}{K}N\right)\left(b-\frac{q}{K}N\right)}\varphi_{m^\prime\left(\alpha+a+\frac{\left(K-1-p\right)N}{K}\right)\left(\beta+b+\frac{\left(K-1-q\right)N}{K}\right)}{\bar{W}}_{{mm}^\prime}}\right)\\
&=\frac{\mathrm{\Theta}_{N,K,M}}{3M^2K^4}\sum_{m^\prime=1}^{M^2}\sum_{m=1}^{M^2}\left(\sum_{p=0}^{K-1}\sum_{q=0}^{K-1}{\psi_{m\left(a-\frac{p}{K}N\right)\left(b-\frac{q}{K}N\right)}\varphi_{m^\prime\left(\alpha+a+\frac{\left(K-1-p\right)N}{K}\right)\left(\beta+b+\frac{\left(K-1-q\right)N}{K}\right)}}\right)^2
\end{aligned}
\end{equation}

\begin{figure}[h]%
\centering
\includegraphics[width=0.8\textwidth]{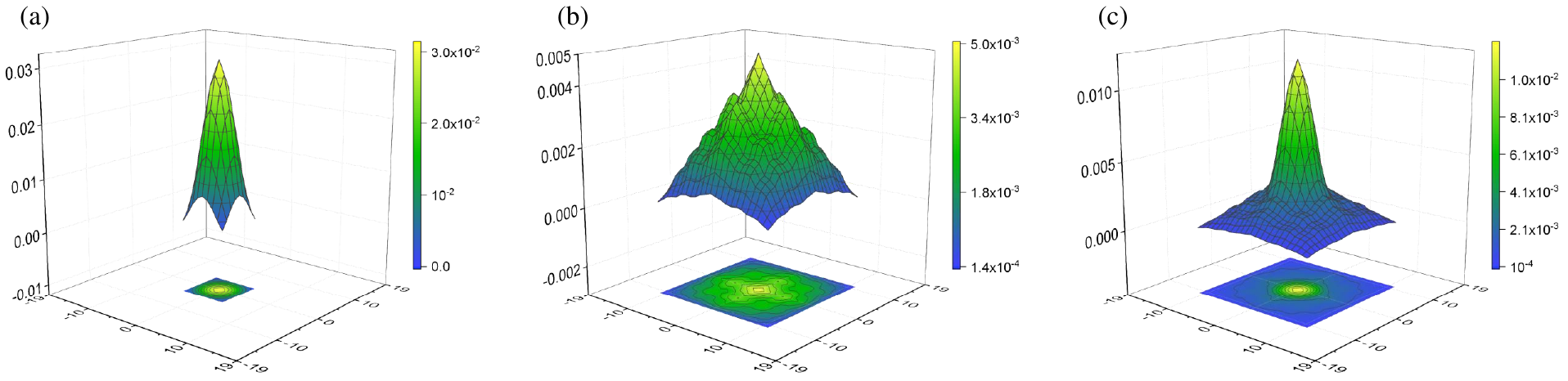}
\caption{Initial receptive field of components in one-block LNO ($N,K,M=12,2,6$). 
(a) Physical path; 
(b) Spectral path;
(c) One block with two paths.}\label{fig:receptivefield_path}
\end{figure}

\begin{figure}[h]%
\centering
\includegraphics[width=\textwidth]{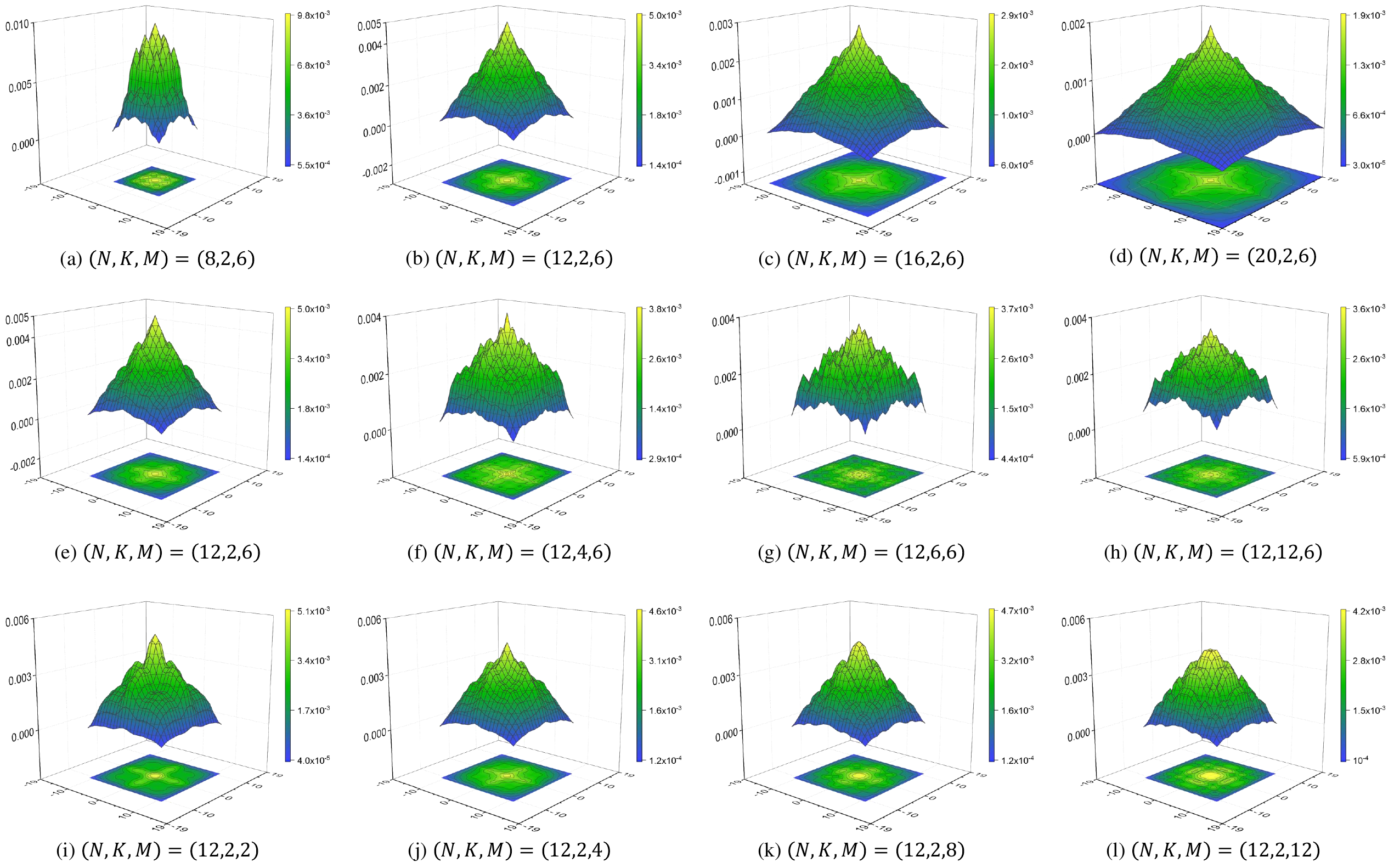}
\caption{The initial receptive field of one-block LNO with different $N,K,M$. 
The three rows show the change of $N,K,M$, respectively.}\label{fig:receptivefield_parameter}
\end{figure}

\vspace{5mm}

To show how the two paths contribute to the overall local-related pattern between the input and output of LNO, Figure~\ref{fig:receptivefield_path} presents three receptive fields $\bar{F}_0$ that,
the first two are fields of solely the physical path or spectral path, and the last is the complete one for the one-block LNO ($N=12,K=2,M=6$). 
Going deeper into the spectral path, which is the distinguishing feature of the present LNO architecture, Figure \ref{fig:receptivefield_parameter} shows receptive fields $\bar{F}_0$ solely for the spectral path with different parameters. 
Three groups of the contours respectively present how the receptive field change with $N$, $K$, and $M$. 
In these figures, the local-related range grows larger as $N$ becomes bigger. 
Though $K$ and $M$ perform minor effects on the receptive range, they change the shape of the receptive field distinctively.
As the number of repetitions $K$ and the number of reserved modes $M$ are greater, the receptive field becomes elaborate. 
One may intuitively think that the elaborate receptive field is better, but in practice, the computational costs grow significantly with $K$ and $M$.
It brings out a trade-off between the performance and the computational costs, and we found it is fairly enough to set $K=2, M=6$ for practices herein. 
Finally, $\varrho_\textup{init}$ of the one-block LNO is obtained by Eq. (\ref{eq:ERRdef}) with the present receptive field $\bar{F}_0$ in Eq. (\ref{eq:F_bar_oneblock2}). 
Table \ref{tab:initialERR_block} lists $\varrho_\textup{init}$ of LNOs with variant $N, K, M$. 
It is consistent with the observation from Figure \ref{fig:receptivefield_parameter} that the width of local spectral transform $N$ affects $\varrho_\textup{init}$ primarily, while  $\varrho_\textup{init}$ is with minor relevance to $K$, $M$.

\begin{table}[]
\footnotesize
\renewcommand{\arraystretch}{0.8}
\centering
    \caption{Initial effective receptive range $\varrho_\textup{init}$ of LNO (the number of blocks $n=1$) with randomly initialized weights regarding the parameters $N$,$K$, and $M$ of LNO. ($\times\Delta x$)\label{tab:initialERR_block}}
\begin{tabular*}{\textwidth}{@{\extracolsep{\fill}}ccccccccc}
\toprule
\multirow{3}{*}{Variant $N$} & $(N,K,M)$ & (8,2,6)  & (10,2,6) & (12,2,6) & (14,2,6) & (16,2,6)  & (18,2,6)  & (20,2,6) \\
\cline{2-9}
                             & analysis  & 4.4608   & 5.7184   & 6.9181   & 8.1223   & 9.3708    & 10.6426   & 11.9264  \\
                             & estimated & 4.4616   & 5.7180   & 6.9200   & 8.1200   & 9.3706    & 10.6368   & 11.9265  \\
                             \midrule
\multirow{3}{*}{Variant $K$} & $(N,K,M)$ & (12,2,6) & (12,3,6) & (12,4,6) & (12,6,6) & (12,12,6) &           &          \\
\cline{2-9}
                             & analysis  & 6.9181   & 7.4164   & 7.6349   & 7.8124   & 7.8719    &           &          \\
                             & estimated & 6.9200   & 7.4186   & 7.6378   & 7.8127   & 7.8702    &           &          \\
                             \midrule
\multirow{3}{*}{Variant $M$} & $(N,K,M)$ & (12,2,2) & (12,2,4) & (12,2,6) & (12,2,8) & (12,2,10) & (12,2,12) &          \\
\cline{2-9}
                             & analysis  & 6.9610   & 7.0230   & 6.9181   & 6.9019   & 6.8092    & 6.8093    &          \\
                             & estimated & 6.9588    & 7.0244   & 6.9200   & 6.9017   & 6.8123   & 6.8075    &    \\     
\bottomrule
\end{tabular*}
\end{table}

With the modern easy-to-use deep learning toolkit such as PyTorch, $\varrho_\textup{init}$ can also be estimated by experiment with the following steps. 
Let $\tilde{u}_{t+\Delta t}\left(a,b\right)=\mathcal{G}_{\theta}\left(u_t\left(i,j\right)\right)\left(a,b\right)$ where $\left(a,b\right)$ and $\left(i,j\right) $ are respectively the discretized positions in input and output domains.
Firstly, set $\tilde{u}_{t+\Delta t}(a,b)$ as the loss function $\mathcal{L}$, i.e., let $\frac{\partial \mathcal{L}}{\partial u_{t+\Delta t}(a,b)}=1$ and $\frac{\partial \mathcal{L}}{\partial u_{t+\Delta t}(a^{\prime},b^{\prime})}=0$,$\forall a^{\prime}\ne a,b^{\prime}\ne b$; 
then, back propagate the gradient to the input and get $\frac{\partial\mathcal{L}}{\partial u_t(i,j)}$, which equals to the desired partial derivative item $\frac{\partial \tilde{u}_{t+\Delta t}(a, b)}{\partial u_t(i, j)}$ because $\frac{\partial \mathcal{L}}{\partial u_t(i, j)}=\sum_{a^{\prime}} \sum_{b^{\prime}} \frac{\partial \mathcal{L}}{\tilde{u}_{t+\Delta t}\left(a^{\prime}, b^{\prime}\right)} \frac{\partial \tilde{u}_{t+\Delta t}\left(a^{\prime}, b^{\prime}\right)}{\partial u_t(i, j)}=\frac{\partial \tilde{u}_{t+\Delta t}(a, b)}{\partial u_t(i, j)} 
$. 
In this way, by repetitive sampling, we can estimate $\mathrm{Var}\left(\frac{\partial \tilde{u}_{t+\Delta t}(a,b)}{\partial u_t(i,j)}\right)$, the receptive field $\bar{F}$, ${\bar{F}}_0$, and $\varrho_\textup{init}$. The estimated $\varrho_\textup{init}$ of LNO with different parameters are listed in Table \ref{tab:initialERR_block} compared to the analytical results.
The estimated ERR matches the analytical results well, which implies that when investigating the locality of LNO with complex structures (e.g., multi-block LNOs), the receptive field and $\varrho_\textup{init}$ obtained via experimental estimation are reliable.

\subsection{Receptive range of multi-block LNOs\label{sec3:5}}
For practical cases of LNO with more than one block, each extra block contributes $\frac{\left(K-1\right)N}{K}\Delta x$ to the MRR $r_\textup{LNO}$. Assuming $\frac{(K-1)N}{K}\geq4$, Eq.~(\ref{eq:MRRdef}) turns to $r_{n=1}=N\Delta x$, then, $r_\textup{LNO}$ for multi-block LNO parameterized by ($n,N,K,M$) is
\begin{equation}
r_\textup{LNO}=r_{n=1}+\left(n-1\right)\frac{\left(K-1\right)N}{K}\Delta x=\frac{nK-n+1}{K}N\Delta x.
\end{equation}
We estimate the initial ERR $\varrho_\textup{init}$ of multi-block LNOs by experiments of repetitive sampling as shown in Table \ref{tab:initialERR_LNO}.
The initial ERR $\varrho_\textup{init}$ of LNOs is shown mainly for variant $N$ since it affects $\varrho_\textup{init}$ primarily.
Results in Table \ref{tab:initialERR_LNO} are also a preparation for the ensuing study in section \ref{sec4} about how the locality affects the performance of LNO and its relation to the transient PDEs to be learned.

\begin{table}[htbp]
\footnotesize
\renewcommand{\arraystretch}{0.8}
\centering
    \caption{Effective receptive range $\varrho_\textup{init}$ of LNO (the number of blocks $n>1$) with randomly initialized weights obtained by repetitive sampling for 300 times. ($\times\Delta x$)\label{tab:initialERR_LNO}}
\begin{tabular*}{\textwidth}{@{\extracolsep{\fill}}ccccccc}
\toprule
\multirow{2}{*}{$n=2$} & ($N,K,M$)               & (8,2,6)  & (12,2,6) & (16,2,6) & (20,2,6) &          \\
                       & $\varrho_\textup{init}$ & 7.2542   & 10.8062  & 14.4610  & 18.1402  &          \\
                       \midrule
\multirow{2}{*}{$n=3$} & ($N,K,M$)               & (8,2,6)  & (12,2,6) & (16,2,6) & (20,2,6) &          \\
                       & $\varrho_\textup{init}$ & 9.3203   & 13.6009  & 18.1527  & 22.6259  &          \\
                       \midrule
\multirow{4}{*}{$n=4$} & ($N,K,M$)               & (6,2,6)  & (8,2,6)  & (10,2,6) & (12,2,6) & (14,2,6) \\
                       & $\varrho_\textup{init}$ & 8.4122   & 11.1483  & 13.6026  & 16.0459  & 18.6843  \\
                       \cline{2-7}
                       & ($N,K,M$)               & (16,2,6) & (18,2,6) & (20,2,6) & (24,2,6) &          \\
                       & $\varrho_\textup{init}$ & 21.3316  & 23.9083  & 26.4803  & 31.6503  &          \\
                       \midrule
\multirow{2}{*}{$n=5$} & ($N,K,M$)               & (8,2,6)  & (12,2,6) & (16,2,6) & (20,2,6) &          \\
                       & $\varrho_\textup{init}$ & 12.8785  & 18.1962  & 24.0824  & 29.7323  &         \\
                       \bottomrule
\end{tabular*}
\end{table}

\section{How the locality acts in LNO learning\label{sec4}}
In this section, we take the task of approximating the time-marching operator of fluid dynamics as an example to monitor, understand, and explain the behavior of LNO regarding the locality. 
Specifically, the investigation in this section looks for possible answers to the following questions in view of the locality: \emph{Why do some LNOs outperform others?} and \emph{how to design a proper network architecture that can perform well for a new learning task?}

\subsection{Performance of LNO in learning fluid dynamics\label{sec4:1}}
The learning tasks here are the fluid dynamics governed by Eq. (\ref{eq:NSequation}) with parameters ($Re,Ma,\Delta t$).
As mentioned in Section \ref{sec2:3}, there are seven tasks with different parameters: with baseline ($Re,Ma,\Delta t$)=(100,0.2,0.05), learning tasks including variant $Re$ as $\{20,100,500\}$, variant $Ma$ as $\{0.1,0.2,0.4\}$, and variant $\Delta t$ as $\{0.03,0.05,0.07\}$. 
LNOs are separately trained to learn the seven tasks.
For each task, LNOs with variant number of blocks $n$ and the width of local spectral transform $N$ are trained and validated following an identical schedule introduced in Section \ref{sec2:3}. 
The other two parameters $K=2,M=6$ are set fixed.
The detailed parameter settings for each task are listed in Table \ref{tab:validationresults}. 
LNO of one parameter setting is trained three times, and the results of averaged errors are summarized in Table \ref{tab:validationresults}. 
The evolution history of loss function $\mathcal{L}$ during training (Eq. (\ref{eq:loss})) of three typical LNOs is depicted in Figure \ref{fig:trainingcurve}(a). 
After training, each LNO is validated by recurrently predicting the flow from unseen initial conditions until $t\approx5$. The time history of validation errors is in Figure~\ref{fig:trainingcurve}(b-d), and contours are shown in Figure~\ref{fig:validation}. 
Though these networks only show quantitative differences in the training loss, they perform quite diversely in the long-term prediction of validation that some successfully predict, while some produce unreasonable oscillation.

The accuracy of LNO prediction is evaluated by the mean $L_2$ error of density $e_t^\rho$, temperature $e_t^T$, and velocities $e_t^v$ (Eqs. (\ref{eq:error_rho}-\ref{eq:error_v})).
The time history of the errors is shown in Figure \ref{fig:trainingcurve}(b-d). 
The models show similar accuracy in the beginning, but as $t$ (the number of iterations) grows, their errors grow at a different speed: some stay nearly constant or grow linearly with low speed, and some grow rapidly to an excessive level, which is consistent to the contours of Figure \ref{fig:validation}. 
To further investigate the performance of these LNOs, the errors $e_t^\rho$,$e_t^T$,$e_t^v$ are averaged over $t=0\sim5$ ($t=0\sim2.5$ for $Re=20$) as ${\bar{e}}^\rho$,${\bar{e}}^T$,${\bar{e}}^v$ to be the representation of performance for each LNO. 
Table \ref{tab:validationresults} exhibits the error of all the trained LNOs on all the learning tasks. 
It can be found that LNOs with different parameters lead to very different accuracy in the same learning task, and LNOs with the same parameter result in different accuracy in different tasks as well.
This could be a great trouble when designing LNO for a new learning task, as the proper parameters of LNO seem to vary from task to task. 
Then, we naturally wonder if there is an explanation for why some parameter settings outperform others.

\begin{figure}[]%
\centering
\includegraphics[width=0.72\textwidth]{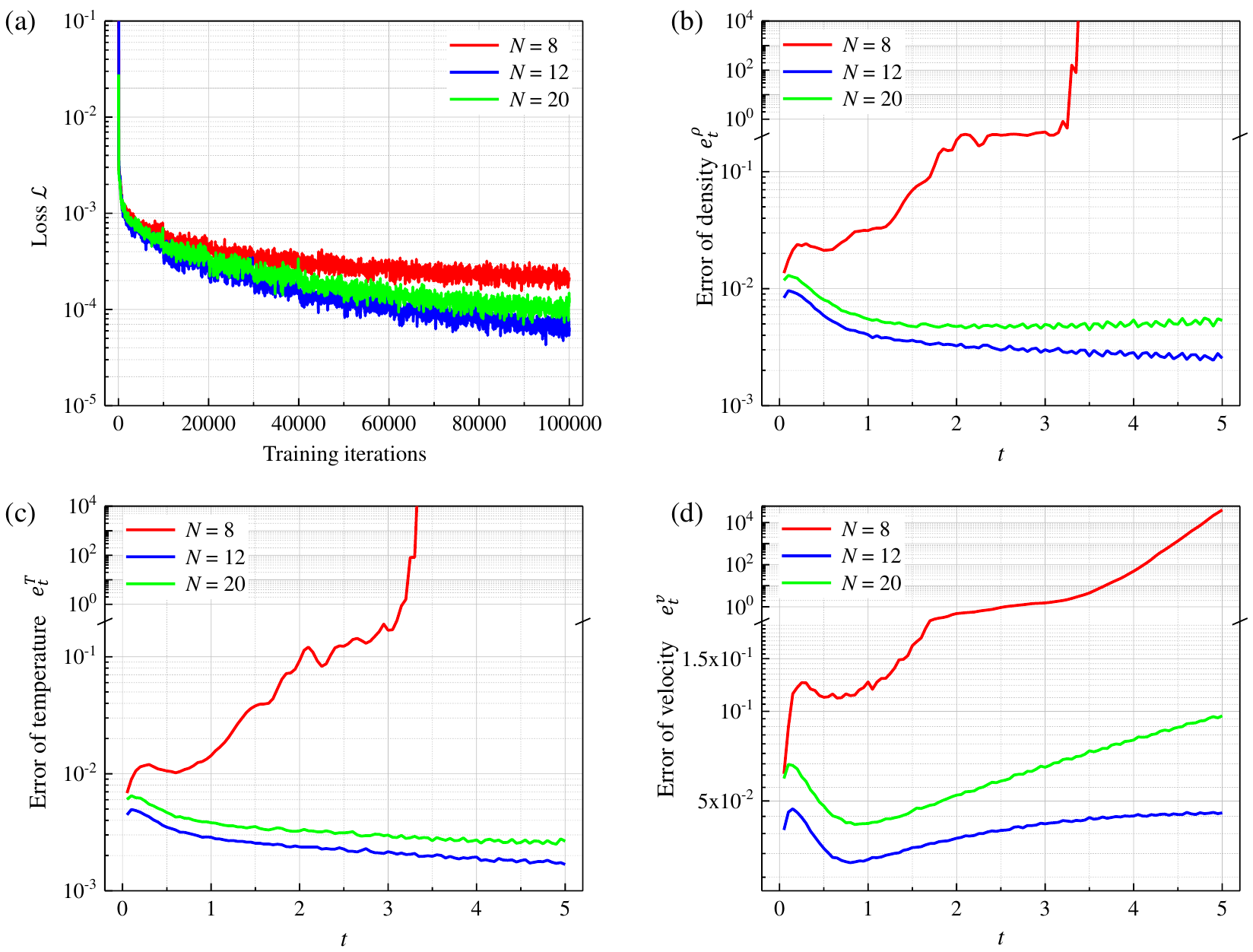}
\caption{Detailed information for LNO training and validation. (a) Evolution of loss $\mathcal{L}$ during the training process and time history of mean $L_2$ error for (b) density, (c) temperature, and (d) velocity in predicting the flow from an unseen initial condition by three LNOs with $n=4,K=2,M=6$ and different $N$.}\label{fig:trainingcurve}
\end{figure}

\begin{figure} 
\ContinuedFloat*
  \centering 
  \subfigure[Density $\rho$]{%
    \includegraphics[width=0.9\textwidth]{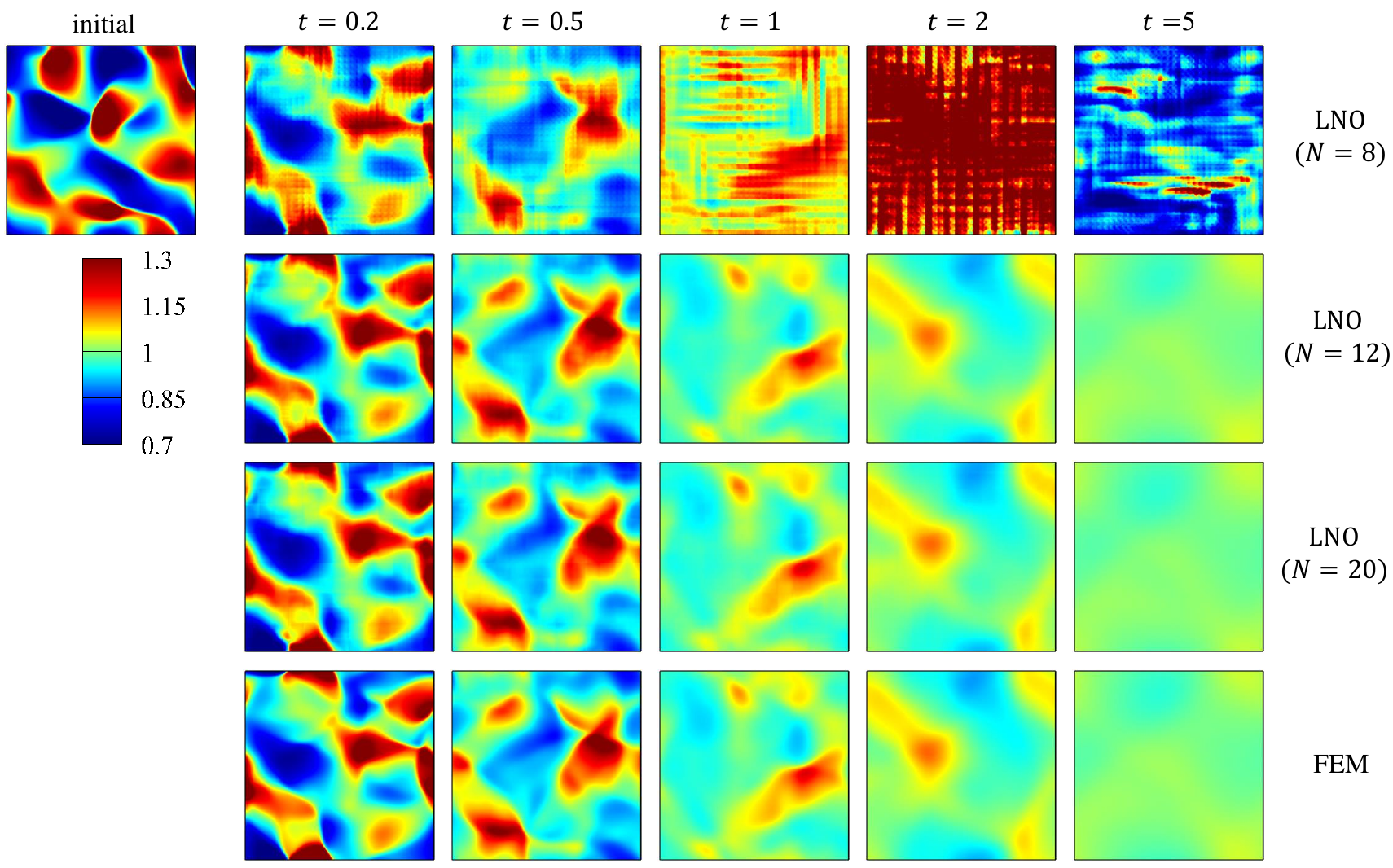}}%
    \\
  \subfigure[Temperature $T$]{%
    \includegraphics[width=0.9\textwidth]{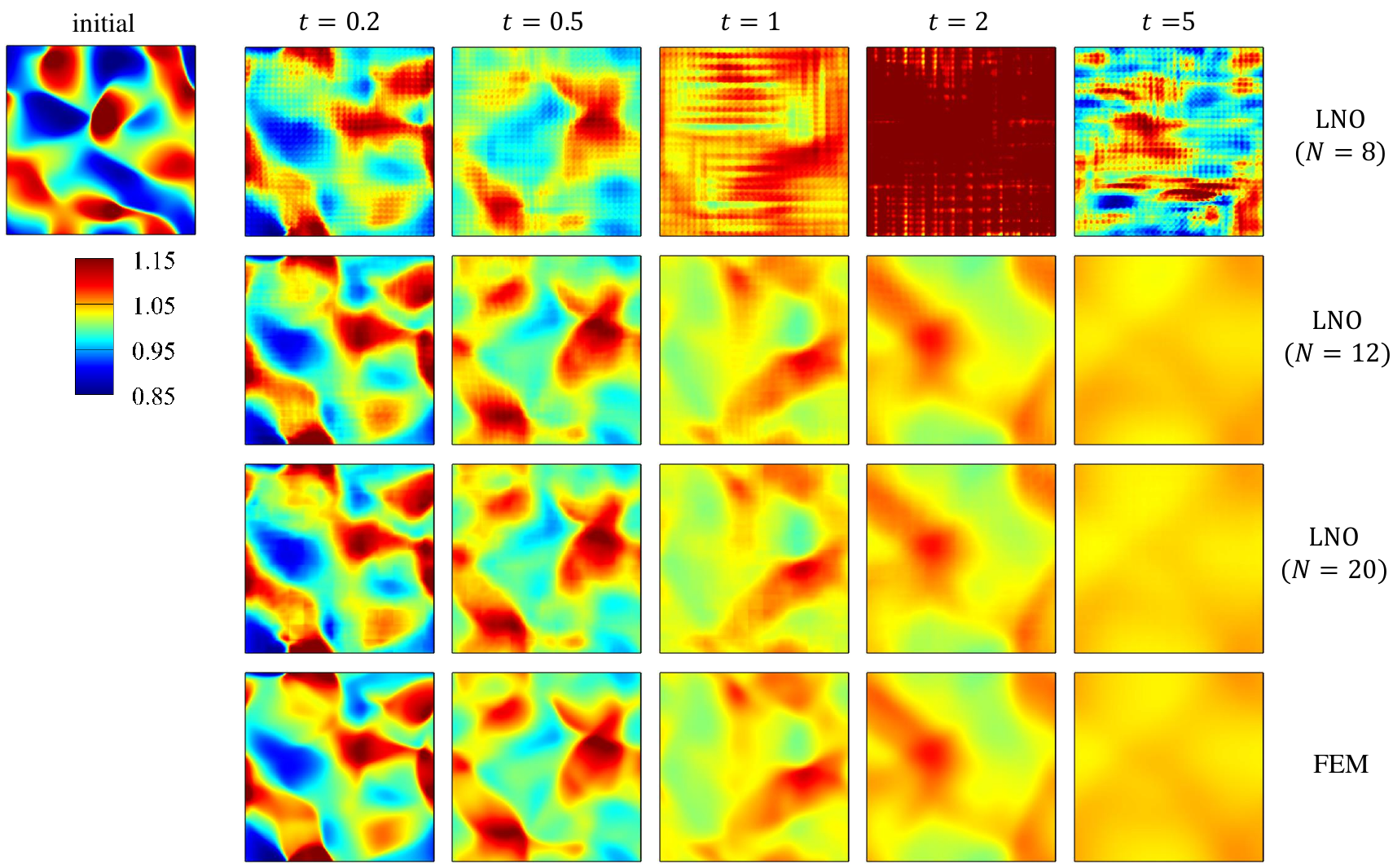}}%
  \caption{Contours of predicted fields by trained LNOs from an unseen initial condition.
In each subfigure, the first three rows are predictions by LNOs with $n=4,K=2,M=6$ and different $N$, while the last row is a reference solution calculated by FEM.}%
  \label{fig:validation}
\end{figure} 
\begin{figure} 
\ContinuedFloat
  \centering 
  \subfigure[Velocity $v_x$]{%
    \includegraphics[width=0.9\textwidth]{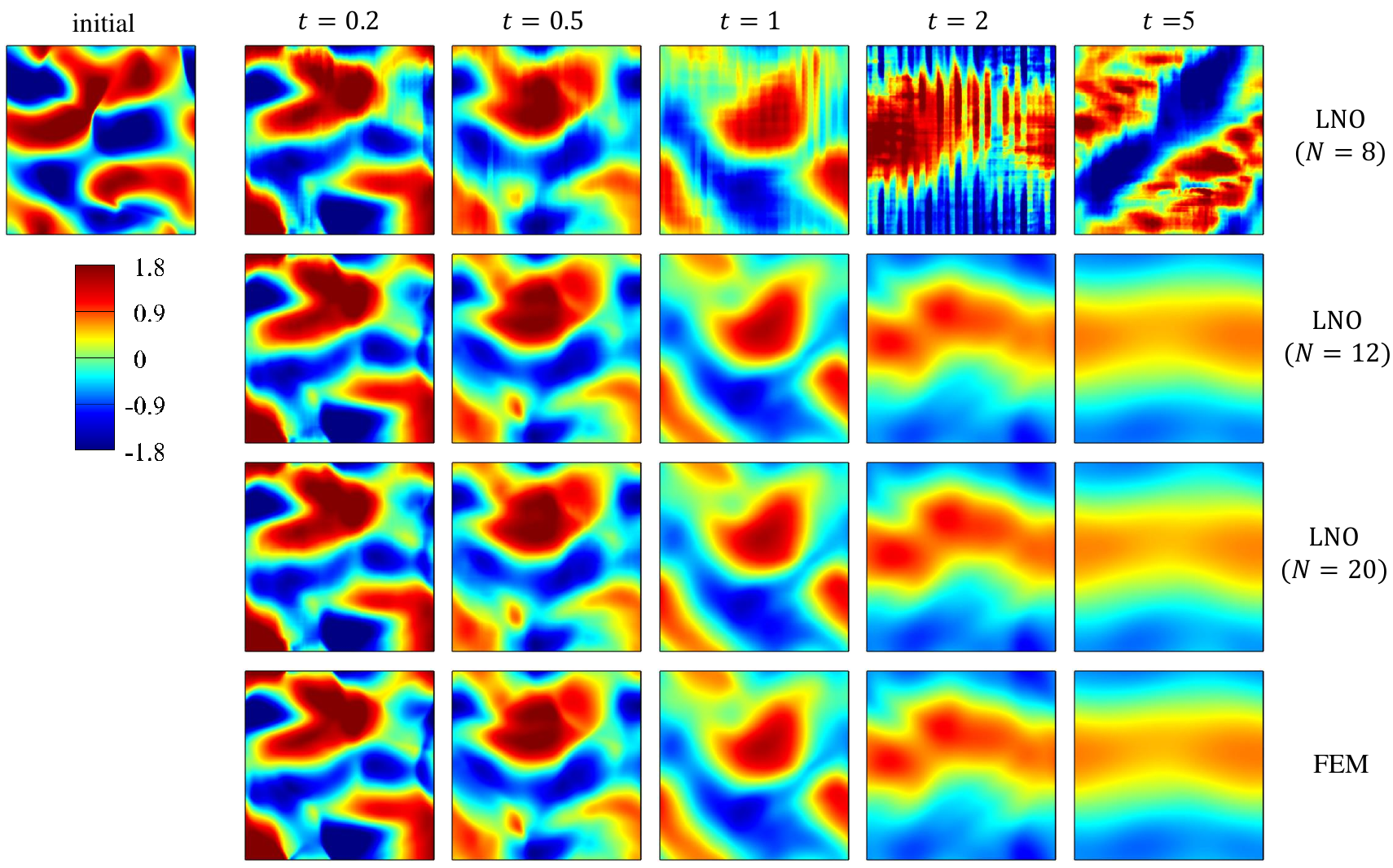}}%
    \\
    \subfigure[Velocity $v_y$]{%
    \includegraphics[width=0.9\textwidth]{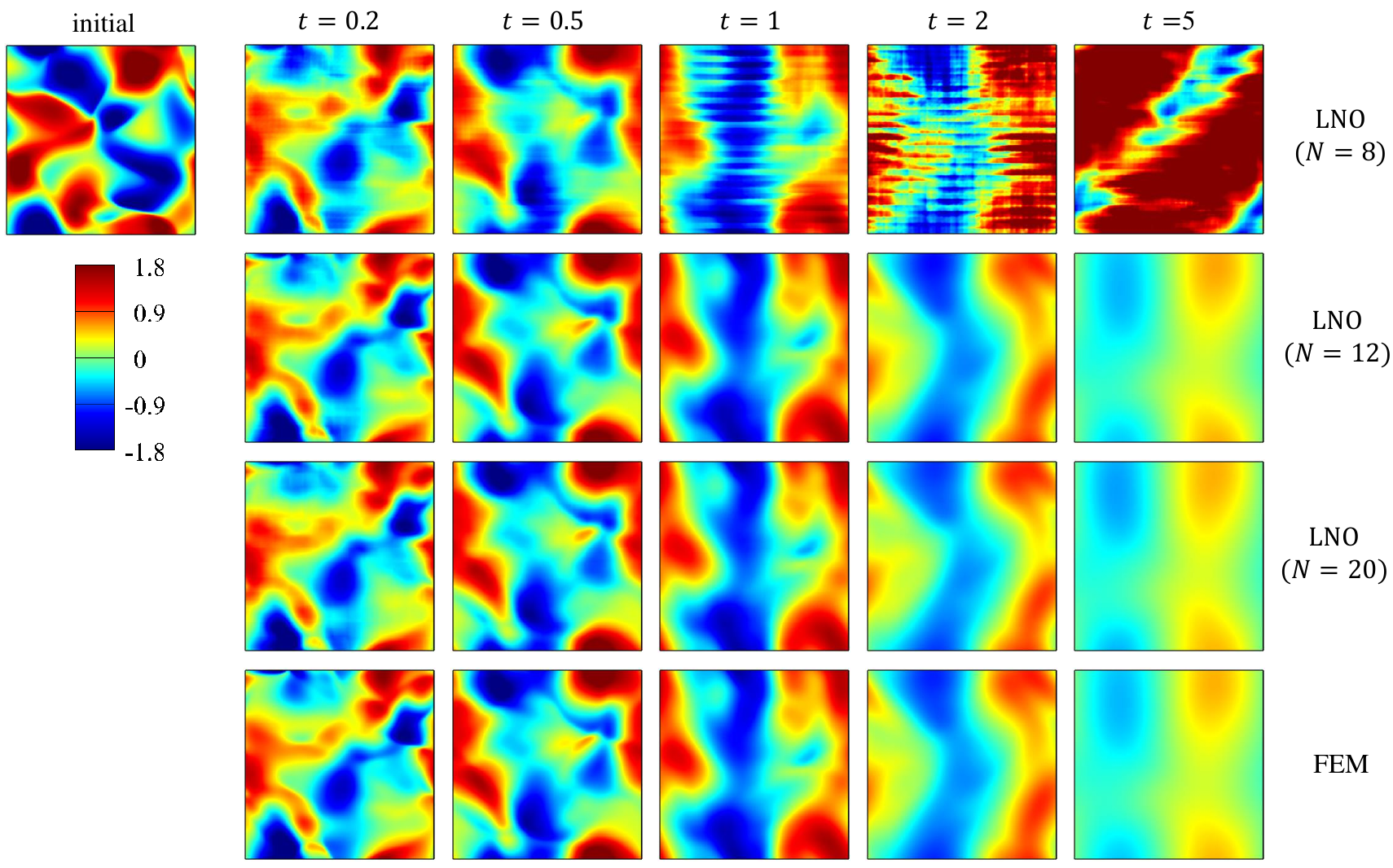}}%
  \caption{Contours of predicted fields by trained LNOs from an unseen initial condition.
In each subfigure, the first three rows are predictions by LNOs with $n=4,K=2,M=6$ and different $N$, while the last row is a reference solution calculated by FEM.}%
\end{figure}

\begin{table}[htbp]
\footnotesize
\renewcommand{\arraystretch}{0.8}
\centering
    \caption{LNO settings and validation results on 7 learning tasks. 
    The symbol ‘$\infty$’ marks values over the upper limit of float-type variables.\label{tab:validationresults}}
\begin{tabular*}{\textwidth}{@{\extracolsep{\fill}}cccccccc}
\toprule
{\makecell{Learning task\\ ($Re,Ma,\Delta t$)}} & Error            & \multicolumn{6}{c}{LNO($n,N,K=2,M=6$)}                             \\
\midrule
\multirow{16}{*}{(100,0.2,0.05)} &                  & (2,8)     & (2,12) & (2,16) & (2,20) &        &        \\
\cline{2-8}
                                 & ${\bar{e}}^\rho$ & 0.0392    & 0.0225 & 0.0129 & 0.0087 &        &        \\
                                 & ${\bar{e}}^T$    & 0.0238    & 0.0129 & 0.0077 & 0.0054 &        &        \\
                                 & ${\bar{e}}^v$    & 0.4665    & 0.1948 & 0.1124 & 0.0884 &        &        \\
                                 \cline{2-8}
                                 &                  & (3,8)     & (3,12) & (3,16) & (3,20) &        &        \\
                                 \cline{2-8}
                                 & ${\bar{e}}^\rho$ & 0.0160    & 0.0093 & 0.0055 & 0.0056 &        &        \\
                                 & ${\bar{e}}^T$    & 0.0114    & 0.0067 & 0.0031 & 0.0037 &        &        \\
                                 & ${\bar{e}}^v$    & 0.1508    & 0.0981 & 0.0476 & 0.0608 &        &        \\
                                 \cline{2-8}
                                 &                  & (4,8)     & (4,10) & (4,12) & (4,14) & (4,16) & (4,20) \\
                                 \cline{2-8}
                                 & ${\bar{e}}^\rho$ & $\infty$  & 0.0041 & 0.0105 & 0.0040 & 0.0046 & 0.0052 \\
                                 & ${\bar{e}}^T$    & $\infty$  & 0.0029 & 0.0030 & 0.0027 & 0.0026 & 0.0030 \\
                                 & ${\bar{e}}^v$    & 417.11    & 0.0395 & 0.0440 & 0.0381 & 0.0535 & 0.0562 \\
                                 \cline{2-8}
                                 &                  & (5,8)     & (5,12) & (5,16) & (5,20) &        &        \\
                                 \cline{2-8}
                                 & ${\bar{e}}^\rho$ & 0.0050    & 0.0037 & 0.0046 & 0.0054 &        &        \\
                                 & ${\bar{e}}^T$    & 0.0033    & 0.0030 & 0.0028 & 0.0029 &        &        \\
                                 & ${\bar{e}}^v$    & 0.0489    & 0.0366 & 0.0444 & 0.0471 &        &        \\
                                 \midrule
\multirow{4}{*}{(20,0.2,0.05)}   &                  & (4,8)     & (4,12) & (4,14) & (4,16) & (4,18) & (4,20) \\
\cline{2-8}
                                 & ${\bar{e}}^\rho$ & 0.1813    & 0.0010 & 0.0010 & 0.0011 & 0.0012 & 0.0013 \\
                                 & ${\bar{e}}^T$    & 0.0332    & 0.0009 & 0.0008 & 0.0008 & 0.0009 & 0.0010 \\
                                 & ${\bar{e}}^v$    & 0.6211    & 0.0062 & 0.0054 & 0.0057 & 0.0062 & 0.0073 \\
                                 \midrule
\multirow{4}{*}{(500,0.2,0.05)}  &                  & (4,8)     & (4,10) & (4,12) & (4,14) & (4,16) & (4,20) \\
\cline{2-8}
                                 & ${\bar{e}}^\rho$ & 0.0882    & 0.0156 & 0.0109 & 0.0111 & 0.0141 & 0.0152 \\
                                 & ${\bar{e}}^T$    & 0.0450    & 0.0103 & 0.0050 & 0.0053 & 0.0066 & 0.0072 \\
                                 & ${\bar{e}}^v$    & 0.6816    & 0.1995 & 0.1787 & 0.1771 & 0.2222 & 0.2667 \\
                                 \midrule
\multirow{4}{*}{(100,0.1,0.05)}  &                  & (4,8)     & (4,12) & (4,16) & (4,18) & (4,20) & (4,24) \\
\cline{2-8}
                                 & ${\bar{e}}^\rho$ & $\infty$  & 0.0487 & 0.0084 & 0.0027 & 0.0023 & 0.0029 \\
                                 & ${\bar{e}}^T$    & $\infty$  & 0.0240 & 0.0084 & 0.0022 & 0.0019 & 0.0033 \\
                                 & ${\bar{e}}^v$    & $\infty$  & 0.3286 & 0.1818 & 0.0685 & 0.0632 & 0.0744 \\
                                 \midrule
\multirow{4}{*}{(100,0.4,0.05)}  &                  & (4,6)     & (4,8)  & (4,10) & (4,12) & (4,16) & (4,20) \\
\cline{2-8}
                                 & ${\bar{e}}^\rho$ & $\infty$  & 0.0083 & 0.0091 & 0.0100 & 0.0111 & 0.0139 \\
                                 & ${\bar{e}}^T$    & $\infty$  & 0.0064 & 0.0059 & 0.0067 & 0.0080 & 0.0083 \\
                                 & ${\bar{e}}^v$    & 0.8141    & 0.0402 & 0.0516 & 0.0551 & 0.0517 & 0.0520 \\
                                 \midrule
\multirow{4}{*}{(100,0.2,0.03)}  &                  & (4,6)     & (4,8)  & (4,10) & (4,12) & (4,16) & (4,20) \\
\cline{2-8}
                                 & ${\bar{e}}^\rho$ & 0.0057    & 0.0041 & 0.0031 & 0.0035 & 0.0039 & 0.0048 \\
                                 & ${\bar{e}}^T$    & 0.0041    & 0.0028 & 0.0023 & 0.0022 & 0.0026 & 0.0033 \\
                                 & ${\bar{e}}^v$    & 0.0597    & 0.0442 & 0.0343 & 0.0370 & 0.0457 & 0.0437 \\
                                 \midrule
\multirow{4}{*}{(100,0.2,0.07)}  &                  & (4,8)     & (4,12) & (4,16) & (4,18) & (4,20) & (4,24) \\
\cline{2-8}
                                 & ${\bar{e}}^\rho$ & 0.0145    & 0.0096 & 0.0053 & 0.0062 & 0.0064 & 0.0111 \\
                                 & ${\bar{e}}^T$    & 0.0099    & 0.0054 & 0.0034 & 0.0035 & 0.0035 & 0.0055 \\
                                 & ${\bar{e}}^v$    & 0.1075    & 0.0835 & 0.0552 & 0.0598 & 0.0606 & 0.0651\\
                                 \bottomrule
\end{tabular*}
\end{table}

\subsection{How the receptive field changes and affects the performance of LNO\label{sec4:2}}
The receptive field may give a hint on this topic. 
The target time-marching operator $\mathcal{G}_\mathrm{L}$ has an unknown but certain receptive field. Thereby, we have an inference that, as the training goes on, the receptive field of LNO $\mathcal{G}_\theta$ gets closer to the receptive field of $\mathcal{G}_\mathrm{L}$. 
In the ideal case, a well-trained LNO should own the same receptive field as $\mathcal{G}_\mathrm{L}$.
On the contrary, deviations from the target receptive field may lead to the poor performance of a trained LNO, that is, the rise of error. 
It prompts us that there is a direct connection between the receptive field and the performance of LNO.
Therefore, we use the receptive field as a window to monitor the behavior of LNO in training and explain its performance.

\begin{figure}[htbp]%
\centering
\includegraphics[width=\textwidth]{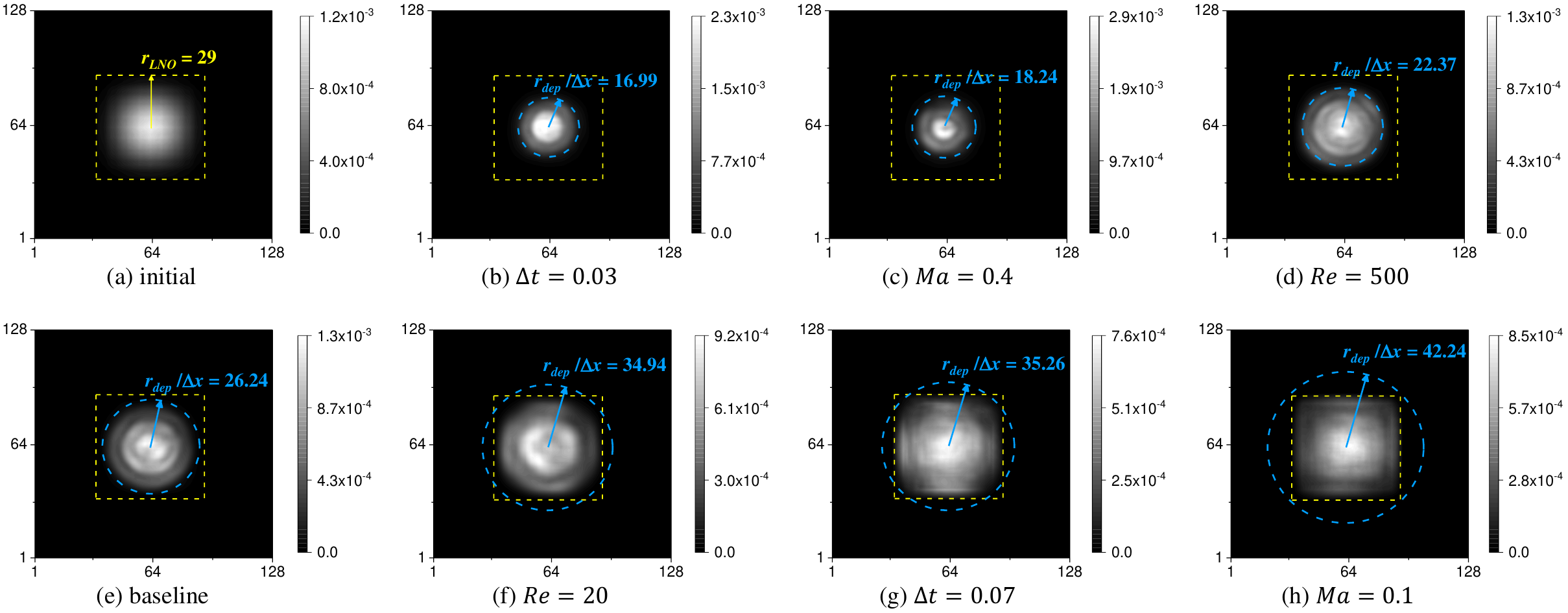}
\caption{The initial receptive field (subfigure a) and the receptive fields after training on 7 different tasks (subfigures b-h) of LNO ($n=4,N=12,K=2,M=6$), in which the maximum receptive field of LNO (according to the MRR defined in Eq. (\ref{eq:MRRdef})) and the analytical range of dependence (according to the results in Table \ref{tab:DEPrange}) are respectively marked in dotted yellow squares and blue circles.
Subfigures b-h are sorted by the range of dependence. The baseline task is with parameters $Re=100,Ma=0.2,\Delta t=0.05$. 
Titles of subfigures b-d, f-h mark the difference from the baseline parameters. }\label{fig:receptivefield1}
\end{figure}

\begin{figure}[htbp]%
\centering
\includegraphics[width=\textwidth]{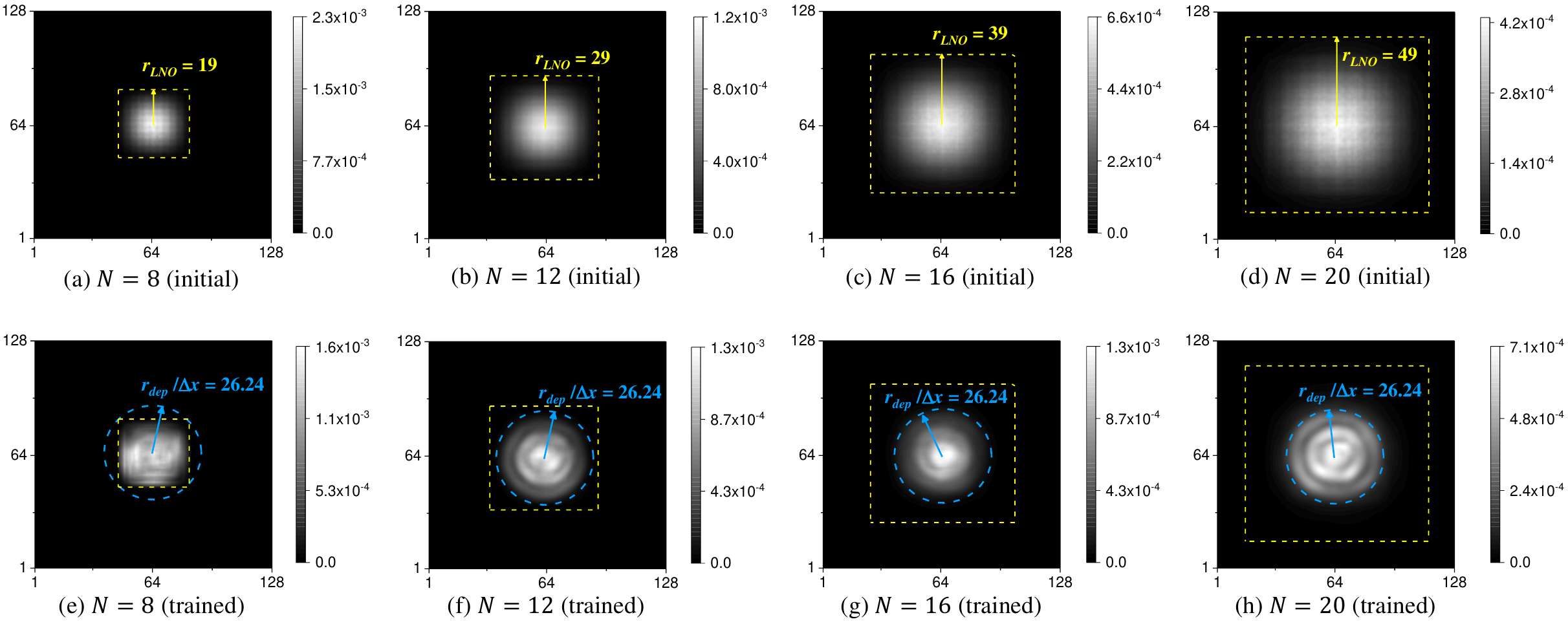}
\caption{ The initial receptive field (subfigures a-d) and the receptive fields after training (subfigures e-h) on the baseline task $Re=100,Ma=0.2,\Delta t=0.05$, in which the maximum receptive field of LNO (according to the MRR defined in Eq. (\ref{eq:MRRdef})) and the analytical range of dependence (according to the results in Table \ref{tab:DEPrange}) are respectively marked in dotted yellow squares and blue circles. The LNO architectures are with parameters $n=4, K=2, M=6$ while the different $N$ is marked as the subfigure titles. }\label{fig:receptivefield2}
\end{figure}

We start by investigating the receptive field of the initial and trained LNOs.
Figure \ref{fig:receptivefield1} depicts the receptive fields of LNOs with the same parameters ($n=4,N=12,K=2,M=6$) before and after training on seven learning tasks.
We mark the MRR $r_\mathrm{LNO}$ for the LNO architecture with a dotted yellow square, and the range of dependence $r_\mathrm{dep}$ for the learning task with a dotted blue circle as the representative of the receptive field of the real operator $\mathcal{G}_\mathrm{L}$ according to Table~\ref{tab:DEPrange}.
There are two basically consistent things: i) the nonzero area in the receptive field of LNO learned from training data and ii) the area enclosed by $r_\mathrm{dep}$ of the learning task.
It implies that LNOs successfully learned varied features of different tasks, and some of the features are reflected in the change of receptive field.
For tasks with smaller $r_\mathrm{dep}$ (large $Ma$, $Re$ and small $\Delta t$), the receptive fields of trained LNOs become more concentrated in the center area.
In other tasks with larger $r_\mathrm{dep}$, the receptive fields diffuse outward. 
However, not every LNO approaches the target operator successfully.
For example, in Figure~\ref{fig:receptivefield1}(f), \ref{fig:receptivefield1}(g), and \ref{fig:receptivefield1}(h), the receptive field seems to be cut off compulsorily. 
The cut-off receptive field appears when the blue circle for $r_\mathrm{dep}$ overflows out the yellow square for MRR ($r_\mathrm{LNO}$).
The receptive fields of different LNOs on the same learning task in Figure \ref{fig:receptivefield2} further confirm the idea. 
Though these LNOs try to approach the same target operator, the receptive fields after training are diverse with a distinct correlation to the initial receptive field. 

\begin{figure}[]%
\centering
\includegraphics[width=0.95\textwidth]{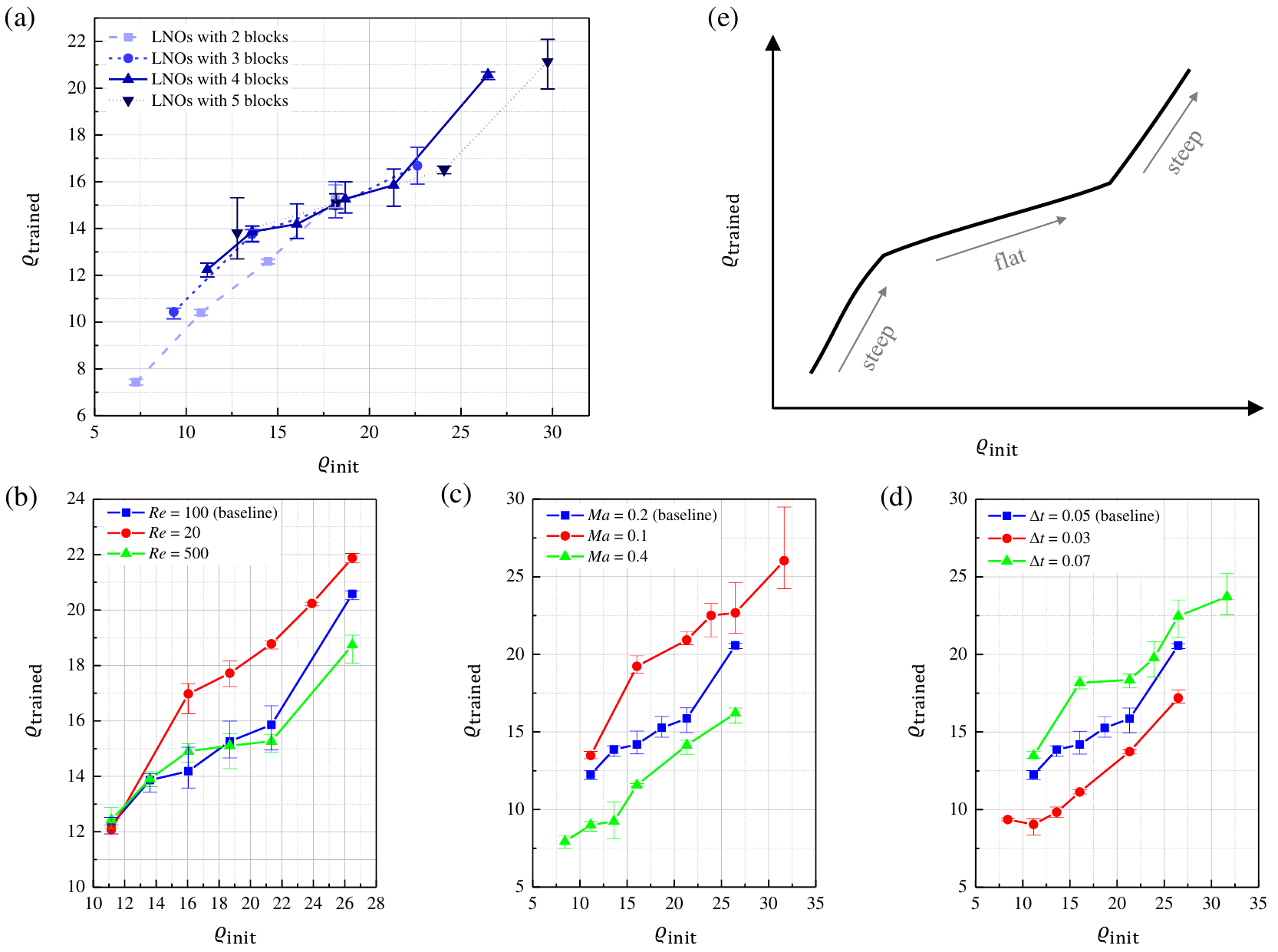}
\caption{The effective receptive range before and after training on (a) the baseline task; (b) tasks with different $Re$; (c) tasks with different $Ma$; (d) tasks with different $\Delta t$. (e) A schematic diagram for the common trend of the curves.}\label{fig:ERRcurve}
\end{figure}

The change in receptive fields is measured quantitatively by relating the initial ERR ($\varrho_\mathrm{init}$) and ERR after training ($\varrho_\mathrm{trained}$) in Figure~\ref{fig:ERRcurve}.
The curves of different learning tasks show a similar tendency that, as $\varrho_\mathrm{init}$ increases, $\varrho_\mathrm{trained}$ increases monotonically with a relatively flat part in the middle of the curve. 
The flat part occurs because ERR of all LNOs tries to get closer to one value (which could be regarded as the ERR of the target operator).
When $\varrho_\mathrm{init}$ is small, it grows larger after training, and vice versa. 
However, the monotonical tendency implies that the change of ERR is limited by the initial ERR, which is decided by the architecture of LNO. 
The limited approach is possibly why LNOs with different parameters perform differently.

\begin{figure}[]%
\centering
\includegraphics[width=0.9\textwidth]{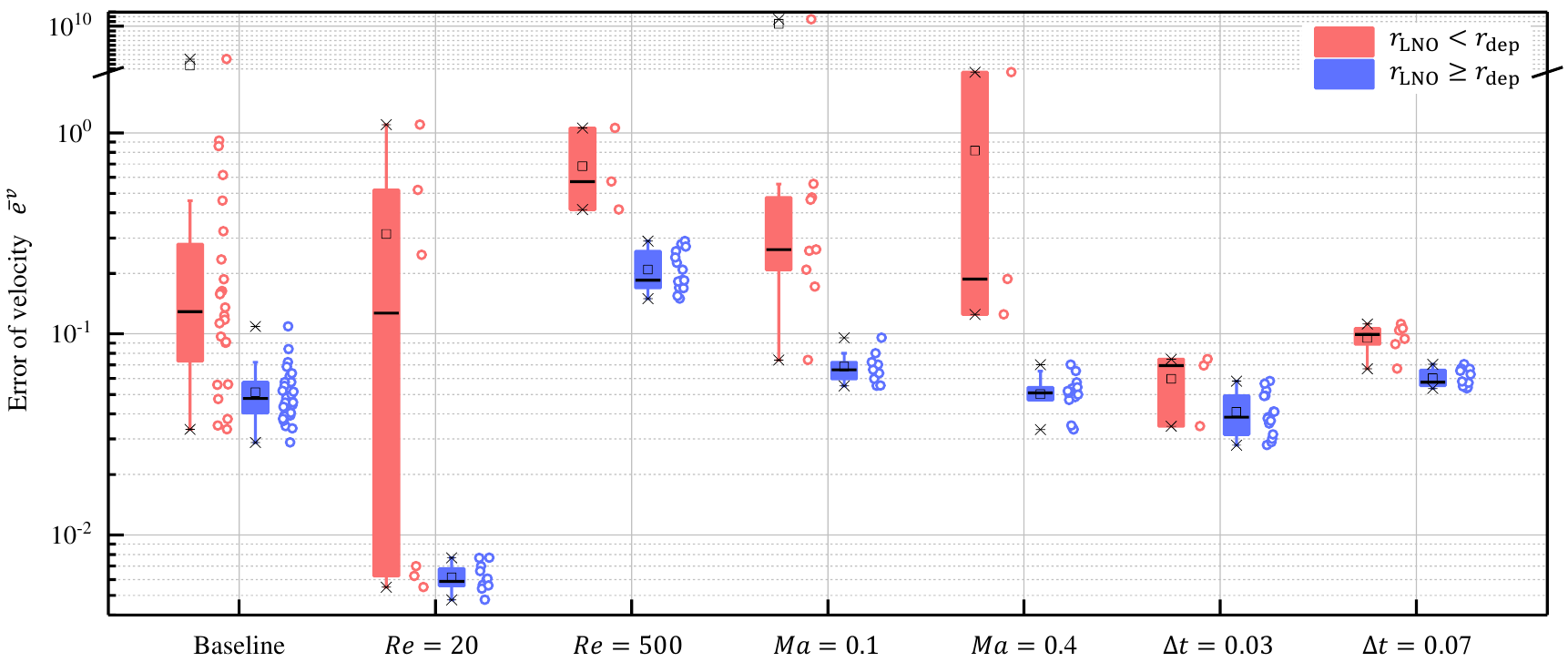}
\caption{Box chart for error of velocity ${\bar{e}}^v$ of seven learning tasks (only the different parameter from the baseline is marked in the title of each task).
Three times of trained LNOs for each parameter setting are all included in the chart.}\label{fig:MRRboxchart}
\end{figure}

Next, we investigate the relation between the performance (represented by the mean $L_2$ error) and the receptive field (represented by MRR and ERR) of LNO.
First, we group the errors in Table~\ref{tab:validationresults} by whether $r_\textup{LNO}\geq r_\textup{dep}$ is satisfied for each learning task and draw the result in a box chart Figure~\ref{fig:MRRboxchart}. 
It is found that the models with extremely large errors all belong to the group $r_\textup{LNO}< r_\textup{dep}$, implying that when MRR is insufficient, it fatally harms the performance.

\begin{figure}[]%
\centering
\includegraphics[width=0.92\textwidth]{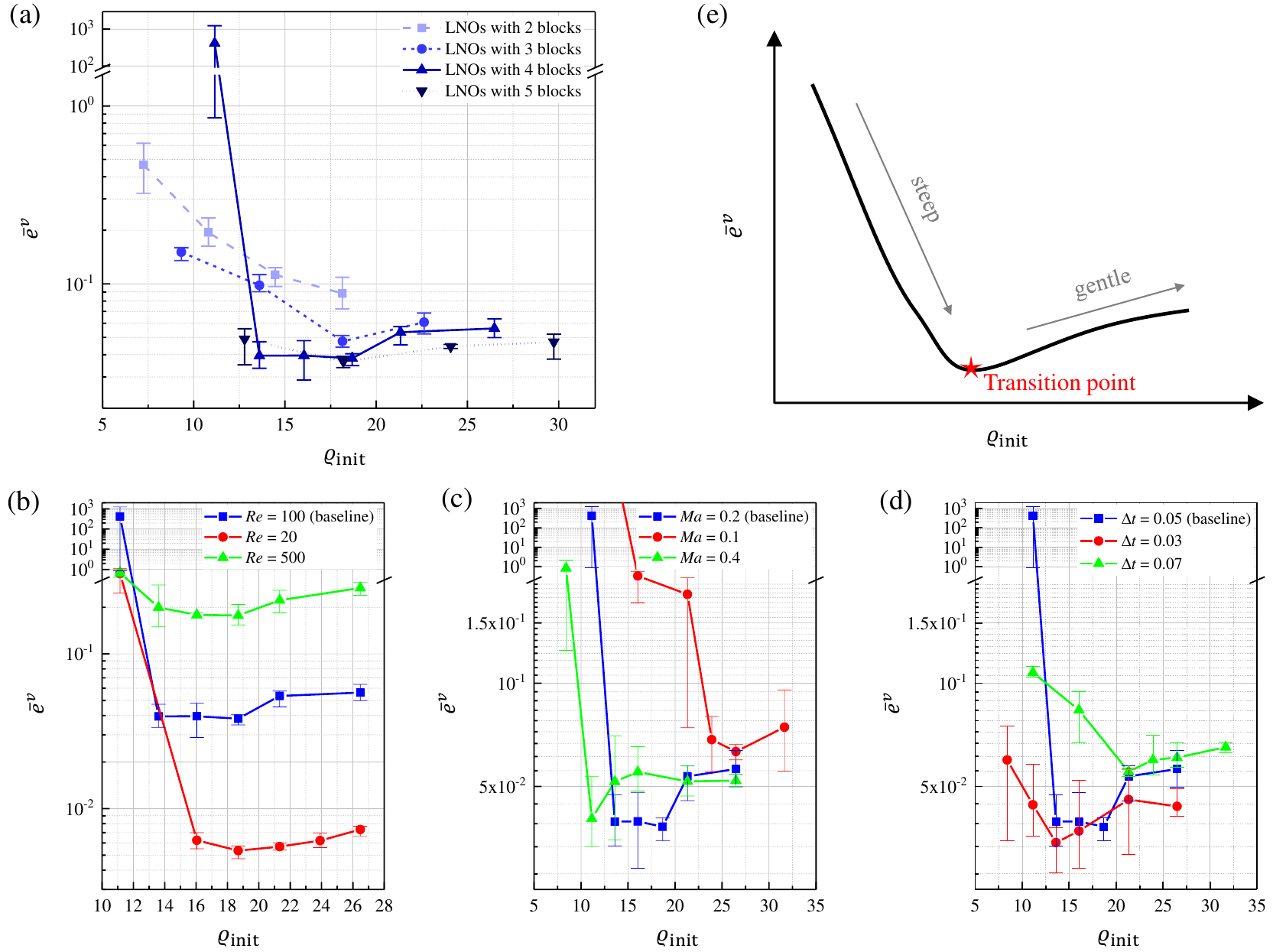}
\caption{The averaged error of velocity ${\bar{e}}^v$ – initial effective receptive range $\varrho_\textup{init}$ curve for LNOs trained on (a) the baseline task; (b) tasks with different $Re$; (c) tasks with different $Ma$; (d) tasks with different $\Delta t$. (e) A schematic diagram for the common trend of the curves.}\label{fig:error-ERR}
\end{figure}

Then, we reexamine the results in Table~\ref{tab:validationresults} concerning their ERRs.
Figure~\ref{fig:error-ERR}(a) shows the ${\bar{e}}^v-\varrho_\textup{init}$ curves for the task ($Re,Ma,\Delta t$)=(100,0.2,0.05), and Figure~\ref{fig:error-ERR}(b-d) shows the curves for all the seven tasks. 
The ${\bar{e}}^v-\varrho_\textup{init}$ curves share a similar tendency as illustrated in Figure~\ref{fig:error-ERR}(e): they change in U-shape and reach the bottom at a transition point in the middle of the curve.
The curve is steeper on the left of the transition point and gentler on the right. 
It implies that the LNO at the transition point with minimal error is the closest one to the real time-marching operator to be learned, and all the LNOs try to approach it during training.
We therefore term the LNO with $\varrho_\mathrm{init}$ close to the optimal is \emph{compatible} with the learning task, in other words, the LNO is with \emph{a compatible locality} to the learning task. 
Otherwise, the LNO is \emph{incompatible} with the learning task. 

\begin{figure}[t!]%
\centering
\includegraphics[width=0.9\textwidth]{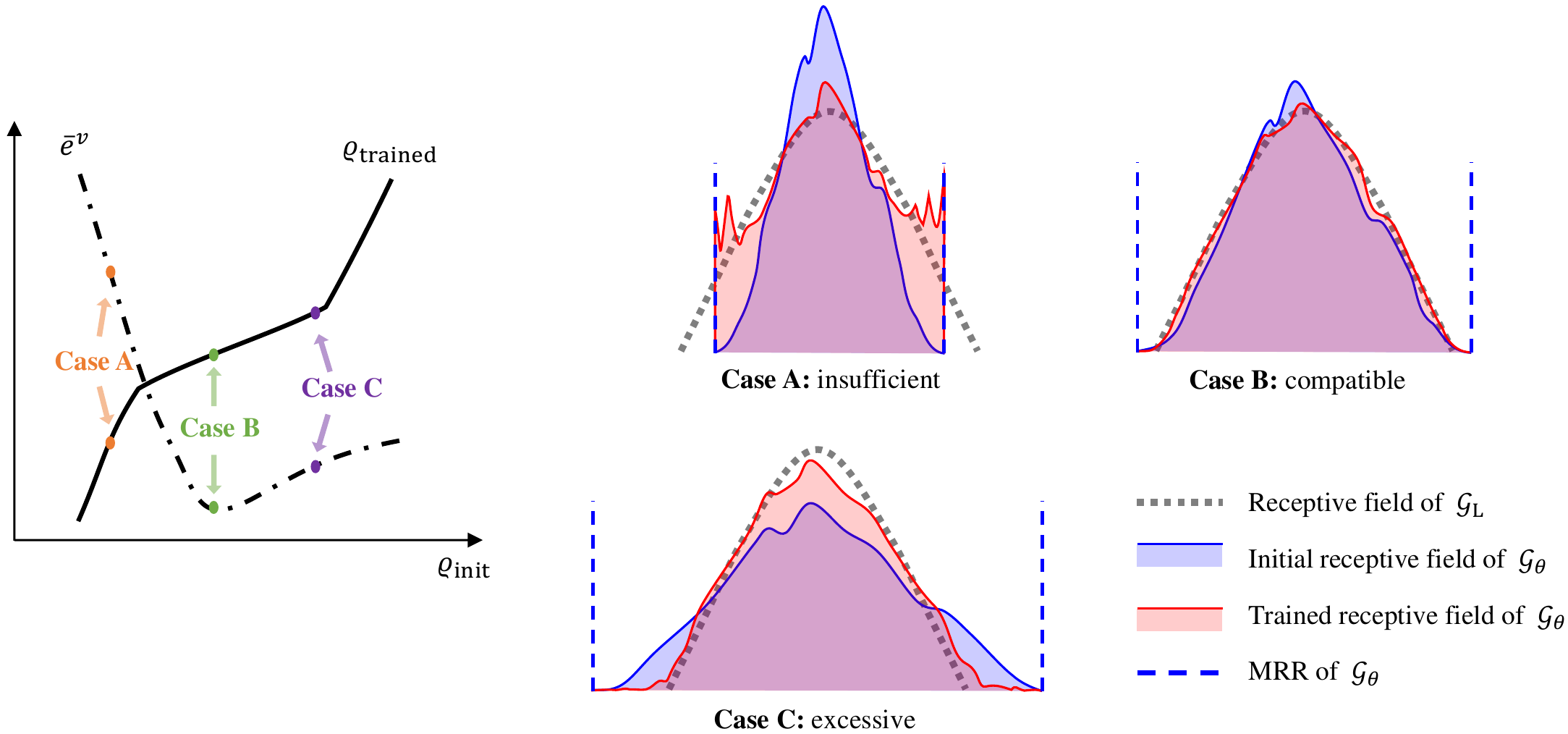}
\caption{The receptive field and the performance of LNO: the schematic diagram for $\varrho_\mathrm{trained}- \varrho_\mathrm{init}$ and ${\bar{e}}^v-\varrho_\mathrm{init}$ curves, and the receptive fields for LNOs with insufficient, compatible, and excessive local-related range. The receptive fields are simplified 1-D schematic diagrams.}\label{fig:ReceptiveSchematic}
\end{figure}

We summarize the above results as schematics in Figure~\ref{fig:ReceptiveSchematic} to go deeper into the mechanism of how the receptive field affects the performance of LNO.
It depicts the schematic of $\varrho_\mathrm{trained}-\varrho_\mathrm{init}$ and ${\bar{e}}^v-\varrho_\mathrm{init}$ curves together with the simplified 1-D diagram of receptive fields before and after training on the same learning task.
LNOs try to approach the same target operator $\mathcal{G}_\mathrm{L}$ and approximate the receptive field of $\mathcal{G}_\mathrm{L}$ with a limitation by the initial receptive field, which is determined by the network architecture. 
When the initial receptive field is insufficient (Case A), the growth of the receptive field is confined by MRR, leading to the cut-off and oscillation near MRR, as appeared in the cases shown in Figures~\ref{fig:receptivefield1}(f), \ref{fig:receptivefield1}(g), \ref{fig:receptivefield1}(h), and \ref{fig:receptivefield2}(e).
It results in huge errors and small $\varrho_\mathrm{trained}$.
When the initial receptive field is compatible with the learning task (Case B), LNO can approximate $\mathcal{G}_\mathrm{L}$ well with only a slight change in the receptive field.
It results in a minimal error (the transition point in ${\bar{e}}^v-\varrho_\mathrm{init}$ curve) and a minor change in ERR. 
When the initial receptive field is too large (Case C), though MRR does not limit the reduction of the receptive field, the excess parts on both sides ranged by MRR over the receptive field of $\mathcal{G}_\mathrm{L}$ are redundant for the approximation.
On the one hand, it slightly limits the decrease of $\varrho_\mathrm{trained}$, resulting in the flat but monotonically increasing part in $\varrho_\mathrm{trained}-\varrho_\mathrm{init}$ curve; on the other hand, it wastes some weights of LNO to approximate unnecessary zeros in this region, resulting in the rise of error and the right half of the U-shape ${\bar{e}}^v-\varrho_\mathrm{init}$ curve.

\begin{table}[h]
\footnotesize
\renewcommand{\arraystretch}{0.8}
\centering
    \caption{The compatible LNO architecture and the effective receptive range after training for different learning tasks. All the listed LNOs are with $n=4,M=6,K=2$.
    \label{tab:compatibleLNO}}
\begin{tabular*}{\textwidth}{@{\extracolsep{\fill}}ccccc}
\toprule
                                    &{\makecell{Learning task \\($Re,Ma,\Delta t$)}}  & {\makecell{Compatible $N$}} & $\varrho_\textup{trained}$ & Tendency                        \\
                                    \midrule
\multirow{3}{*}{Variant $Re$}       & (20,0.2,0.05)                    & 12                                 & 16.9778                    & \multirow{3}{*}{$Re\uparrow,\mu\downarrow,$ then $ \varrho_\textup{trained}\downarrow$} \\
                                    & (100,0.2,0.05)                   & 14                                 & 15.2645                    &                                 \\
                                    & (500,0.2,0.05)                   & 14                                 & 15.1058                    &                                 \\
                                    \midrule
\multirow{3}{*}{Variant $Ma$}       & (100,0.1,0.05)                   & 20                                 & 22.6692                    & \multirow{3}{*}{$Ma\uparrow,c\downarrow,$ then $ \varrho_\textup{trained}\downarrow$}               \\
                                    & (100,0.2,0.05)                   & 14                                 & 15.2645                    &                                 \\
                                    & (100,0.4,0.05)                   & 8                                  & 8.9966                     &                                 \\
                                    \midrule
\multirow{3}{*}{Variant $\Delta t$} & (100,0.2,0.03)                   & 10                                 & 9.8342                     & \multirow{3}{*}{$\Delta t\uparrow,$ then $ \varrho_\textup{trained}\uparrow$}               \\
                                    & (100,0.2,0.05)                   & 14                                 & 15.2645                    &                                 \\
                                    & (100,0.2,0.07)                   & 16                                 & 18.3590                    &         \\
                                    \bottomrule
\end{tabular*}
\end{table}

Overall, the analysis reveals that a compatible locality to the learning task is crucial for LNO to achieve good performance. 
Our findings could be helpful in the design of LNO architecture: 
Firstly, architectures with too small MRR should be avoided as they perform poorly; 
Secondly, though the large receptive range of LNO seems to cover the learning task with small ranges of dependence, unthinkingly choosing architectures with a large receptive range may not lead to an ideal model as desired while costing a lot in training.

In addition, Table \ref{tab:compatibleLNO} lists the compatible LNO for different tasks. 
By comparison among different tasks, it is found that the change of $\varrho_\textup{trained}$ follows the tendency for the range of dependence mentioned in Section \ref{sec2:2}.
Specifically, larger $Re$ (smaller viscosity), larger $Ma$ (smaller sound speed), and smaller $\Delta t$ result in smaller $r_\textup{dep}$ as well as smaller $\varrho_\textup{trained}$.

\begin{figure}[]%
\centering
\includegraphics[width=0.95\textwidth]{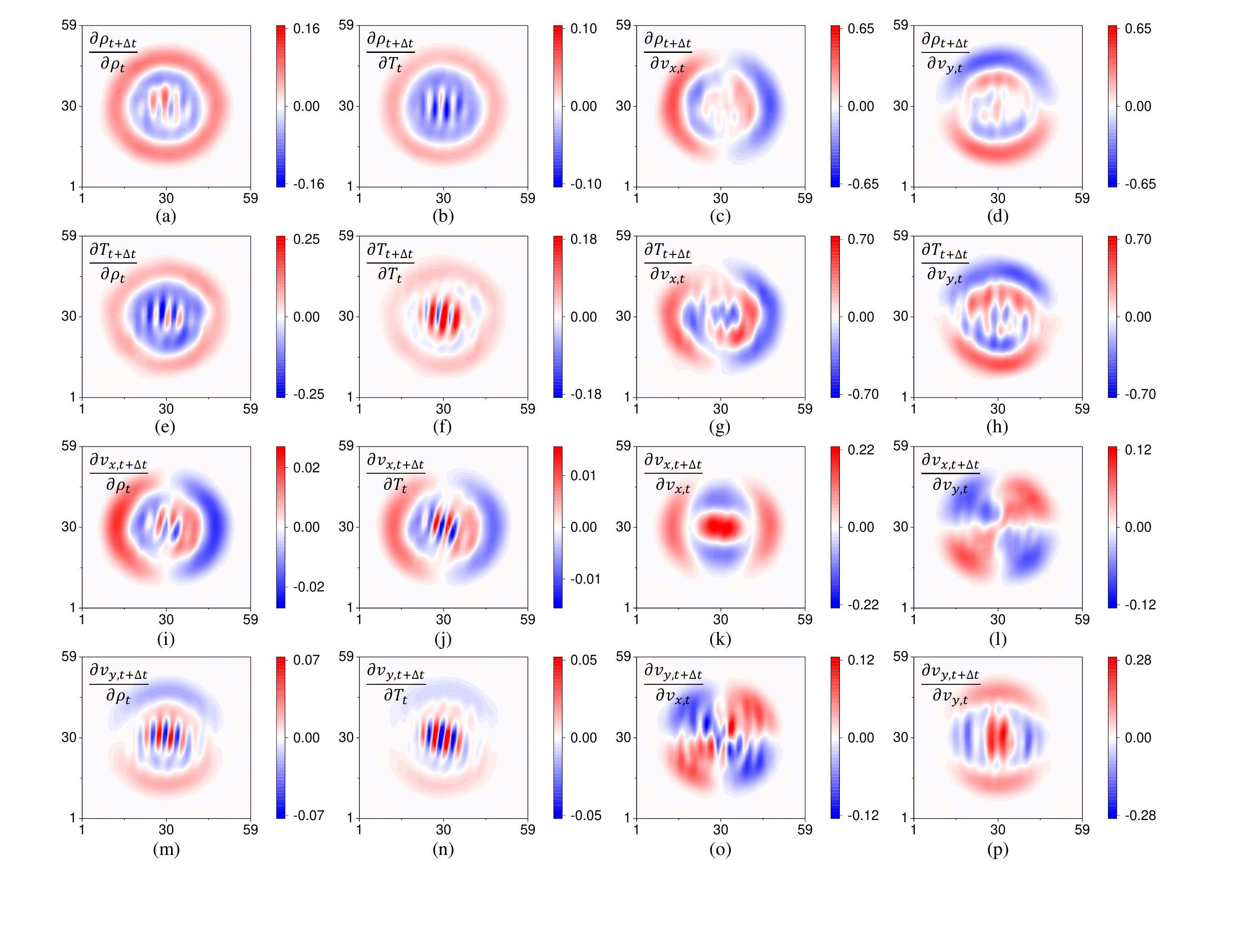}
\caption{16 components of the receptive field for LNO ($n=4,N=12,K=2,M=6$) trained on task with $Re=100,Ma=0.2,\Delta t=0.05$. }\label{fig:receptivecomponent}
\end{figure}

\subsection{More discoveries\label{sec4:3}}
The above results show that LNO learns the time-marching operators well with relatively small errors. 
Then, one may be curious about what is learned by LNO in training. 
There could be clues by monitoring the components that constitute the receptive field. 
For the compressible fluid dynamics problem with four channels ($\rho,T,v_x,v_y$), there are 16 components of receptive fields representing the time-marching relationship between each two of the channels. 
These components of one trained LNO are depicted in Figure \ref{fig:receptivecomponent}. 
For example, Figure \ref{fig:receptivecomponent}(a) shows the contour of $\frac{\partial \rho_{t+\Delta t}}{\partial \rho_t}$ and Figure \ref{fig:receptivecomponent}(l) shows $\frac{\partial v_{x,t+\Delta t}}{\partial v_{y,t}}$. 
The primary trend of these receptive fields can be interpreted according to their physical meanings.
In general, the fields related to $v_x$ (subfigures c, g, i, j, k) all show the left-right pattern, which is in the same direction as the $x$-axis;
the fields related to $v_y$ (subfigures d, h, m, n, p) all show the up-down pattern;
and the cross fields of $v_x$ and $v_y$ (subfigures l, o) show the pattern divided by quadrants. 
Because velocities have a direction, any change in velocity directly affects the direction of information transport. 
For example, in Figure \ref{fig:receptivecomponent}(c) for $\frac{\partial \rho_{t+\Delta t}}{\partial v_{x,t}}$, more fluids flow away as $v_{x,t}$ in the right increases, then $\rho_{t+\Delta t}$ around the central area decreases.
This process results in negative $\frac{\partial \rho_{t+\Delta t}}{\partial v_{x,t}}$ on the right side. 
The positive value in the left side could be interpreted similarly. 
On the contrary, the fields related to $\rho$ and $T$ (subfigures a, b, e, f) all show an isotropic pattern. 
Any increase in $\rho$ or $T$ in the central area results in an outward pressure gradient that forces $\rho$ or $T$ in this region to decrease. 
As these patterns all have physical interpretations, beyond showing that LNO does learn the basic law of fluid dynamics successfully, it could guide us to improve LNO in the future.
For example, the difference between Figure \ref{fig:receptivecomponent}(l) and (o) implies that velocity $v_x$ and $v_y$ are not fully symmetrical to each other in the trained LNO. 
Strategies to preserve the symmetry in LNO must be developed to elevate LNO further.

\section{Practical Examples\label{sec5}}
In this section, the pre-trained LNOs are applied as CFD solvers to predict several practical flow problems to show the non-negligible effect of the locality in applications.
Three LNOs with parameter $n=4,K=2,M=6$ and $N=\left\{8,\ 12,\ 20\right\}$ trained on the baseline learning task ($Re=100,Ma=0.2,\Delta t=0.05$) from Section \ref{sec4:1} represent LNOs with insufficient, compatible, and excessive receptive range. 
The LNOs respectively predict the solution of each example problem, and we simultaneously provide numerical solutions by conventional finite element method (FEM) for reference.

To solve practical problems, the pre-trained LNO predicts the solution away from the boundaries itself and collaborates with boundary treatments for the near-boundary areas.
According to \cite{LiHongyu2022}, the boundary conditions are divided into artificial ones (e.g., the far-field or periodic BCs) and real ones (mainly the solid wall in this work).
For the former, the computational domain of input is extended by padding operation before sending into LNO.
For the latter, the no-slip condition is introduced on the output of LNO by immersed boundary method (IBM).

\subsection{Flow around a circular cylinder\label{sec5:1}}
\begin{figure}[t]%
\centering
\includegraphics[width=\textwidth]{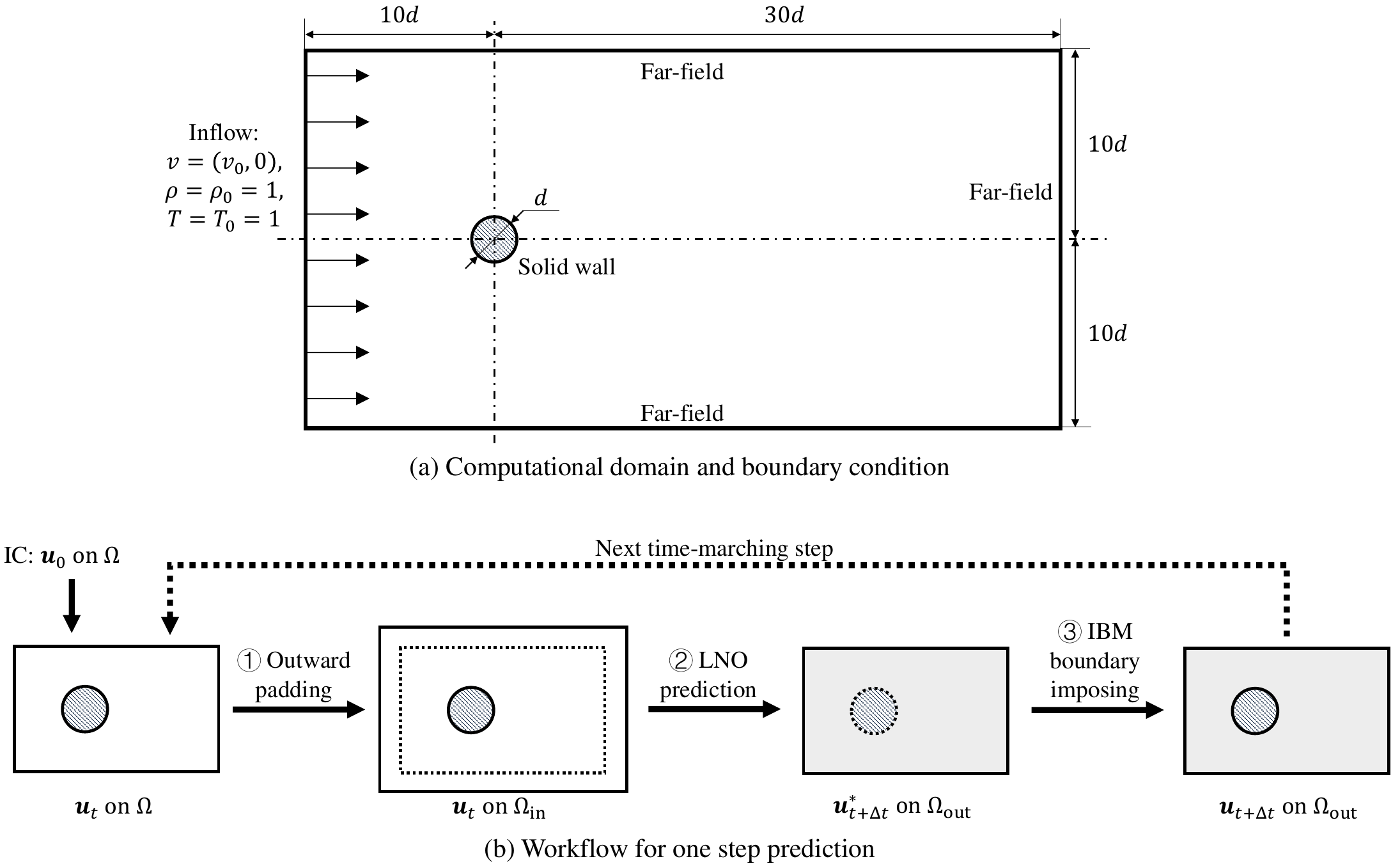}
\caption{Schematic diagram for solving flow around a circular cylinder.}\label{fig:CCschematic}
\end{figure}

The first application example is the flow around a circular cylinder, a commonly used benchmark to test the performance of numerical methods in simulating unsteady flows \cite{Lazar2013,Mittal1998}. 
The schematic diagram is shown in Figure~\ref{fig:CCschematic}(a). 
A circular cylinder with diameter $d$ is placed in a 2-D infinite plane. 
The uniform flow comes from the left is with velocity $v_0$, density $\rho_0$, and temperature $T_0$. 
The computational domain is $\left[-10d,30d\right]\times\left[-10d,10d\right]$.
The far-field BC is set on the four sides of the domain.
The solid wall boundary condition is set on the cylinder.
The characteristic Reynolds number for this problem is ${Re}_d=\rho_0v_0d\cdot Re$, where $Re=100$ is the Reynolds number of the learning task that LNOs were trained on.
The inflow density and temperature are set as constant $\rho_0=1$, $T_0=1$, then ${Re}_d$ can be altered by setting different $v_0$ and $d$.

The concrete schedule for applying the pre-trained LNO to solve this problem is illustrated in Figure~\ref{fig:CCschematic}(b).
Firstly, the domain $\Omega$ is extended to $\Omega_\mathrm{in}$ by constant padding operation to treat the far-field BC on the four sides of $\Omega$. 
Then, the input on $\Omega_\mathrm{in}$ is sent to pre-trained LNO to obtain the output $\boldsymbol{u}_{t+\Delta t}^*$.
The last step is to introduce the solid wall BC by imposing a velocity correction on $\boldsymbol{u}_{t+\Delta t}^*$ by IBM to obtain the final prediction $\boldsymbol{u}_{t+\Delta t}$, which also serves as the input for the next time-marching step.
With the initial condition $\boldsymbol{u}_0$ as the first input, the schedule is repeated to predict the long-term solution of the problem.

\begin{figure}[htbp]%
\centering
\includegraphics[width=\textwidth]{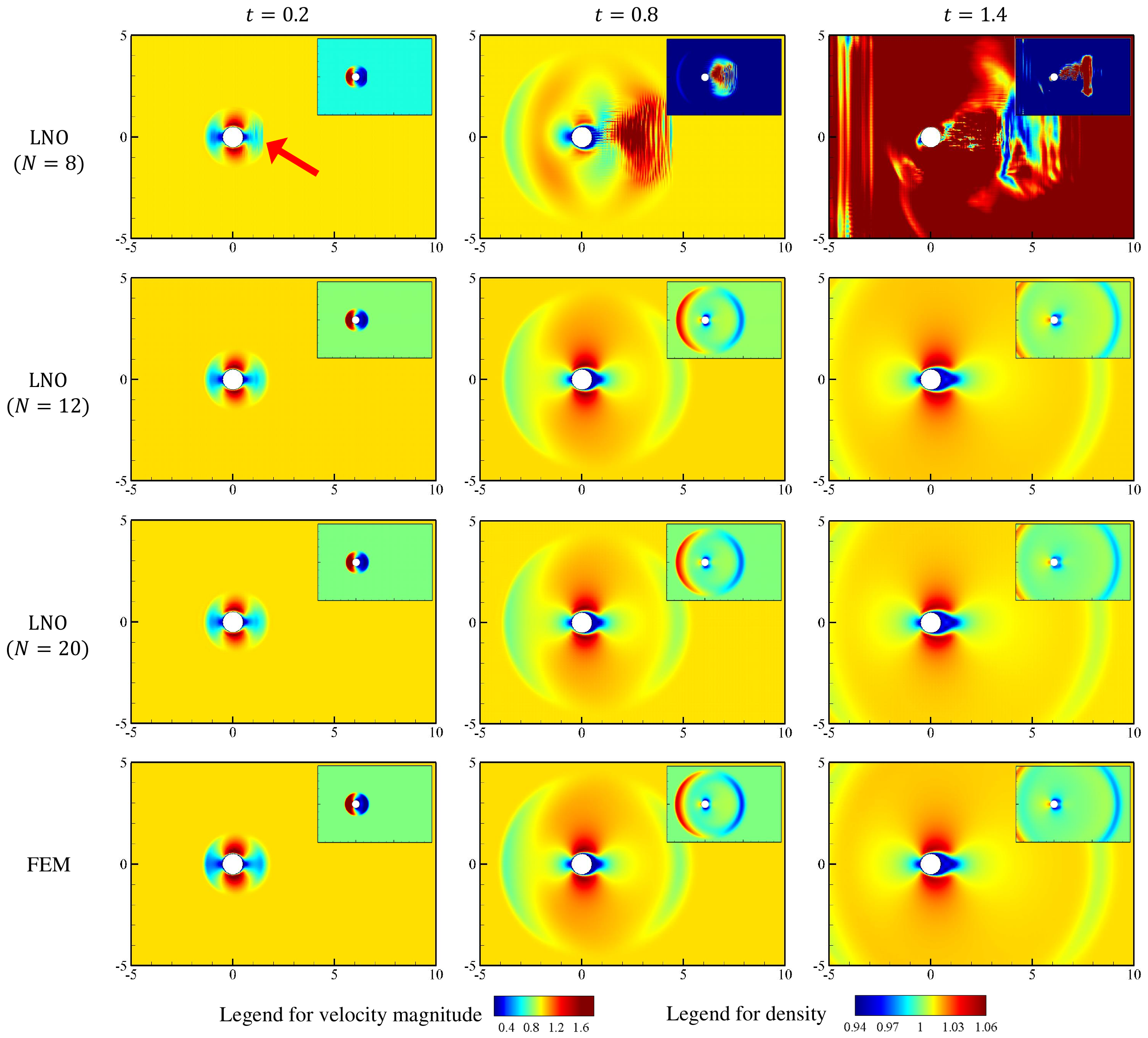}
\caption{ The initial stage of flow around a circular cylinder. The contours of velocity magnitude are shown with contours of density in the upper right corner.}\label{fig:CCinitial}
\end{figure}

\begin{figure}[]%
\centering
\includegraphics[width=0.85\textwidth]{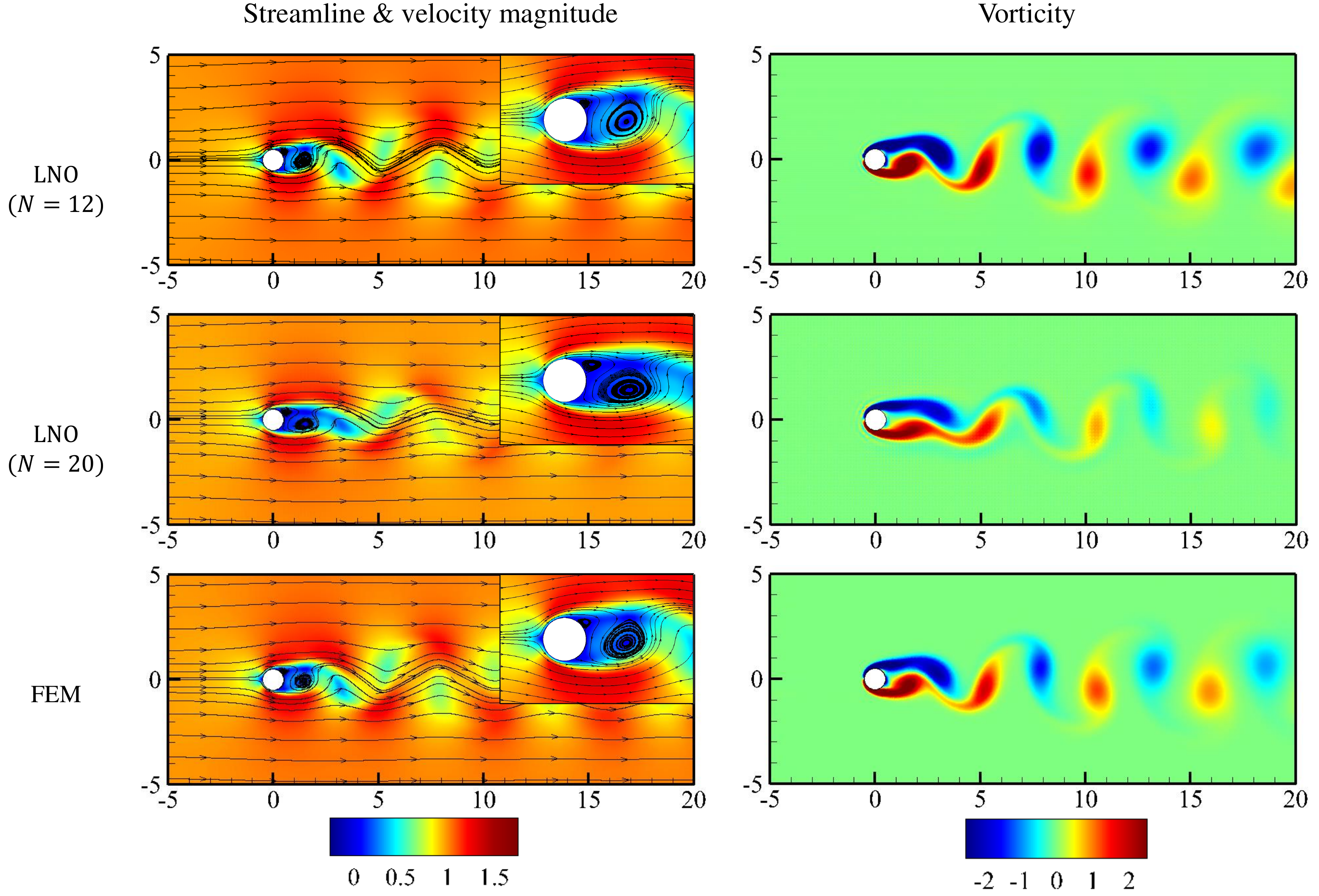}
\caption{The fully developed stage of flow around a circular cylinder.
The left column is the contours of velocity magnitude and streamlines. 
The right column is contours of vorticity.}\label{fig:CCfully}
\end{figure}

First, we employ LNO to predict the flow with $u_0=1$ and $d=1$, i.e., ${Re}_d=100$. 
Figure \ref{fig:CCinitial} shows the contours of density and velocity magnitude at the early stage ($t\le1.4$), and Table \ref{tab:CCerror} lists the error of predicted variables. 
Figure \ref{fig:CCfully} shows the solution predicted by LNOs after reaching the fully developed state ($t\geq120$).
The results show that, due to the effect of the circular cylinder, there are high-speed regions generated on the upper and lower sides and low-speed regions upstream and downstream of the cylinder. 
These regions grow larger as $t$ increases. 
Eventually, the solution field shows periodic vortex shedding i.e., the Von Kármán Vortices, which is expected to appear as ${Re}_d=100$ \cite{Williamson1988}. 
Among the early-stage results in Figure \ref{fig:CCinitial}, an independent wave can be found propagating around in a faster speed than the regions mentioned above.
It is the sound wave generated by the cylinder at the beginning of the prediction.
By comparing the contours with the reference solution, it is found that all these phenomena are captured correctly by LNOs (except for $N=8$ which is discussed later). 
It is worth noting that during the training process, the information of the sound wave mode and the convection mode is provided together by the training samples without any separation or extra mark.
The correct prediction of the initial sound wave suggests LNO successfully learns the intrinsic law from these random and disorganized samples. 

\begin{table}[]
\footnotesize
\renewcommand{\arraystretch}{0.8}
\centering
    \caption{Mean $L_2$ error for predicting flow around a circular cylinder with pre-trained LNOs. The reference solution is calculated by FEM.\label{tab:CCerror}}
\begin{tabular*}{\textwidth}{@{\extracolsep{\fill}}ccccc}
\toprule
\multirow{2}{*}{LNO}    & \multirow{2}{*}{Error} & \multicolumn{3}{c}{Mean $L_2$ error at different $t$} \\
\cline{3-5}
                        &                        & $t=0.2$     & $t=0.8$     & $t=1.4$     \\
                        \midrule
\multirow{3}{*}{$N=8$}  & $e_t^\rho$             & 0.0089      & 0.0652      & 0.1956      \\
                        & $e_t^T$                & 0.0043      & 0.0216      & 0.0599      \\
                        & $e_t^v$                & 0.0365      & 0.0975      & 1.1971      \\
                        \midrule
\multirow{3}{*}{$N=12$} & $e_t^\rho$             & 0.0012      & 0.0024      & 0.0035      \\
                        & $e_t^T$                & 0.0004      & 0.0011      & 0.0018      \\
                        & $e_t^v$                & 0.0279      & 0.0322      & 0.0354      \\
                        \midrule
\multirow{3}{*}{$N=20$} & $e_t^\rho$             & 0.0006      & 0.0009      & 0.0011      \\
                        & $e_t^T$                & 0.0004      & 0.0006      & 0.0008      \\
                        & $e_t^v$                & 0.0311      & 0.0392      & 0.0360      \\
                        \bottomrule
\end{tabular*}
\end{table}

These results also intuitively show how the locality of LNO impacts its performance in applications. 
In Figure \ref{fig:CCinitial}, the LNOs with compatible or large ERR ($N=12$ and $20$) carry out similar results with the reference solution, while LNO with small ERR ($N=8$) leads to oscillation and eventually blows up. 
It is not only because the large prediction error accumulates during the time marching process but more importantly due to the insufficient receptive range of LNO ($N=8$). 
Clear evidence is that in the contour of $t=0.2$, the right side of the sound wave is cut off non-physically (marked with a red arrow in Figure \ref{fig:CCinitial}). 
Further, Figure \ref{fig:CCfully} visualizes the difference between LNOs with compatible and large ERR ($N=12$ and $20$).
Although both $N=12$ and $20$ successfully predict the vortex street with clear streamlines presented, the intensity of vortex shedding of $N=20$ is lower than $N=12$ and the reference solution. 
It may be because only the first $M$ low-order modes are used in the spectral path of LNO, $N=20$ means that more modes and energy are abandoned than smaller $N$, leading to the loss of flow details.
This problem may be relieved by choosing a larger $M$, but it would exponentially increase the number of trainable weights and render the network training much harder.

\begin{figure}[]%
\centering
\includegraphics[width=0.75\textwidth]{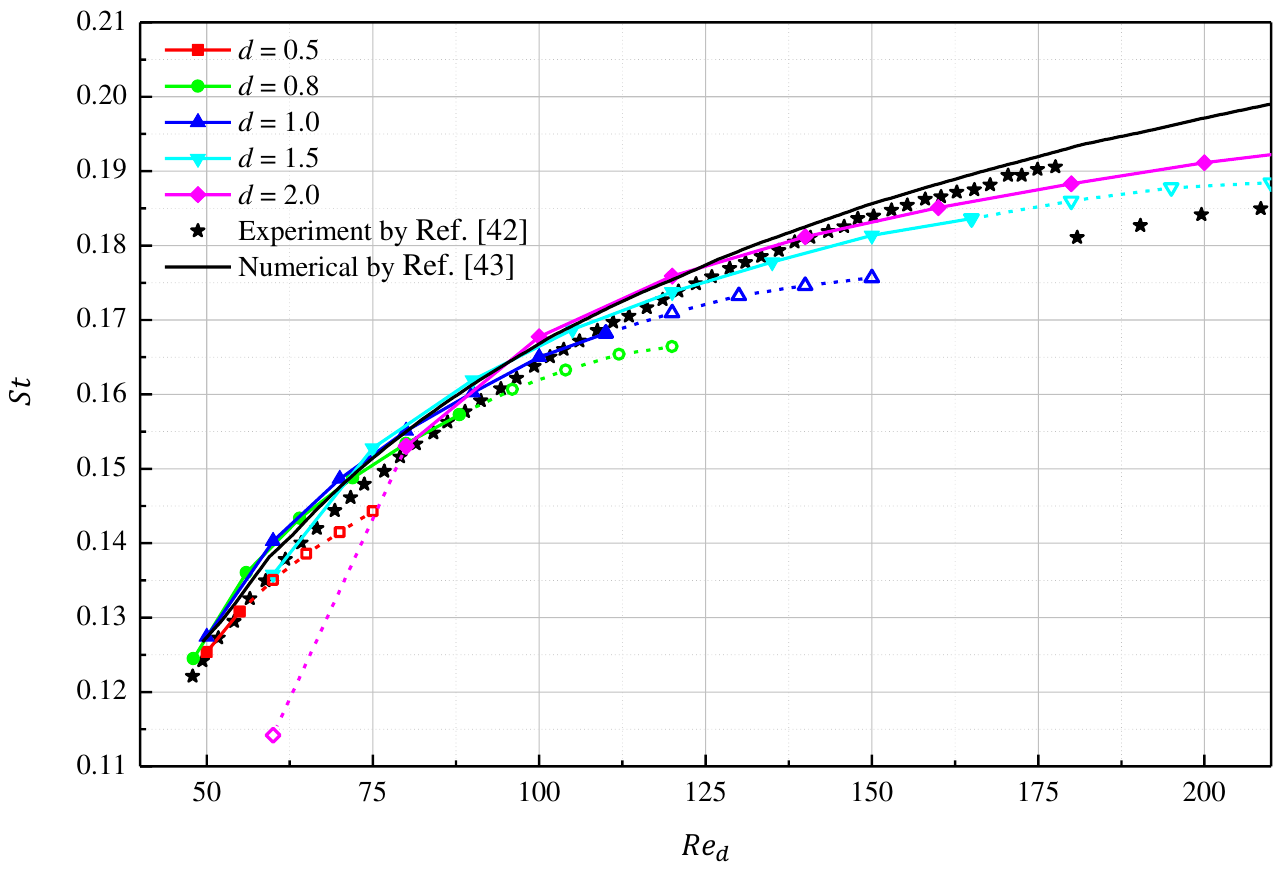}
\caption{The Strouhal-Reynolds number curve for flow around a circular cylinder. 
The dotted lines with hollow symbols denote cases with inflow velocity $v_0$ out of the optimal range.}\label{fig:StRe}
\end{figure}

To further show the performance of LNO, we employ the pre-trained LNO ($N=12$) to predict the $St-{Re}_d$ curve of this problem.
$St=\frac{fd}{v_0}$ is the Strouhal number for describing the frequency of fluctuation of fluids (in this case, the vortex shedding). 
$f$ is the frequency. 
Early studies about the flow around a circular cylinder find that $St$ is a single-valued function of ${Re}_d$ in a certain range \cite{Williamson1988,Henderson1997}.
Here, we change the diameter of the cylinder $d$ and the inflow velocity $v_0$ to achieve LNO prediction and obtain the corresponding $St$ for ${Re}_d=50\sim 200$.
Concretely, we set $d=0.5,0.8,1,1.5,2$ and for each $d$ the inflow velocity $v_0$ varies from $\max\left(0.3,\frac{50}{Re\times d}\right)$ to $\min\left(1.5,\frac{200}{Re\times d}\right)$ (note again that $Re$ and $Re_d$ are the characteristic Reynolds number of the learning task with characteristic length 1 and that of the application case with length $d$, respectively).
$f$ is obtained by recording the time history of $v_x$ and $v_y$ in the downstream of the cylinder and calculating the period of fluctuation after the flow reaches the fully developed state.
The $St-{Re}_d$ curves predicted by LNO ($N=12$) are shown in Figure \ref{fig:StRe}.
Each curve represents results with constant $d$ and changing $v_0$.
It is seen that results from LNO are in good agreement with the results from \cite{Williamson1988,Henderson1997}, showing the capability of LNO for predicting the dynamic process of fluid flow.

A further finding from Figure 16 is that, results at two ends of each curve representing different $d$ deviate from the reference slightly.
It suggests LNO performs best within a specific range of input variables (velocities).
For the present example case, the optimal range for inflow velocity $v_0$ is around $[0.5,1.1]$ which could be related to the velocity distribution in the training data samples (shown in Figure~\ref{fig:datadistribut}).
The optimal range may differ depending on the specific conditions of the problem to be solved, the training data, and the normalization as mentioned in \ref{secA5}, which remains further investigation for establishing a uniform criterion.

\subsection{The flow around a vehicle in a tunnel\label{sec5:2}}
\begin{figure}[htbp]%
\centering
\includegraphics[width=0.8\textwidth]{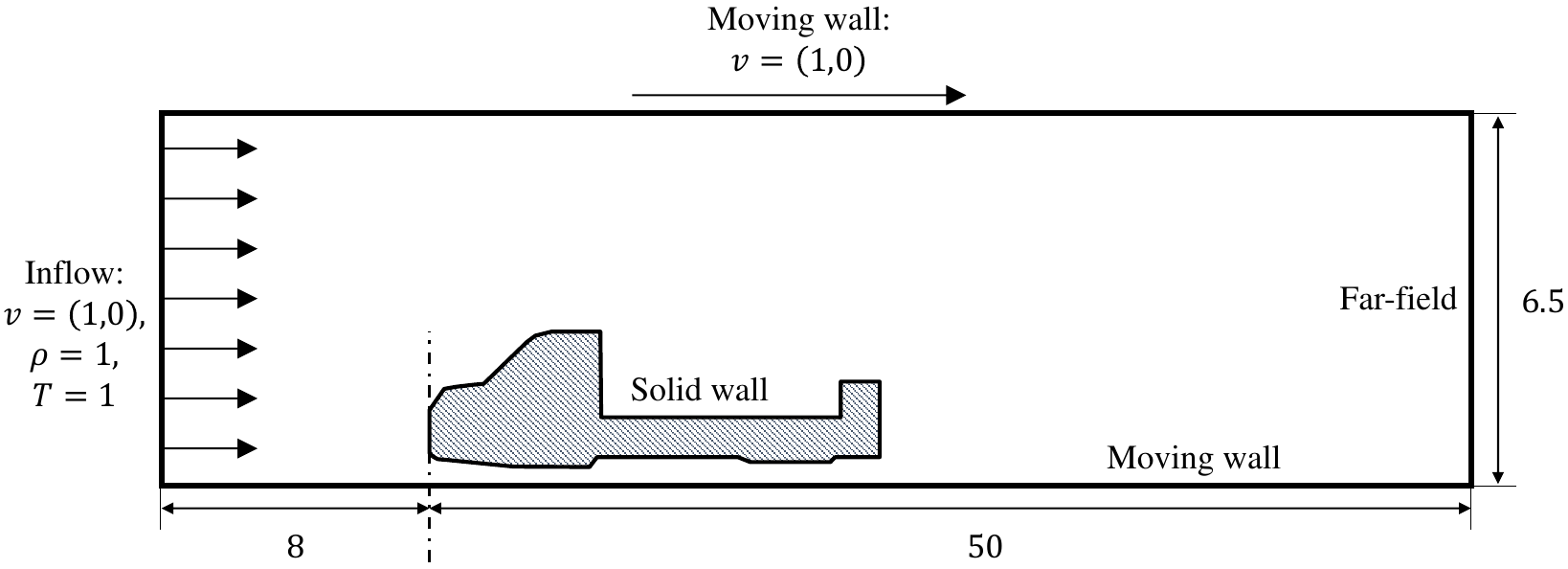}
\caption{Schematic diagram for a vehicle in a tunnel.}\label{fig:Vehicleschematic}
\end{figure}

We next apply the pre-trained LNO in predicting the flow around objects with more complex geometries and different boundary conditions. 
The schematic diagram is in Figure \ref{fig:Vehicleschematic}, a vehicle is placed in a tunnel, with far-field BC on the left and right sides and solid wall BC on the upper and lower sides. 
Set the vehicle as the reference frame, the velocity BC on all four sides is $v_x=v_0=1$, $v_y=0$. 
Three different types of vehicles are chosen as examples, including the sportscar, the sports utility vehicle (SUV), and the truck. 
Their sizes and geometry complexity increase one by one, leading to increasingly complicated flow fields and challenges for LNO prediction.
The schedule to predict this problem is similar to the circular cylinder in Section~\ref{sec5:1}. 
The only difference is that the upper and lower sides are solid wall BCs, so they are treated along with the interior solid wall after one-step LNO prediction rather than with the far-filed BCs.

\begin{figure}[t]%
\centering
\includegraphics[width=\textwidth]{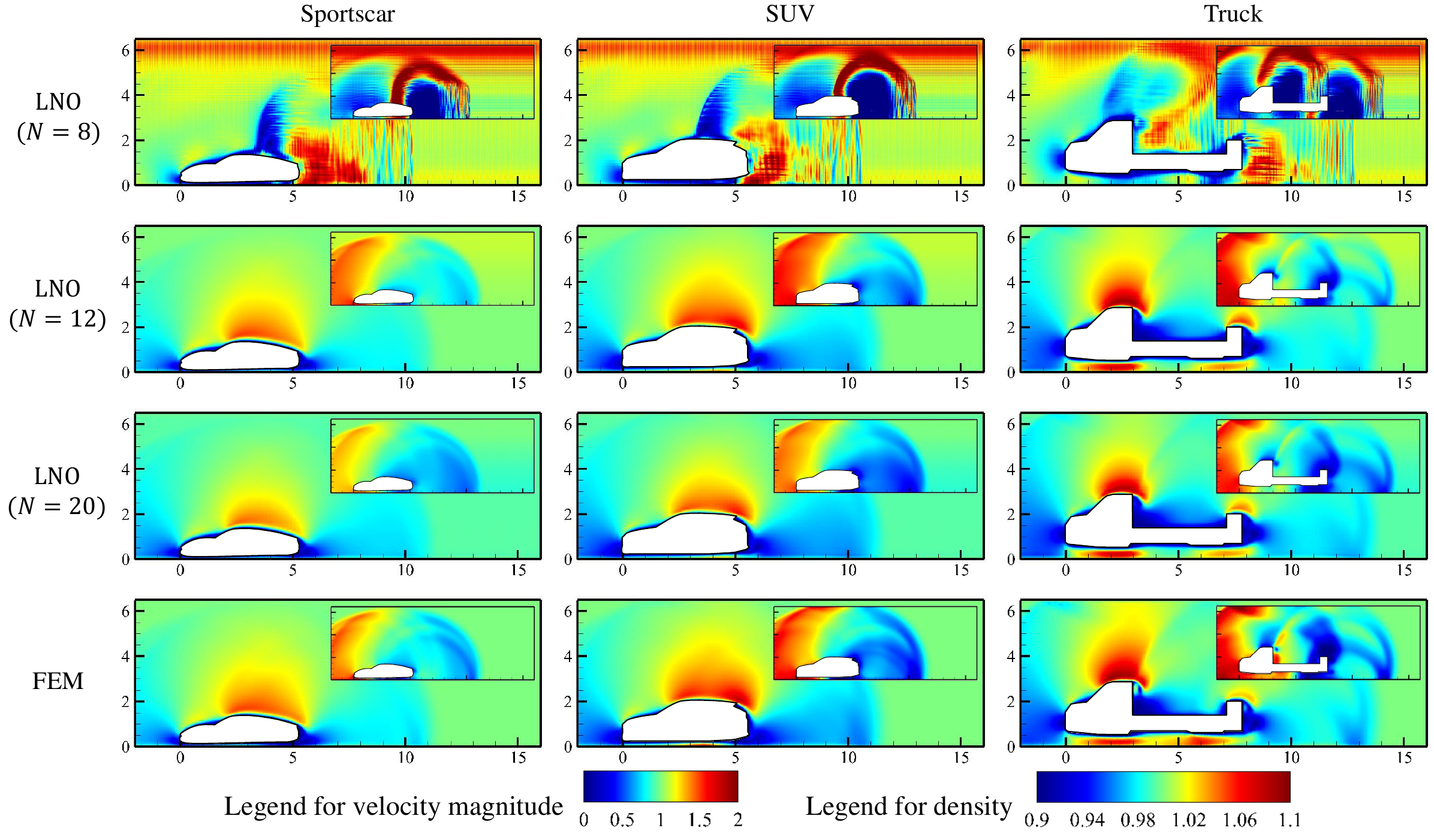}
\caption{The initial stage of flow around a vehicle in a tunnel ($t=1$). 
The contours of velocity magnitude are shown with contours of density in the upper right corner.}\label{fig:Vehicleinitial}
\end{figure}

Figure \ref{fig:Vehicleinitial} shows the flow field at $t=1$ for the three vehicles by LNOs as well as the reference solution by FEM, and Table \ref{tab:Vehicleerror} lists the error of predicted variables. 
The results are similar to the case of the circular cylinder in that LNO ($N=8$) leads to nonphysical oscillation downstream of the vehicle. 
The prediction of LNO ($N=20$) is smooth but not very accurate, especially the high-speed region on the upper side of the vehicle is clearly smaller than the reference solution. 
Among all the results, LNO $\left(N=12\right)$ with the most compatible ERR carries out the prediction closest to the reference solution. 
Not only the initial sound wave but also the reflection of the wave caused by the solid walls are captured correctly.

\begin{table}[t]
\centering
    \caption{Mean $L_2$ error for predicting flows around a vehicle ($t=1$) with LNOs. The reference solution is calculated by FEM.\label{tab:Vehicleerror}}
\footnotesize
\renewcommand{\arraystretch}{0.8}
\begin{tabular*}{\textwidth}{@{\extracolsep{\fill}}ccccc}
\toprule
\multirow{2}{*}{LNO}    & \multirow{2}{*}{Error} & \multicolumn{3}{c}{Mean $L_2$ error at $t=1$} \\
\cline{3-5}
                        &                        & Sportscar     & SUV        & Truck      \\
                        \midrule
\multirow{3}{*}{$N=8$}  & $e_t^\rho$             & 0.0341        & 0.0370     & 0.0385     \\
                        & $e_t^T$                & 0.0176        & 0.0182     & 0.0189     \\
                        & $e_t^v$                & 0.2379        & 0.2467     & 0.2500     \\
                        \midrule
\multirow{3}{*}{$N=12$} & $e_t^\rho$             & 0.0079        & 0.0085     & 0.0095     \\
                        & $e_t^T$                & 0.0016        & 0.0020     & 0.0024     \\
                        & $e_t^v$                & 0.0126        & 0.0153     & 0.0227     \\
                        \midrule
\multirow{3}{*}{$N=20$} & $e_t^\rho$             & 0.0066        & 0.0071     & 0.0083     \\
                        & $e_t^T$                & 0.0020        & 0.0023     & 0.0027     \\
                        & $e_t^v$                & 0.0570        & 0.0576     & 0.0616     \\
                        \bottomrule
\end{tabular*}
\end{table}

\begin{figure}[htbp]%
\centering
\includegraphics[width=\textwidth]{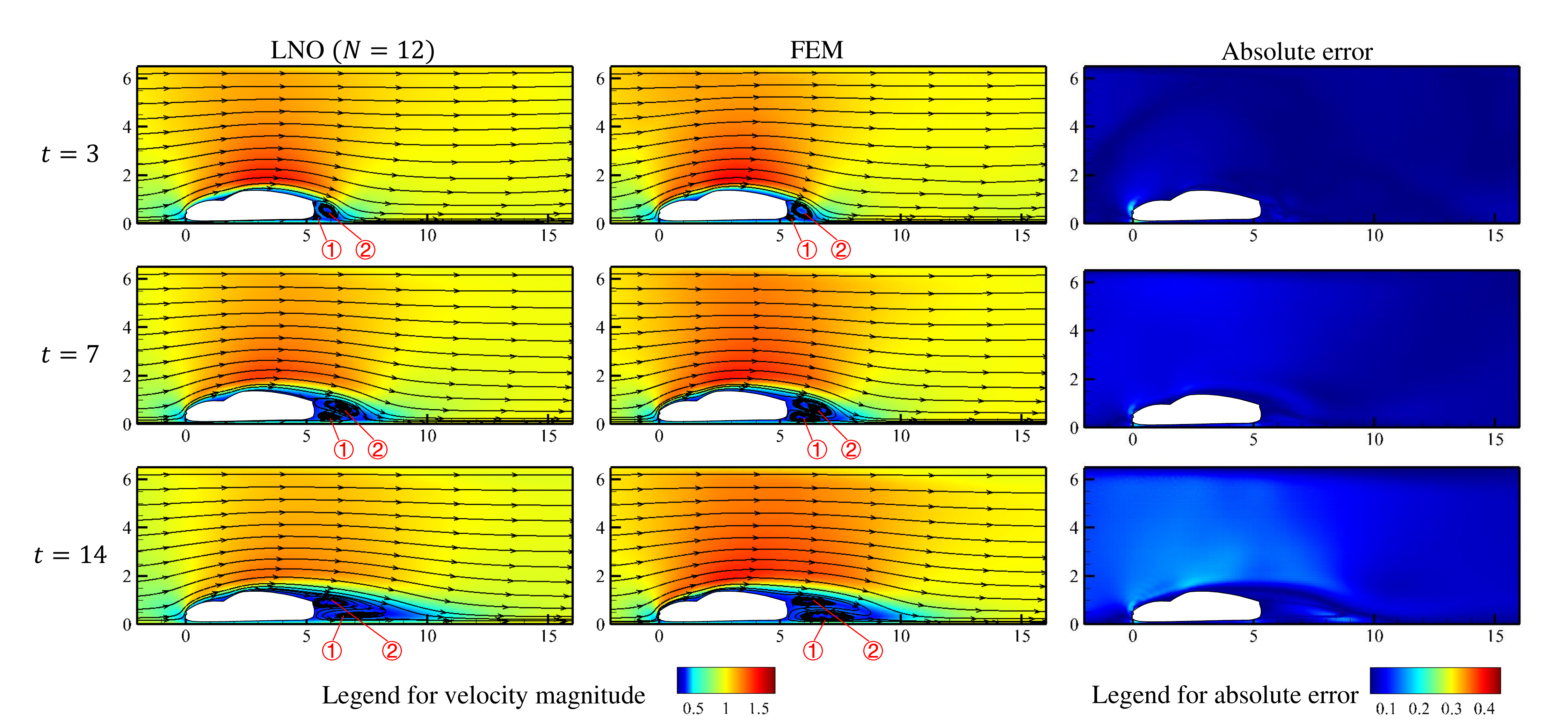}
\caption{The developing process of flow around sportscar: streamline and contours of velocity magnitude (the first and second columns), and absolute error of velocity magnitude (the third column).}\label{fig:sportscar}
\end{figure}

\begin{figure}[htbp]%
\centering
\includegraphics[width=\textwidth]{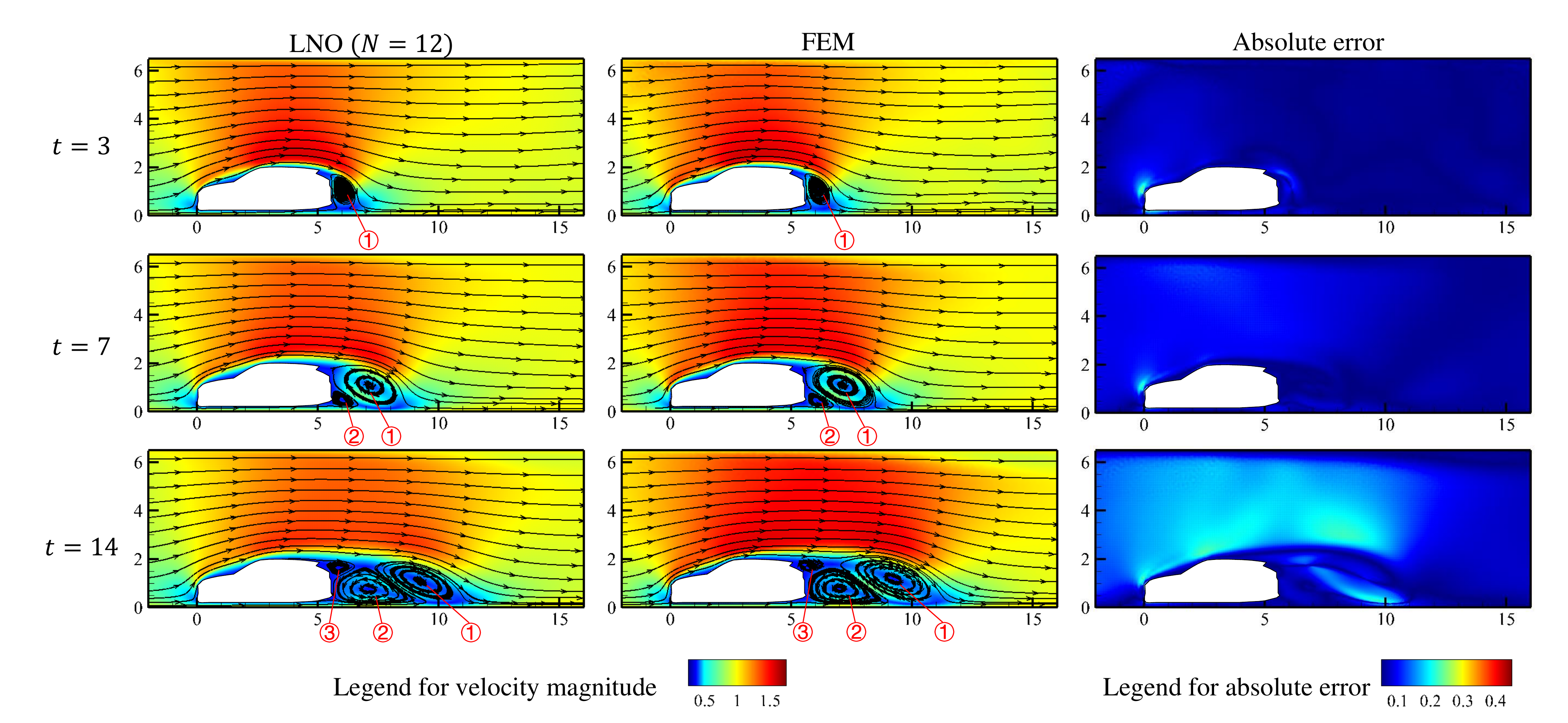}
\caption{The developing process of flow around SUV: streamline and contours of velocity magnitude (the first and second columns), and absolute error of velocity magnitude (the third column).}\label{fig:SUV}
\end{figure}

\begin{figure}[htbp]%
\centering
\includegraphics[width=\textwidth]{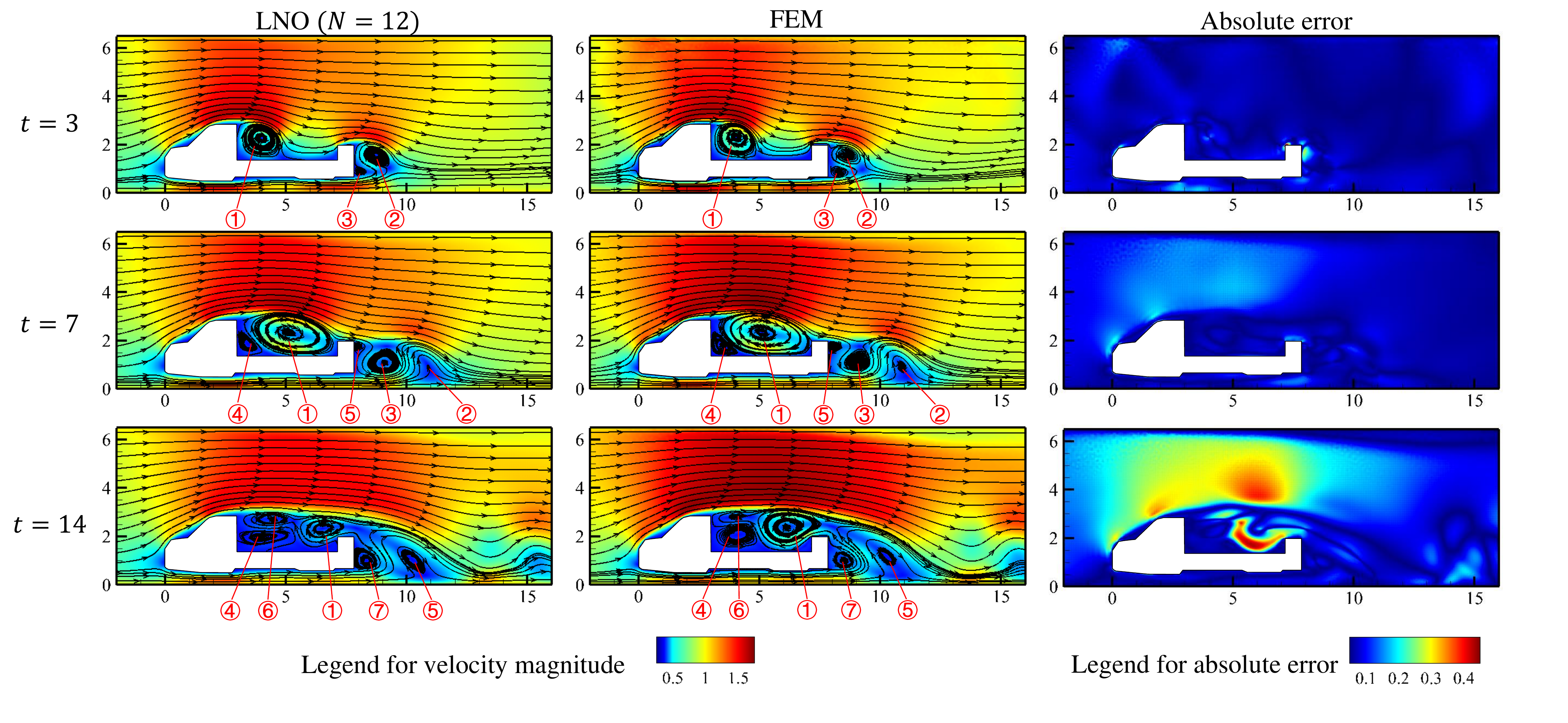}
\caption{The developing process of flow around truck: streamline and contours of velocity magnitude (the first and second columns), and absolute error of velocity magnitude (the third column).}\label{fig:truck}
\end{figure}

We further use LNO ($N=12$) to predict the follow-up dynamic process.
The streamlines and contours of vorticity for three vehicles at $t=3,7,14$ are given in Figures \ref{fig:sportscar}-\ref{fig:truck}. 
For the sportscar (Figure \ref{fig:sportscar}), a pair of vortices are generated at the end of the vehicle and gradually elongated. 
For the SUV (Figure \ref{fig:SUV}), two similar vortices are generated and convected downstream, then, a third vortex is generated at the upper rear of the vehicle. 
For the truck (Figure \ref{fig:truck}), the condition is more complicated. 
Vortices are generated at both the vehicle’s rear and the cargo box.
At $t=14$, there are totally 5 vortices observed. 
All the vortices are captured successfully by LNO. 

In summary, the practical examples shown here support the discovery in Section \ref{sec4} that the locality of LNO and its compatibility with the learning task (the PDEs) strongly affect the performance in predicting the fluid dynamics. 
When the receptive range is too small to cover the range of dependence, nonphysical oscillation may occur and easily lead to divergence, as physical phenomena propagating at high speed can hardly be predicted. 
Conversely, when the receptive range is much larger than the need of learning task, the general flow pattern can be predicted, but the flow details may be lost. 
When the locality of LNO and the learning task is fairly compatible, the pre-trained LNO can predict complex fluid flows with commendable accuracy.

With the essential accuracy guaranteed, the efficiency of the pre-trained LNO becomes the following concern. 
As a reference, here compares the time consumption of LNO prediction and the conventional FEM. 
We provide primary parameter settings as follows to ensure a fair comparison as far as possible. 
The FEM is with linear triangular elements and unstructured meshes. The mesh is coarser in the far-field regions and finer near the solid wall boundaries with the smallest size of $\frac{1}{64}$ to ensure the mesh size is no less than that in LNO prediction.
The total number of nodes is 127443 and 99344 for the case of circular cylinder ($d=1$) and truck, respectively. 
The time discretization adopts the explicit fourth-order four-stage Runge-Kutta scheme with $\Delta t=0.001$, which is the maximum allowable time interval according to our numerical experiment. 
To predict the flow around the circular cylinder ($d=1$) until $t=0.05$, LNO costs 0.425 seconds, while FEM costs 43.722 seconds. 
The speedup ratio is 102.8.
To predict the flow around the truck until $t=0.05$, LNO and FEM cost 0.210 seconds and 28.374 seconds, respectively, so the speedup ratio is 135.1. 
Generally, LNO shows superior computational efficiency compared with conventional numerical schemes.

\section{Conclusions \label{sec6}}
This study focused on the essential characteristic of local neural operator (LNO), i.e., the locality.
Two measurements are raised from the receptive field for describing the locality: the maximum receptive range (MRR) and effective receptive range (ERR), which are the maximum related distance and the second-order central moment of the receptive field.
Then, we trained and validated LNOs with different hyper-parameters on the learning task of compressible fluid dynamics to investigate how the locality acts in LNO learning. 
The results show that ERR of LNO changes during training to approach a special value. 
We deem this special value an intrinsic character of the real target time-marching operator, which denotes its effective local-related range. 
In view of the LNO performance, the LNO with proper initial ERR (referred to as the LNO with compatible locality to the learning task) achieves the best performance. 
LNO architecture with improper (whether insufficient or excessive) locality limits the approach and negatively affects the accuracy. 
At last, the pre-trained LNOs with different compatibilities were applied to predict two unsteady flows to further confirm our findings and show the ability of pre-trained LNOs as practical CFD solvers in complex problems.

The present work introduced a new perspective for explaining the performance and guiding the design of neural operators regarding the locality.
Investigation in this work reveals that a compatible locality is a primary requirement for LNO learning.
Towards a compatible locality, we should first clarify the local-related range of the real time-marching operator to be learned concerning the equations of physics and the selected time interval $\Delta t$.
The other thing to do is about LNO architecting.
We should adequately design and choose the architecture and hyperparameters of LNO to ensure that i) the MRR defined in Eq. (\ref{eq:MRRdef}) is greater than the upper bound of the local-related range of the learning task, ii) the ERR defined in Eq. (\ref{eq:ERRdef}) is as close as possible to the real one. 
Though the ERR of the real operator is unknown at first, we can still obtain an approximation of it through several trial training, and we can obtain a direction for optimizing toward the real ERR by monitoring the change of ERR before and after LNO training. 
Moreover, the analysis and conclusion are general for not only the present fluid dynamics and N-S equations but also other transient PDEs describing the physics. 
This work is the first time to introduce the locality of both the physical problem and the transient PDE into the architecture design of neural operators as a guiding principle, which which could be helpful for improving the performance and interpretability of neural operators.

Beyond the locality of LNO, there are many worth-exploration topics in the future. 
First, more attempts on various network architectures beyond the present physical path and spectral path are feasible based on the LNO definition.
The present analysis and principle on locality are still applicable. 
Second, the interpretability of LNO could be an interesting topic, and efforts on it can help to improve LNO and our understanding of it, e.g., the contours of receptive fields (Figure \ref{fig:receptivecomponent}) may provide information more than locality.
Overall, the LNO framework is worth further developing to approach a possible revolution of AI-powered numerical computation.

\section*{Acknowledgements}

This work was supported by the National Natural Science Foundation of China (Grant No. 52176043).

\section*{Code availability}
The code is available at https://github.com/PPhub-hy/torch-lno-compressible-fluid-dynamics.

\bibliographystyle{elsarticle-num}

\nomenclature[C,01]{$\boldsymbol{v}=\{v_x,v_y\}$}{Velocity with its two components }
\nomenclature[C,02]{$\rho$}{Density}
\nomenclature[C,03]{$T$}{Temperature}
\nomenclature[C,04]{$p$}{Pressure}
\nomenclature[C,05]{$\boldsymbol{\tau}$}{Tensor of viscous stress}
\nomenclature[C,06]{$E$}{Total energy}
\nomenclature[C,07]{$\mu$}{Viscosity}
\nomenclature[C,08]{$R$}{Gas constant}
\nomenclature[C,09]{$C_\mathrm{v}$}{Heat capacity}
\nomenclature[C,10]{$\kappa$}{Thermal conductivity}
\nomenclature[C,11]{$Re$}{Reynolds number}
\nomenclature[C,12]{$Ma$}{Mach number}
\nomenclature[C,13]{$Pr$}{Prandtl number}
\nomenclature[C,14]{$\gamma$}{Specific heat ratio}
\nomenclature[C,15]{$r_\textup{dep},r_\textup{inv},r_\textup{visc}$}{Overall range of dependence, range of dependence for the inviscid step and viscous step}

\nomenclature[L,01]{$\mathcal{G}_\mathrm{L}$}{Target local-related time-marching operator of transient PDEs}
\nomenclature[L,02]{$\mathcal{G}_\theta$}{LNO for approximating $\mathcal{G}$ with  the set $\theta$ of trainable weights}
\nomenclature[L,03]{$\boldsymbol{u}_t$}{Physical fields at time $t$}
\nomenclature[L,04]{$\tilde{\boldsymbol{u}}_{t+\Delta t}$}{The predicted physical fields at time $t+\Delta t$ by LNO}
\nomenclature[L,05]{$D_1,D_2$}{Representational output/input domain}
\nomenclature[L,06]{$\Omega_1,\Omega_2$}{Complete output/input domain}
\nomenclature[L,07]{$\mathcal{L}$}{Loss function}
\nomenclature[L,08]{$\mathcal{P}$}{Pointwise operation}
\nomenclature[L,09]{$\mathcal{A}$}{Activation function}
\nomenclature[L,10]{$\mathcal{C}$}{Convolutional operation}
\nomenclature[L,11]{$\mathcal{T},\mathcal{T}^{-1}$}{Legendre transform and its inverse}
\nomenclature[L,12]{$\mathcal{W},\boldsymbol{W}$}{Linear transform and its learnable weight}
\nomenclature[L,13]{$n$}{Block number}
\nomenclature[L,14]{$N$}{Width of local spectral transform}
\nomenclature[L,15]{$K$}{Number of repetitions for geometry decomposition}
\nomenclature[L,16]{$M$}{Number of spectral modes adopted}
\nomenclature[L,17]{$\boldsymbol{v}^{(i)}$}{The $i^\textup{th}$ intermediate tensors}
\nomenclature[L,18]{$\boldsymbol{\psi},\boldsymbol{\varphi}, \psi_{mij},\varphi_{mij}$}{Forward and backward Legendre kernel with their components}
\nomenclature[L,19]{$e^\rho_t,e^T_t,e^v_t$}{Error for density, temperature, and velocity at time $t$ in LNO prediction}
\nomenclature[L,20]{${\bar{e}}^\rho,{\bar{e}}^T,{\bar{e}}^v$}{Time-averaged error for density, temperature, and velocity in LNO prediction}
\nomenclature[L,21]{$F,\bar{F},\bar{F}_0$}{Two-point Receptive fields, receptive fields of one output point, and that averaged in the output unit domain}
\nomenclature[L,22]{$r_\textup{LNO},r_{n=1}$}{Maximum receptive range for LNO and one block}
\nomenclature[L,23]{$\varrho,\varrho_\textup{init},\varrho_\textup{trained}$}{Effective receptive range, and that of the initialized or trained LNO}

\nomenclature[W,01]{$\sigma^2$}{Variance of random input}
\nomenclature[W,02]{$c_\textup{in},c_\textup{out}$}{Number of input and output channels for one layer}
\nomenclature[W,03]{$k$}{Size of the convolutional kernels}
\nomenclature[W,04]{$n_d$}{Number of dimensions}
\nomenclature[W,05]{$\mathrm{\Theta}_{N,K,M}$}{Initialization factor of the spectral path regarding the LNO parameters $N$,$K$,and $M$}

\printnomenclature
\emph{Remark: Italic symbols are variables; bold symbols are tensors (including vectors and higher-order ones). }

\appendix

\section{The derivation of solution for the viscous step\label{secA2}}
The partial differential equation to be solved in the viscous step is:
\begin{equation}
    \frac{\partial v_x}{\partial t}=\frac{\mu}{\rho}\left(\frac{\partial \tau_{x x}}{\partial x}+\frac{\partial \tau_{x y}}{\partial y}\right)=\frac{\mu}{\rho}\left(\frac{4}{3} \frac{\partial^2 v_x}{\partial x^2}+\frac{\partial^2 v_x}{\partial y^2}+\frac{1}{3} \frac{\partial^2 v_y}{\partial x \partial y}\right),
    \label{eq:A1}
\end{equation}
\begin{equation}
    \frac{\partial v_y}{\partial t}=\frac{\mu}{\rho}\left(\frac{\partial \tau_{y x}}{\partial x}+\frac{\partial \tau_{y y}}{\partial y}\right)=\frac{\mu}{\rho}\left(\frac{\partial^2 v_y}{\partial x^2}+\frac{4}{3} \frac{\partial^2 v_y}{\partial y^2}+\frac{1}{3} \frac{\partial^2 v_x}{\partial x \partial y}\right),
    \label{eq:A2}
\end{equation}
with initial condition:
\begin{equation}
    v_x=\delta(x,y),\quad v_y=0.
    \label{eq:A3}
\end{equation}
Noting that no viscous term presents in the continuity equation, $\frac{\partial\rho}{\partial t}=0$, i.e., $\rho=\textup{const}$ during the viscous step, $\rho$ is moved to the right-hand side of the equation.

According to Fourier transform, $v_x$, $v_y$ can be written as:
\begin{equation}
v_x(x, y)=\int_{-\infty}^{\infty} \int_{-\infty}^{\infty} \hat{v}_x\left(\omega_x, \omega_y\right) e^{i\left(\omega_x x+\omega_y y\right)} d \omega_x d \omega_y,
\label{eq:A4}
\end{equation}
\begin{equation}
v_y(x, y)=\int_{-\infty}^{\infty} \int_{-\infty}^{\infty} \hat{v}_y\left(\omega_x, \omega_y\right) e^{i\left(\omega_x x+\omega_y y\right)} d \omega_x d \omega_y.
\label{eq:A5}
\end{equation}
Substitute Eqs. (\ref{eq:A4}-\ref{eq:A5}) into Eqs. (\ref{eq:A1}-\ref{eq:A2}), then the following equation should be satisfied for any $\omega_x$, $\omega_y$:
\begin{equation}
\frac{d \hat{v}_x}{d t}=-\frac{\mu}{\rho}\left[\left(\frac{4}{3} \omega_x^2+\omega_y^2\right) \hat{v}_x+\frac{1}{3} \omega_x \omega_y \hat{v}_y\right],
\label{eq:A6}
\end{equation}
\begin{equation}
\frac{d \hat{v}_y}{d t}=-\frac{\mu}{\rho}\left[\frac{1}{3} \omega_x \omega_y \hat{v}_x+\left(\omega_x^2+\frac{4}{3} \omega_y^2\right) \hat{v}_y\right].
\label{eq:A7}
\end{equation}
Substitute Eq. (\ref{eq:A7}) into Eq. (\ref{eq:A6}) to remove all the terms containing ${\hat{v}}_x$ and get an ordinary differential equation for ${\hat{v}}_y$:
\begin{equation}
    \frac{d^2 \hat{v}_y}{d t^2}+\frac{7 \mu}{3 \rho}\left(\omega_x^2+\omega_y^2\right) \frac{d \hat{v}_y}{d t}+\frac{4 \mu^2}{3 \rho^2}\left(\omega_x^2+\omega_y^2\right)^2 \hat{v}_y=0.
    \label{eq:A8}
\end{equation}
The general solution for Eq. (\ref{eq:A8}) is:
\begin{equation}
    \hat{v}_y=c_1 e^{-\frac{\mu}{\rho}\left(\omega_x^2+\omega_y^2\right) t}+c_2 e^{-\frac{4 \mu}{3 \rho}\left(\omega_x^2+\omega_y^2\right) t},
    \label{eq:A9}
\end{equation}
where $c_1$, $c_2$ are undetermined constants. 
Then, the general solution of ${\hat{v}}_x$ is obtained by substituting Eq. (\ref{eq:A9}) into Eq. (\ref{eq:A7}):
\begin{equation}
    \hat{v}_x=\frac{1}{\omega_x \omega_y}\left[-\omega_y^2 c_1 e^{-\frac{\mu}{\rho}\left(\omega_x^2+\omega_y^2\right) t}+\omega_x^2 c_2 e^{-\frac{4 \mu}{3 \rho}\left(\omega_x^2+\omega_y^2\right) t}\right] .
    \label{eq:A10}
\end{equation}
The initial condition Eq. (\ref{eq:A3}) in Fourier space is:
\begin{equation}
    \hat{v}_x(t=0)=1,\quad \hat{v}_y(t=0)=0.
\end{equation}
Thus $c_1$, $c_2$ are determined as:
\begin{equation}
    c_1=-\frac{\omega_x\omega_y}{\omega_x^2+\omega_y^2},\quad
    c_2=\frac{\omega_x\omega_y}{\omega_x^2+\omega_y^2}.
    \label{eq:A12}
\end{equation}
Combining Eq. (\ref{eq:A9}), Eq. (\ref{eq:A10}), and Eq. (\ref{eq:A12}), we obtain the solution for the viscous step:
\begin{equation}
\hat{v}_x=\frac{\omega_y^2}{\omega_x^2+\omega_y^2} e^{-\frac{\mu}{\rho}\left(\omega_x^2+\omega_y^2\right) t}+\frac{\omega_x^2}{\omega_x^2+\omega_y^2} e^{-\frac{4 \mu}{3 \rho}\left(\omega_x^2+\omega_y^2\right) t}, 
\end{equation}
\begin{equation}
\hat{v}_y=\frac{\omega_x \omega_y}{\omega_x^2+\omega_y^2}\left[-e^{-\frac{\mu}{\rho}\left(\omega_x^2+\omega_y^2\right) t}+e^{-\frac{4 \mu}{3 \rho}\left(\omega_x^2+\omega_y^2\right) t}\right],
\end{equation}
\begin{equation}
v_x=\frac{1}{(2 \pi)^2} \int_{-\infty}^{\infty} \int_{-\infty}^{\infty} \hat{v}_x e^{i\left(\omega_x x+\omega_y y\right)} d \omega_x d \omega_y,
\end{equation}
\begin{equation}
v_y=\frac{1}{(2 \pi)^2} \int_{-\infty}^{\infty} \int_{-\infty}^{\infty} \hat{v}_y e^{i\left(\omega_x x+\omega_y y\right)} d \omega_x d \omega_y .
\end{equation}

\begin{figure}[]%
\centering
\includegraphics[width=0.45\textwidth]{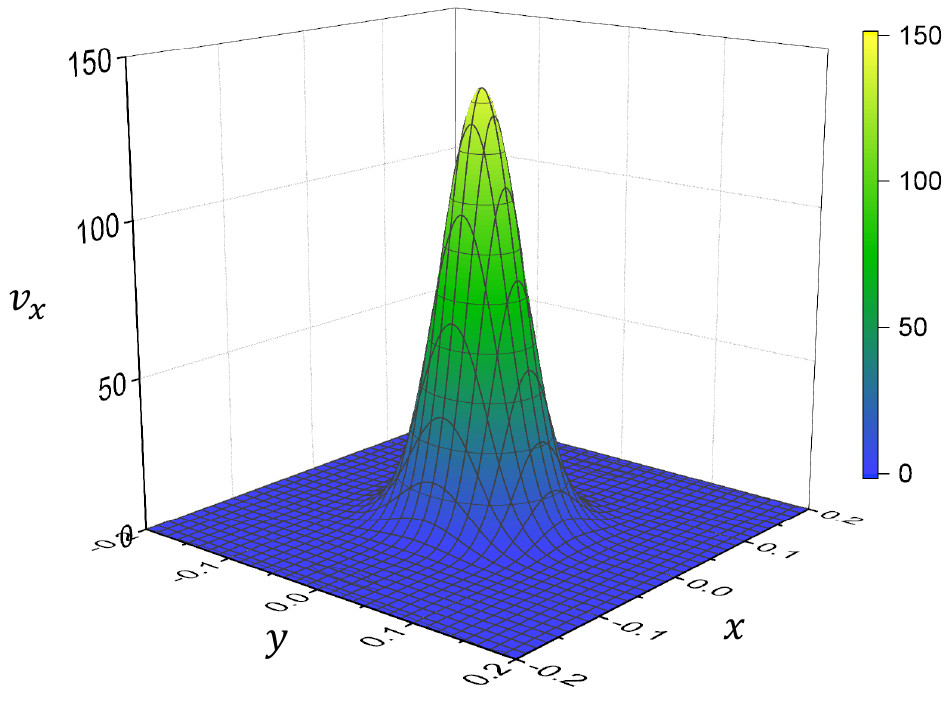}
\caption{The distribution of $v_x$ at $t=0.05$ with $\mu=0.01$, $\rho=1$ generated by an initial disturbance $v_x=\delta\left(x,y\right)$ at $t=0$.}\label{fig:viscousdist}
\end{figure}

As an example, the distribution of $v_x$ with $\mu=0.01,\rho=1,t=0.05$ is shown in Figure \ref{fig:viscousdist}, here the effect of non-constant density is ignored.
It can be found that the value decreases rapidly as the distance to $x=0$ increases, which allows us to define the range of dependence in a numerically truncated way.

Here we choose the effect on velocity by the disturbance of $v_x$ as the measure of $r_\textup{visc}$.
It is also feasible to choose the disturbance of $v_y$ or $E$ and the effect on other physical values, which does not affect the main conclusion of this part, that is, to help us estimate the range of dependence of the viscous terms.

\section{Proof for formula (\ref{eq:DEPrange})\label{secA3}}
First, we define the range of dependence for the proof. 1-D condition is considered here for convenience.
A disturbance $u=\delta(x)$ is produced at $t=0$ and propagates till $t=\tau$ in domain $\Omega$.
The distribution of $\left |u_{\tau}(x)\right |$ is assumed to be continuous and monotonic decreasing.
Additionally, $\left ||u_{\tau}(x)\right |\rightarrow 0$ when $x\rightarrow\infty$.
Here only $x\geq0$ is considered for example. 
Then the range of dependence $r_{\textup{dep}}(\tau)$ is the position that $\left |u_{\tau}\right |$ decreases to a certain value $\eta$:
\begin{equation}
    \left |u_{\tau}(r_{\textup{dep}(\tau)})\right |=\eta.
\end{equation}

For the split solution $\tilde{u}_{\tau}$ obtained by the time-splitting method, there is a ‘splitting’ range of dependence ${\tilde{r}}_\textup{dep}(\tau)$ that
\begin{equation}
    \left |\tilde{u}_{\tau}(\tilde{r}_\textup{dep}(\tau))\right |=\eta.
\end{equation}
${\tilde{r}}_\textup{dep}$ is equivalent to $r_\textup{visc}+r_\textup{inv}$ in Formula (\ref{eq:DEPrange}).

According to the theory of time-splitting method \cite{Guermond2006}, the error between the splitting solution $\tilde{u}_{\tau}$ and the analytical solution ${u}_{\tau}$ can be bounded by:
\begin{equation}
    \left |\tilde{u}_{\tau}(x)-u_{\tau}(x)\right |<\lambda\tau,\quad \lambda>0,\forall x\in \Omega.
\end{equation}
Thus, the value of ${u}_{\tau}$ at ${\tilde{r}}_\textup{dep}$ can be bounded by
\begin{equation}
    \eta-\lambda\tau<\left |u_{\tau}(\tilde{r}_{\textup{dep}})\right |<\eta+\lambda\tau.
    \label{eq:B4}
\end{equation}
Formula (\ref{eq:DEPrange}) asks for the bound of $\tau$ for a fixed $\varepsilon$ to satisfy
\begin{equation}
    \left |\tilde{r}_{\textup{dep}}(\tau)-{r}_{\textup{dep}}(\tau)\right |<\varepsilon,\quad \forall \tau<\Delta t,
    \label{eq:B5}
\end{equation}
i.e.,
\begin{equation}
    {r}_{\textup{dep}}(\tau)-\varepsilon<\tilde{r}_{\textup{dep}}(\tau)<{r}_{\textup{dep}}(\tau)+\varepsilon.
\end{equation}
Recall that $\left |u\right |$ is monotonic decreasing, $\left |u\right |$ satisfies
\begin{equation}
    \left |u_{\tau}({r}_{\textup{dep}}+\varepsilon)\right |<\left |u_{\tau}(\tilde{r}_{\textup{dep}})\right |<\left |u_{\tau}\left (\max({r}_{\textup{dep}}-\varepsilon,0)\right )\right |.
    \label{eq:B7}
\end{equation}
By comparing Eq. (\ref{eq:B4}) and Eq. (\ref{eq:B7}), Eq. (\ref{eq:B5}) is satisfied when
\begin{equation}
    \eta+\lambda\tau\leq \left |u_{\tau}\left (\max({r}_{\textup{dep}}-\varepsilon,0)\right )\right |,
\end{equation}
\begin{equation}
    \eta-\lambda\tau\geq \left |u_{\tau}({r}_{\textup{dep}}(\tau)+\varepsilon)\right |.
\end{equation}
This leads to the bound of $\tau$ that
\begin{equation}
    \tau\leq \frac{\left |u_{\tau}\left (\max({r}_{\textup{dep}}-\varepsilon,0)\right )\right |-\eta}{\lambda},
    \label{eq:tau_bound1}
\end{equation}
\begin{equation}
    \tau\leq \frac{\eta-\left |u_{\tau}({r}_{\textup{dep}}(\tau)+\varepsilon)\right |}{\lambda}.
    \label{eq:tau_bound2}
\end{equation}
The left and right hands of Eqs. (\ref{eq:tau_bound1})-(\ref{eq:tau_bound2}) are both related to $\tau$.
According to the physical feature of $u$, $\left |u_{\tau}({r}_{\textup{dep}}(\tau)-\varepsilon)\right |$ decreases and $\left |u_{\tau}({r}_{\textup{dep}}(\tau)+\varepsilon)\right |$ grows as $\tau$ grows up.
So there exists an upper bound of $\tau$ denoted as $\Delta t$ to make Eqs. (\ref{eq:tau_bound1})-(\ref{eq:tau_bound2}) hold.
Thus, for any $\varepsilon>0$, there exists a certain value of $\Delta t$ by
\begin{equation}
    \Delta t=\min\left(\frac{\left |u_{\tau}\left (\max({r}_{\textup{dep}}-\varepsilon,0)\right )\right |-\eta}{\lambda},\frac{\eta-\left |u_{\Delta t}({r}_{\textup{dep}}+\varepsilon)\right |}{\lambda}\right)>0
\end{equation}
to satisfy
\begin{equation}
    \left |\tilde{r}_{\textup{dep}}(\tau)-{r}_{\textup{dep}}(\tau)\right |<\varepsilon,\quad \forall \tau<\Delta t.
\end{equation}

\section{Formulas of operations in LNO\label{secA4}}
Here introduces formulas of $\mathcal{A},\mathcal{P},\mathcal{C},\mathcal{T},\mathcal{W},\mathcal{T}^{-1}$ in LNO architecture.
We consider a 2-D grid discretization case in the present study.
\begin{itemize}
    \item $\mathcal{A}$: 
    D. Hendrycks et al. \cite{Hendrycks2016} designed the GELU activation by softening the ReLU activation with randomization introduced by the Bernoulli distribution. 
    They give an approximation of GELU as
    \begin{equation}
        \mathcal{A}(x)=0.5 x\left(1+\tanh \left[\sqrt{\frac{2}{\pi}}\left(x+0.044715 x^3\right)\right]\right) .
    \end{equation}
    \item $\mathcal{P}$: 
    With input $\boldsymbol{z}$ of $c_\textup{in}$ channels, output $\boldsymbol{z}^\prime$ of $c_\textup{out}$ channels, and the learnable weight $W\in\mathbb{R}^{c_\textup{in}\times c_\textup{out}}$, the formula of a pointwise operation $\mathcal{P}^{\left(c_\textup{in},c_\textup{out}\right)}$ from $c_\textup{in}$ to $c_\textup{out}$ channel is
    \begin{gather}
    \label{eq:C2}
        \mathcal{P}^{\left(c_\textup{in},c_\textup{out}\right)}:\ \left\{z_m\right\}_{m=1}^{c_\textup{in}}=\boldsymbol{z}\mapsto\boldsymbol{z}^\prime=\left\{z_l^\prime\right\}_{l=1}^{c_\textup{out}},\\
        \textup{i.e.},\quad z_l^\prime=\mathcal{P}^{\left(c_\textup{in},c_\textup{out}\right)}\left(\left\{z_m\right\}_{m=1}^{c_\textup{in}}\right)=\sum_{m=1}^{c_\textup{in}}{W_{ml}z_m},\quad l=1,2,...,c_\textup{out}.\nonumber
    \end{gather}
    \item $\mathcal{C}$: 
    According to \cite{LiHongyu2022}, the physical layers in LNO realized by discretized convolutional layers. 
    With input $\boldsymbol{z}\in\mathbb{R}^{c_\textup{in}\times\left(A+k-1\right)\times\left(B+k-1\right)}$, output $\boldsymbol{z}^\prime\in\mathbb{R}^{c_\textup{out}\times A\times B}$, and the learnable weight $\boldsymbol{W}\in\mathbb{R}^{c_\textup{in}\times c_\textup{out}\times k\times k}$, the formula of a convolutional layer $\mathcal{C}^{\left(c_\textup{in},c_\textup{out}\right)}$ from $c_\textup{in}$ to $c_\textup{out}$ channel is
    \begin{gather}
    \label{eq:C3}
        \mathcal{C}^{\left(c_\textup{in},c_\textup{out}\right)}:\mathbb{R}^{c_\textup{in}\times\left(A+k-1\right)\times\left(B+k-1\right)}\ni\boldsymbol{z}\mapsto\boldsymbol{z}^\prime\in\mathbb{R}^{c_\textup{out}\times A\times B}, \\
        \textup{i.e.},\quad z_{lxy}^\prime=\mathcal{C}^{\left(c_\textup{in},c_\textup{out}\right)}\left(\left\{z_{m(x+i)(y+j)}|1\leq m\leq c_\textup{in}, 1\leq i\leq k, 1\leq j\leq k\right\}\right)=\sum_{m=1}^{c_\textup{in}}\sum_{i=1}^{k}\sum_{j=1}^{k}{W_{mlij}z_{m\left(x+i\right)\left(y+j\right)}},\nonumber \\
        l=1,2,...,c_\textup{out},\quad x=1,2,...,A,\quad y=1,2,...,B. \nonumber
    \end{gather}

    \item $\mathcal{T}^{-1}\circ\mathcal{W}\circ\mathcal{T}$: 
    According to \cite{LiHongyu2022}, the operations in the spectral path are comprised of the Legendre transform $\mathcal{T}$, the linear layer $\mathcal{W}$, and the inverse Legendre transform $\mathcal{T}^{-1}$. 
    These operations are channel-separated, while each channel still owns its independent weight. 
    Considering one representative unit with input $\boldsymbol{z}\in\mathbb{R}^{\frac{N\left(2K-1\right)}{K}\times\frac{N\left(2K-1\right)}{K}}$ of $c$ channels, output $\boldsymbol{z}^\prime\in\mathbb{R}^{\frac{N}{K}\times\frac{N}{K}}$ of $c$ channels, the interior tensors $ \hat{\boldsymbol{z}}\in\mathbb{R}^{M^2\times K\times K}$ and $\hat{\boldsymbol{z}}^\prime\in\mathbb{R}^{M^2\times K\times K}$, the 2-D forward and backward Legendre kernel $\boldsymbol{\varphi}$ and $\boldsymbol{\psi}$ of size $M^2\times N\times N$ (refer to \cite{LiHongyu2022} for detailed expression about $\boldsymbol{\varphi}$ and $\boldsymbol{\psi}$), and the learnable weight $\boldsymbol{W}\in\mathbb{R}^{c\times M^2\times M^2}$, the formulas for $\mathcal{T}^{-1},\mathcal{W},\mathcal{T}$ respectively are
    \begin{gather}
    \label{eq:C4}
        \mathcal{T}:\mathbb{R}^{\frac{N\left(2K-1\right)}{K}\times\frac{N\left(2K-1\right)}{K}}\ni\boldsymbol{z}\mapsto\hat{\boldsymbol{z}}\in\mathbb{R}^{M^2\times K\times K},\\
        \textup{i.e.},\quad  {\hat{z}}_{m^\prime p q}=\mathcal{T}\left(\left\{z_{\left(i+\frac{N}{K}p\right)\left(j+\frac{N}{K}q\right)}|1\leq i\leq N, 1\leq j\leq N\right\}\right)=\frac{1}{K^2}\sum_{i=1}^{N}\sum_{j=1}^{N}{\varphi_{m^\prime i j}z_{\left(i+\frac{N}{K}p\right)\left(j+\frac{N}{K}q\right)}}, \nonumber\\
        m^\prime=1,2,...,M^2,\quad  p,q=0,1,...,\ K-1.\nonumber
    \end{gather}
    \begin{gather}
    \label{eq:C5}
        \mathcal{W}:\mathbb{R}^{M^2\times K\times K}\ni\hat{\boldsymbol{z}}\mapsto\ \hat{\boldsymbol{z}}^\prime\in\mathbb{R}^{M^2\times K\times K}, \\
        \textup{i.e.},\quad {\hat{z}^\prime}_{mpq}=\mathcal{W}\left(\left\{\hat{z}_{m^\prime pq}\right\}_{m^\prime=1}^{M^2}\right)
        =\sum_{m^\prime=1}^{M^2}{W_{m{m}^\prime}{\hat{z}}_{m^\prime p q}}, \quad     m=1,2,...,M^2,\quad  p,q=0,1,..., K-1. \nonumber
    \end{gather}
    \begin{gather}
    \label{eq:C6}
        \mathcal{T}^{-1}:\mathbb{R}^{M^2\times K\times K}\ni\hat{\boldsymbol{z}}^\prime\mapsto\boldsymbol{z}^\prime\in\mathbb{R}^{\frac{N}{K}\times\frac{N}{K}},\\
        \textup{i.e.},\quad z_{ab}^\prime=\mathcal{T}^{-1}\left(\left\{\hat{z}^\prime_{mpq}|1\leq m\leq M^2, 0\leq p\leq K-1,0\leq q\leq K-1\right\}\right)
        =\sum_{p=0}^{K-1}\sum_{q=0}^{K-1}\sum_{m=1}^{M^2}{\psi_{m\left(a+\frac{K-1-p}{K}N\right)\left(b+\frac{K-1-q}{K}N\right)}{\hat{z}^\prime}_{mpq}},\nonumber \\ a,b=1,2,...,\frac{N}{K}.\nonumber
    \end{gather}
    By combining the operations in Eqs. (\ref{eq:C4}-\ref{eq:C6}) together, we have
    \begin{gather}
    \label{eq:C7}
        \mathcal{T}^{-1}\circ\mathcal{W}\circ\mathcal{T}:\mathbb{R}^{\frac{N\left(2K-1\right)}{K}\times\frac{N\left(2K-1\right)}{K}}\ni\boldsymbol{z}\mapsto\boldsymbol{z}^\prime\in\mathbb{R}^{\frac{N}{K}\times\frac{N}{K}},
    \end{gather}\vspace{-30pt}
    \begin{align*}
        \textup{i.e.},\quad z_{ab}^\prime &= \mathcal{T}^{-1} \circ\mathcal{W} \circ\mathcal{T}\left(\left\{z_{\left(i+\frac{N}{K}p\right)\left(j+\frac{N}{K}q\right)}|1\leq i,j\leq N, 0\leq p,q\leq K-1\right\}\right),\nonumber \\
        &=\frac{1}{K^2}\sum_{p=0}^{K-1}\sum_{q=0}^{K-1}\sum_{m=1}^{M^2}{\psi_{m\left(a+\frac{K-1-p}{K}N\right)\left(b+\frac{K-1-q}{K}N\right)}\sum_{m^\prime=1}^{M^2}{W_{{mm}^\prime}\sum_{i=1}^{N}\sum_{j=1}^{N}{\varphi_{m^\prime i j}z_{\left(i+\frac{N}{K}p\right)\left(j+\frac{N}{K}q\right)}}}}, \quad a,b=1,2,...,\frac{N}{K}.
    \end{align*}
\end{itemize}
These operations compose the architecture of LNO as 
\begin{gather}
    \mathcal{G}_\theta:\mathbb{R}^{4\times\left[2\frac{n\left(K-1\right)N}{K}+A\frac{N}{K}\right]\times\left[2\frac{n\left(K-1\right)N}{K}+B\frac{N}{K}\right]}\ni\boldsymbol{u}_t\mapsto \boldsymbol{u}_{t+\Delta t}\in\mathbb{R}^{4\times A\frac{N}{K}\times B\frac{N}{K}},\quad A,B\in\mathbb{N}^+,
\end{gather}
where
\begin{gather}
    \mathcal{G}_\theta{:=}\mathcal{P}_\textup{projection}\circ\mathcal{B}_n\circ...\circ\mathcal{B}_1\circ\mathcal{P}_\textup{lifting}, \\
    \mathcal{P}_\textup{lifting}{:=}\mathcal{P}_0^{\left(4,40\right)},\nonumber\\    \mathcal{B}_i{:=}\mathcal{A}\circ\left(\mathcal{T}^{-1}\circ\mathcal{W}_i\circ\mathcal{T}+\mathcal{C}_{i4}^{\left(40,40\right)}\circ\mathcal{A}\circ\mathcal{C}_{i3}^{\left(40,40\right)}\circ\mathcal{A}\circ\mathcal{C}_{i2}^{\left(40,40\right)}\circ\mathcal{A}\circ\mathcal{C}_{i1}^{\left(40,40\right)}\right),\quad i=1,2,...,n,\nonumber\\
    \mathcal{P}_\textup{projection}{:=}\mathcal{P}_2^{\left(128,4\right)}\circ\mathcal{A}\circ\mathcal{P}_1^{\left(40,128\right)}.\nonumber
\end{gather}
The subscripts are used to identify operations with independent trainable weights in different layers.

\section{Variable normalization and weight initialization\label{secA5}}
There are worth mentioning techniques in LNO training to elevate the training performance: variable normalization and weight initialization. 
The two techniques are important but could be unfamiliar to emerging interdisciplinary researchers. 
All the experiments and discussions in this paper are trained upon the variable normalization and weight initialization introduced here. 

\begin{figure}[]%
\centering
\includegraphics[width=0.95\textwidth]{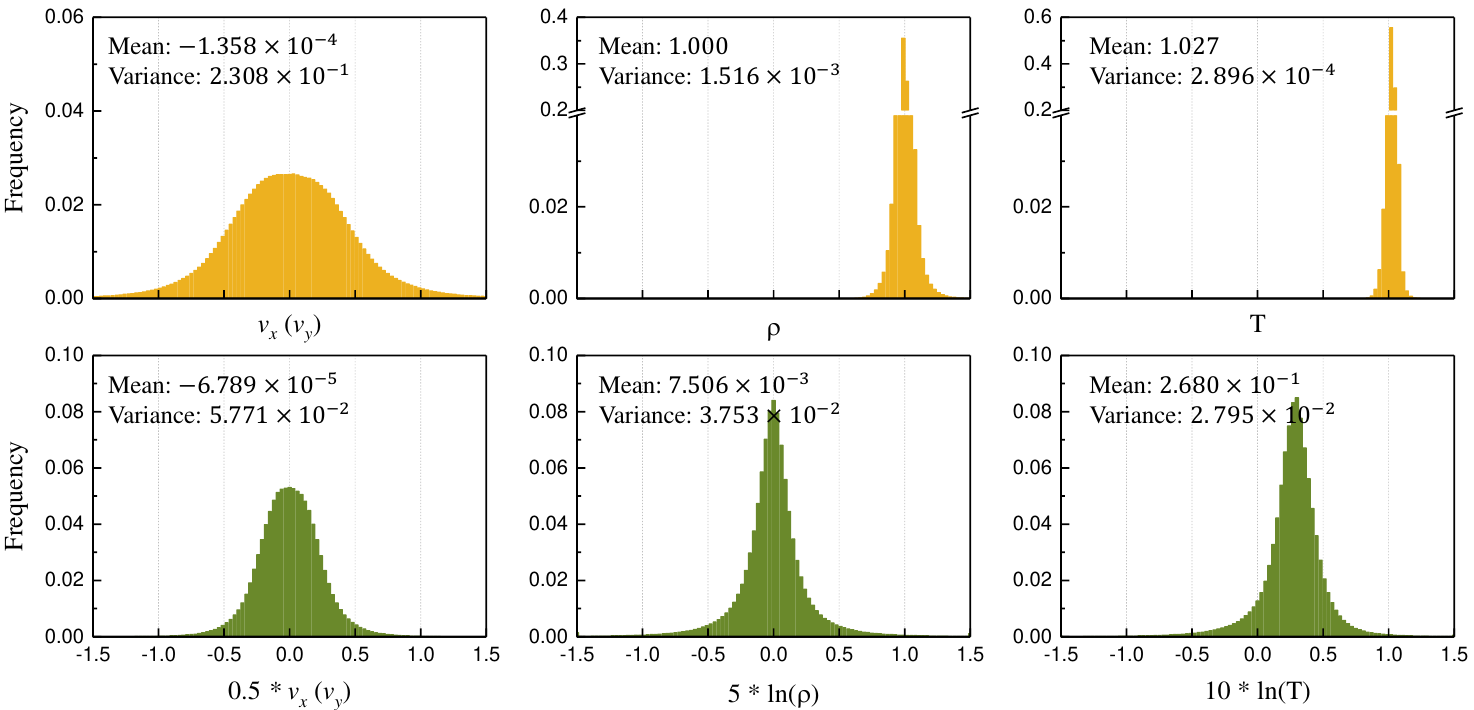}
\caption{Distribution of physical variables $v_x(v_y),\rho,T$ in training samples. 
The first row is the original data, the second row is data after normalization.}\label{fig:datadistribut}
\end{figure}

The normalization of input variable is crucial for making the best of neural networks because the nonlinearity of the activation functions is within a range of small absolute values.
The gradient goes to a constant (0 or 1) when the input absolute value increases, which blocks the convergence of the network in training \cite{Ioffe2015}. 
Herein, the input and output vectors here are with specific physical meanings, thus they are with diverse ranges and distributions. 
$\rho$ and $T$ are defined in $\mathbb{R}^+$ while $v_x$, $v_y$ are in $\mathbb{R}$, and the three different physical fields own diverse distributions, as shown in Figure \ref{fig:datadistribut}.
Simple transformations are applied to get them unified. 
One is to use a natural logarithm transform on $\rho$ and $T$ to change their range from $\mathbb{R}^+$ to $\mathbb{R}$.
The values of $\ln(\rho)$ and $\ln(T)$ are gathering around 0 because their distributions are approximately centered on 1 in the dataset.
The other is to get the variance normalized. 
As presented together in Figure \ref{fig:datadistribut}, the variance is normalized by rescaling the data samples of $\left\{v_x,v_y,\rho,T\right\}$ with the factor of $\left\{0.5,0.5,5,10\right\}$.
In practice, the input is first normalized; then LNO infers the output; finally, the output is rescaled back to the original expression.
The normalization formula and the factors should be kept the same and not be changed for one pre-trained LNO. 

The weight initialization was once thought to be the bottleneck of the neural network to achieve deeper. Xavier et al. \cite{Glorot2010} and He K. et al. \cite{He2015} developed impressive initialization methods to relieve the vanishing or exploding of gradients in the early training stage. 
Following them and initializing the weights properly is necessary for a smooth training process of LNO. 
Ideally, the initialization should ensure that the variance of output is the same as that of input, thus here we go through a variance analysis to figure out the best initialization for weights of LNO.

Prior to the analysis, we put several basic formulas in statistics here as a preliminary. 
For two independent random variables $A$ and $B$, the variance of $A+B$ and $AB$ are:
\begin{equation}
    \mathrm{Var}\left(A+B\right)=\mathrm{Var}\left(A\right)+\mathrm{Var}\left(B\right),
\end{equation}
\begin{equation}
\mathrm{Var}\left(AB\right)=\mathrm{Var}\left(A\right)\mathrm{Var}\left(B\right)+\mathrm{Var}\left(A\right)\mathrm{E}^2\left(B\right)+\mathrm{E}^2\left(A\right)\mathrm{Var}\left(B\right).
\end{equation}
When $\mathrm{E}\left(A\right)=0$, $\mathrm{E}\left(B\right)=0$, we have $\mathrm{Var}\left(AB\right)=\mathrm{Var}\left(A\right)\mathrm{Var}\left(B\right)$.

Then we derive the variation of the output regarding the elements in the input tensor as i.i.d. with zero mean and variance of $\sigma^2$.
\begin{itemize}
    \item For point-wise operations, consider the variance of Eq. (\ref{eq:C2}) as
    \begin{equation}
        \mathrm{Var}(z_l^\prime)=\sum_{m=1}^{c_\textup{in}}\mathrm{Var}(W_{ml})\mathrm{Var}(z_m).
    \end{equation}
    With the default initialization that $W_{ml}\sim U\left(-\sqrt{\frac{1}{c_{in}}},\sqrt{\frac{1}{c_{in}}}\right)$, $\mathrm{Var}\left(W_{ml}\right)=\frac{1}{3c_\textup{in}}$. $\mathrm{Var}\left(z_m\right)=\sigma^2$.
    So, $\mathrm{Var}\left(z_l^\prime\right)=\frac{1}{3}\sigma^2$. 
    Hence, by taking action of $\times\sqrt3$ on $W_{ml}$, the variance of the output keeps the same with the input.
    \item For convolutions, consider the variance of Eq. (\ref{eq:C3}) as
    \begin{equation}
        \operatorname{Var}\left(z_{l x y}^{\prime}\right)=\sum_{m=1}^{c_\textup{in}} \sum_{i=1}^k \sum_{j=1}^k \mathrm{Var}\left(W_{m l i j}\right) \mathrm{Var}\left(z_{m(x+i)(y+j)}\right).
    \end{equation}
    With the default initialization that $W_{mlij}\sim U\left(-\sqrt{\frac{1}{c_\textup{in}k^2}},\sqrt{\frac{1}{c_\textup{in}k^2}}\right)$, $\mathrm{Var}\left(W_{mlij}\right)=\frac{1}{3c_\textup{in}k^2}$. $\mathrm{Var}\left(z_{m\left(x+i\right)\left(y+j\right)}\right)=\sigma^2$. 
    So $\mathrm{Var}\left(z_{lxy}^\prime\right)=\frac{1}{3}\sigma^2$.
    Hence, by taking action of $\times\sqrt3$ on $W_{mlij}$, variance of the output keeps the same with the input.
    \item 	For operations in the spectral path, let $I=i+\frac{N}{K}p$,\ $J=j+\frac{N}{K}q$ and rearrange the summations in Eqs. (\ref{eq:C4}-\ref{eq:C6}) as
    \begin{equation}
        z_{a b}^{\prime}=\frac{1}{K^2} \sum_{I=1}^{\frac{N(2 K-1)}{K}} \sum_{J=1}^{\frac{N(2 K-1)}{K}} \sum_{m=1}^{M^2} \sum_{m^{\prime}=1}^{M^2} \sum_{p=0}^{K-1} \sum_{q=0}^{K-1} \psi_{m\left(a+\frac{K-1-p}{K} N\right)\left(b+\frac{K-1-q}{K} N\right)} \varphi_{m^{\prime}\left(I-\frac{N}{K} p\right)\left(J-\frac{N}{K} q\right)} W_{m m^{\prime}} z_{I J}.
        \label{eq:D5}
    \end{equation}
    Consider the variance of Eq. (\ref{eq:D5}) and with $W_{{mm}^\prime}\sim U\left(-\sqrt{\frac{1}{c_\textup{in}}},\sqrt{\frac{1}{c_\textup{in}}}\right)$, $\mathrm{Var}\left(W_{{mm}^\prime}\right)=\frac{1}{3M^2} \mathrm{Var}\left(z_{IJ}\right)=\sigma^2$, we derive
    \begin{equation}
        \begin{aligned}
            & \mathrm{Var}\left(z_{a b}^{\prime}\right) \\
& =\frac{1}{K^4} \sum_{I=1}^{\frac{N(2 K-1)}{K}}  \sum_{J=1}^{\frac{N(2 K-1)}{K}} \sum_{m=1}^{M^2} \sum_{m^{\prime}=1}^{M^2}\left(\sum_{p=0}^{K-1} \sum_{q=0}^{K-1} \psi_{\left(a+\frac{K-1-p}{K} N\right)\left(b+\frac{K-1-q}{K} N\right) m} \varphi_{m^{\prime}\left(I-\frac{N}{K} p\right)\left(J-\frac{N}{K} q\right)}\right)^2 \mathrm{Var}\left(W_{m m^{\prime}}\right) \mathrm{Var}\left(z_{I J}\right)\\
& =\frac{\sigma^2}{3 M^2 K^4} \sum_{I=1}^{\frac{N(2 K-1)}{K}} \sum_{J=1}^{\frac{N(2K-1)}{N}} \sum_{m=1}^{M^2} \sum_{m^{\prime}=1}^{M^2}\left(\sum_{p=0}^{K-1} \sum_{q=0}^{K-1} \psi_{\left(a+\frac{K-1-p}{K} N\right)\left(b+\frac{K-1-q}{K} N\right) m} \varphi_{m^{\prime}\left(I-\frac{N}{K} p\right)\left(J-\frac{N}{K} q\right)}\right)^2 \\
& \stackrel{\text { average on } a, b}{\longrightarrow} \frac{\sigma^2}{\Theta_{N, K, M}}
        \end{aligned}
    \end{equation}
    where $\mathrm{\Theta}_{N,K,M}$ is the newly defined initialization factor of the spectral path of LNO. 
    $\mathrm{\Theta}_{N,K,M}$ is a constant value related to $N,K,M$ of LNO and the Legendre kernel $\boldsymbol{\psi}$ and $\boldsymbol{\varphi}$.
    We give some of $\mathrm{\Theta}_{N,K,M}$ in Table \ref{tab:normalization} for convenient usage.
    Hence, by taking action of $\times\sqrt{\mathrm{\Theta}_{N,K,M}}$ on $W_{{mm}^\prime}$, variance of the output keeps the same with the input. 
    We write ${\bar{W}}_{{mm}^\prime}=\sqrt{\mathrm{\Theta}_{N,K,M}}W_{{mm}^\prime}$, thus it has $\mathrm{Var}\left({\bar{W}}_{{mm}^\prime}\right)=\frac{\mathrm{\Theta}_{N,K,M}}{3M^2}$.
\end{itemize}

Table \ref{tab:initialization} summarizes the changes for weight initialization from the default value in the deep learning framework PyTorch \cite{Paszke2019} version 1.6.0.

\begin{table}[]
\centering
\caption{Specific values of $\mathrm{\Theta}_{N,K,M}$.\label{tab:normalization}}
\footnotesize
\renewcommand{\arraystretch}{0.8}
\begin{tabular*}{\textwidth}{@{\extracolsep{\fill}}ccccccccc}
\toprule
\multirow{2}{*}{Variant $N$} & $\left(N,K,M\right)$      & (8,2,6)  & (10,2,6) & (12,2,6) & (14,2,6) & (16,2,6)  & (18,2,6)  & (20,2,6) \\
\cline{2-9}
                             & $\mathrm{\Theta}_{N,K,M}$ & 7.7491   & 10.1948  & 11.8781  & 18.5122  & 25.5560   & 31.9326   & 37.6120  \\
                             \midrule
\multirow{2}{*}{Variant $K$} & $\left(N,K,M\right)$      & (12,2,6) & (12,3,6) & (12,4,6) & (12,6,6) & (12,12,6) &           &          \\
\cline{2-9}
                             & $\mathrm{\Theta}_{N,K,M}$ & 11.8781  & 35.4591  & 68.4968  & 176.1236 & 358.5074  &           &          \\
                             \midrule
\multirow{2}{*}{Variant $M$} & $\left(N,K,M\right)$      & (12,2,2) & (12,2,4) & (12,2,6) & (12,2,8) & (12,2,10) & (12,2,12) &          \\
\cline{2-9}
                             & $\mathrm{\Theta}_{N,K,M}$ & 252.3647 & 41.3953  & 11.8781  & 5.1523   & 3.2647    & 2.6566    &        \\
                             \bottomrule
\end{tabular*}
\end{table}

\begin{table}[]
\centering
\caption{Weight initialization for operations in LNO.\label{tab:initialization}}
\footnotesize
\renewcommand{\arraystretch}{0.8}
\begin{tabular*}{\textwidth}{@{\extracolsep{\fill}}cccc}
\toprule
Operations                                                         & Default initialization in Pytorch 1.6.0                                     & Variance change                   & Action on weights                                \\
\midrule
Pointwise ($\mathcal{P}$)                                         & $U\left(-\sqrt{\frac{1}{c_\textup{in}}},\sqrt{\frac{1}{c_\textup{in}}}\right)$       & $\frac{1}{3}\sigma^2$             & $\times\sqrt3$                                   \\
Convolutions ($\mathcal{C}$)                                       & $U\left(-\sqrt{\frac{1}{c_\textup{in}k^{n_d}}},\sqrt{\frac{1}{c_\textup{in}k^{n_d}}}\right)$ & $\frac{1}{3}\sigma^2$             & $\times\sqrt3$                                   \\
Spectral path ($\mathcal{T}^{-1}\circ\mathcal{W}\circ\mathcal{T}$) & $U\left(-\sqrt{\frac{1}{c_\textup{in}}},\sqrt{\frac{1}{c_\textup{in}}}\right)$       & $\mathrm{\Theta}_{N,K,M}\sigma^2$ & $\times\sqrt{\frac{1}{\mathrm{\Theta}_{N,K,M}}}$ \\
GELU activation ($\mathcal{A}$)                                    & \textbackslash{}                                                       & $\frac{1}{2}\sigma^2$             & \makecell[c]{$\times\sqrt2$ (applied to weights \\ of the nearest learnable layer)}                                  \\
\bottomrule
\end{tabular*}
\end{table}

\begin{figure}[]%
\centering
\includegraphics[width=\textwidth]{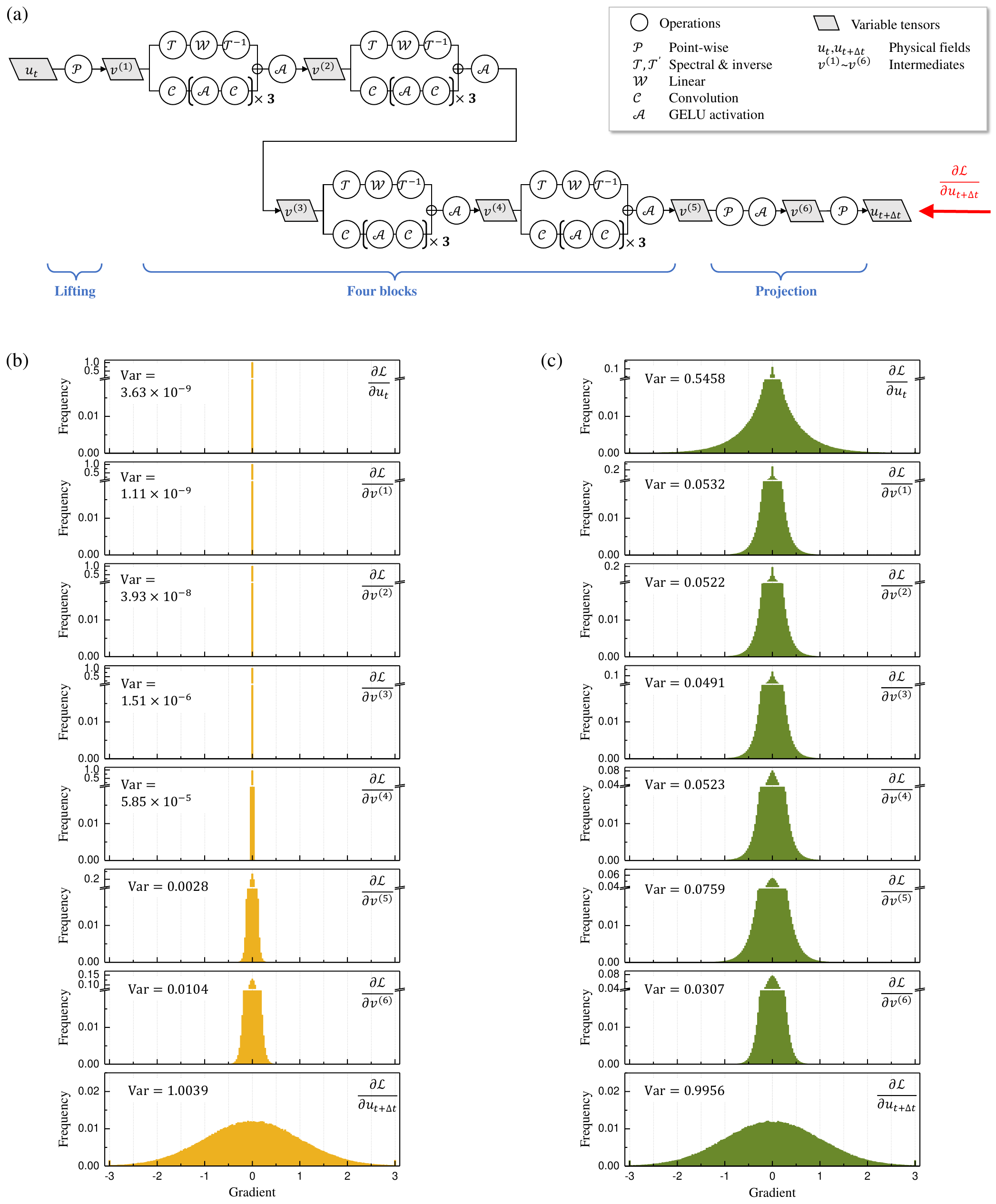}
\caption{The architecture of LNO (subfigure a) and the corresponding distribution of gradient during backpropagation without (subfigure b) and with the normalized initialization (subfigure c).}\label{fig:gradientdistribut}
\end{figure}

Several results are presented to show the necessity of variable normalization and weight initialization. 
In Figure \ref{fig:gradientdistribut}, we conduct a back-propagation test on LNO with initialized weights. 
We let the gradient at the output layer $\frac{\partial \mathcal{L}}{\partial \boldsymbol{u}_{t+\Delta t}}\sim N(0,1)$, then back propagate the gradient to all variable tensors. 
The distributions of the gradient propagate to the intermediates $\frac{\partial\mathcal{L}}{\partial\boldsymbol{v}^{(i)}}$ ($i=1\sim 6$) and the input tensor $\frac{\partial\mathcal{L}}{\partial\boldsymbol{u}_t}$ are summarized in Figure \ref{fig:gradientdistribut} separately for two LNO instances. 
One is initialized following the default setting of PytTorch, and the other is initialized based on our variance analysis. 
It is seen that the variance of gradient for the default initialization declines to ${10}^{-9}$ which makes the LNO training really hard or even unavailable.
In contrast, the variance holds well with our weight initialization.
Results in Table \ref{tab:ablation} present how the present variable normalization and weight initialization affect the performance of the trained LNOs. 
These tests are conducted on LNO with $(n,N,K,M)=(4,12,2,6)$ to learn the fluid dynamics with $(Re,Ma,\Delta t)=(100,0.2,0.05)$. 
It is seen that the techniques introduced here make LNOs more stable to converge in training, and reduce both the mean $L_2$ error and the dispersity of results. 
It suggests the variable normalization and weight initialization are necessary for LNO training.

\begin{table}[]
\centering
\caption{Mean $L_2$ of LNO trained with different techniques. $e_t^v=\infty$ means LNO is failed to train.\label{tab:ablation}}
\footnotesize
\renewcommand{\arraystretch}{0.8}
\begin{tabular*}{\textwidth}{@{\extracolsep{\fill}}cccc}
\toprule
\multirow{2}{*}{Weight initialization} & \multicolumn{2}{c}{Variable normalization} & \multirow{2}{*}{Error $e_t^v$ at $t=5$ (10 runs)} \\
\cline{2-3}
                                       & Range             & Variance            &                                                   \\
                                       \midrule
$\times$                                    &   $\times$               &     $\times$          & $\infty\times10$                                  \\
$\surd$                                &   $\times$              &     $\times$         & $0.293\pm0.104$ with $\infty\times1$                  \\
 $\times$                               & $\surd$              &     $\times$         & $0.118\pm0.056$ with $\infty\times1$                  \\
   $\times$                              & $\surd$              & $\surd$             & $0.102\pm0.029$                                     \\
$\surd$                                & $\surd$              & $\surd$             & $0.076\pm0.017$     \\
\bottomrule
\end{tabular*}
\end{table}








\end{document}